\documentclass[journal]{IEEEtran}

\usepackage{xcolor,soul,framed} 

\colorlet{shadecolor}{yellow}
\usepackage[pdftex]{graphicx}
\graphicspath{{../pdf/}{../jpeg/}}
\DeclareGraphicsExtensions{.pdf,.jpeg,.png}

\usepackage[cmex10]{amsmath}
\usepackage{amsthm}

\usepackage{array}
\usepackage{mdwmath}
\usepackage{mdwtab}
\usepackage{eqparbox}

\usepackage{amssymb}
\usepackage{amsfonts}
\usepackage{mathrsfs}
\usepackage{svg}
\usepackage{hyperref}
\usepackage{cite}

\usepackage{tikz}
\usepackage{pgfplotstable}
\usepackage{rotate}

\definecolor{rubblue}{cmyk}{1,0.5,0,0.6}
\definecolor{rubgreen}{cmyk}{0.5,0,1,0}
\definecolor{rubgray}{cmyk}{0.03,0.03,0.03,0.1}

\usepgfplotslibrary{units}

\usetikzlibrary{%
patterns,%
calc,%
fit,%
arrows,%
plotmarks,%
shadows,%
chains,%
shapes%
}

\usepackage{pgfplots}
\usetikzlibrary{calc}
\usetikzlibrary{positioning}
\usepackage{siunitx}

\tikzset{>=latex'} 
\tikzstyle{every picture}+=[remember picture] 
\pgfdeclarelayer{background}
\pgfdeclarelayer{foreground}
\pgfsetlayers{background,main,foreground}

\tikzstyle{blueblock}=[draw=rubblue, rectangle, thick, drop shadow, minimum width=20mm, minimum height=8mm,fill=rubblue!20, text width=20mm, text centered]
\tikzstyle{bluebox}=[draw=rubblue, rectangle, thick, drop shadow, minimum width=8mm, minimum height=8mm,fill=rubblue!20, text width=8mm, text centered]
\tikzstyle{greenblock}=[draw=rubgreen, rectangle, thick, drop shadow, minimum width=20mm, minimum height=8mm,fill=rubgreen!20, text width=20mm, text centered]
\tikzstyle{dot} = [draw, circle, minimum size=0.2pt,scale=0.3,fill=black,black]
\tikzstyle{smalldot} = [draw, circle, minimum size=0.1pt,scale=0.2,fill=black,black]
\tikzstyle{reddot}  =[draw,circle,minimum size=0.2pt,scale=0.8,fill=red,thin]
\tikzstyle{greendot}  =[draw,circle,minimum size=0.2pt,scale=0.8,fill=Green,thin]
\tikzstyle{bluedot}  =[draw,circle,minimum size=0.2pt,scale=0.8,fill=blue,thin]
\tikzstyle{whitedot}=[draw,circle,minimum size=0.2pt,scale=0.8,fill=white,thin]
\tikzstyle{blackdot} = [draw, circle, minimum size=0.2pt,scale=0.7,fill=black,black]
\tikzstyle{sum} = [drop shadow, draw=rubblue, thick, fill=rubblue!20, circle]
\tikzstyle{relay} = [blueblock, minimum width=5mm, minimum height=20mm, text width=5mm, rounded corners=2pt]
\tikzstyle{relay2} = [blueblock, minimum width=5mm, minimum height=15mm, text width=5mm, rounded corners=2pt]
\tikzstyle{relay3} = [blueblock, minimum width=5mm, minimum height=25mm, text width=5mm, rounded corners=2pt]
\tikzstyle{relay4} = [blueblock, minimum width=5mm, minimum height=10mm, text width=5mm, rounded corners=2pt]
\tikzstyle{relay5} = [blueblock, minimum width=5mm, minimum height=50mm, text width=5mm, rounded corners=2pt]
\tikzstyle{relay6} = [blueblock, minimum width=5mm, minimum height=5mm, text width=5mm, rounded corners=2pt]
\tikzstyle{circgreen} = [draw, circle, inner sep=2pt, fill=rubgreen, drop shadow, thick]
\tikzstyle{circwhite} = [draw, circle, inner sep=2pt, fill=white, drop shadow, thick]
\tikzstyle{circdashed} = [draw, dashed, circle, inner sep=2pt, fill=rubgray, drop shadow, thick]
\tikzstyle{vertbox} = [rectangle, draw=rubblue, thick, rotate=90, text centered, minimum width=16.5mm, minimum height=8mm, text width=16.5mm, inner sep=0pt, fill=rubblue!20, drop shadow]
\tikzstyle{vertboxb} = [rectangle, draw=rubblue, thick, rotate=90, text centered, minimum width=16.5mm, minimum height=8mm, text width=16.5mm, fill=rubblue!20, drop shadow]
\tikzstyle{vertboxshort} = [rectangle, draw=rubblue, thick, rotate=90, text centered, minimum width=10mm, minimum height=8mm, text width=10mm, inner sep=0pt, fill=rubblue!20, drop shadow]
\tikzstyle{smalldotgreen} = [draw=rubgreen, circle, minimum size=0.2pt,scale=0.8,fill=rubgreen!20]
\tikzstyle{antenna} = [regular polygon, regular polygon sides=3, draw, shape border rotate=180, minimum size=0.2pt, scale=0.3]

\tikzstyle{poly} = [regular polygon, regular polygon sides=6, shape aspect=0.5, minimum width=1.5cm, minimum height=0.35cm, draw, dashed]

\definecolor{cff9e00}{RGB}{255,158,0}
\definecolor{c4fff00}{RGB}{79,255,0}
\definecolor{cff0012}{RGB}{255,0,18}
\definecolor{c00c5ff}{RGB}{0,197,255}
\definecolor{c046f00}{RGB}{4,111,0}
\definecolor{c004b9d}{RGB}{0,75,157}

\newlength{\mylen}
\usetikzlibrary{petri}
\usetikzlibrary{shapes}
\usetikzlibrary{positioning,arrows,patterns}
\usepackage{scalefnt}
\usetikzlibrary{calc,decorations.markings}
\pgfplotsset{compat=1.10}
\usetikzlibrary{arrows.meta}
\usepackage[columnwise,switch,mathlines]{lineno} 

\usepackage[linesnumbered,ruled,vlined]{algorithm2e}
\definecolor{darkgreen}{rgb}{0.0, 0.7, 0.0} 
\definecolor{mypurple}{RGB}{160,32,240}
\definecolor{mydarkred}{RGB}{255,102,102}
\definecolor{mylightblue}{RGB}{0,102,204}
\definecolor{mydarkgreen}{RGB}{0,100,0}
\usepackage{cancel}  

\newtheoremstyle{nonitalic}
  {3pt} 
  {3pt} 
  {\upshape} 
  {} 
  {\bfseries} 
  {.} 
  {.5em} 
  {} 

\theoremstyle{nonitalic} 
\newtheorem{definition}{Definition}
\newtheorem{theorem}{Theorem}
\newtheorem{lemma}{Lemma}
\newtheorem{assumption}{Assumption}
\newtheorem{corollary}{Corollary}

\begin{document}
\bstctlcite{IEEEexample:BSTcontrol}
    \title{UAV-assisted Unbiased Hierarchical Federated Learning: Performance and Convergence Analysis}
  \author{Ruslan Zhagypar,~\IEEEmembership{Student Member,~IEEE,}
        Nour~Kouzayha,~\IEEEmembership{Member,~IEEE,}
		Hesham~ElSawy,~\IEEEmembership{Senior Member,~IEEE,}
		Hayssam~Dahrouj,~\IEEEmembership{Senior Member,~IEEE,}
		and~Tareq Y. Al-Naffouri,~\IEEEmembership{Senior Member,~IEEE}
  
}


\maketitle

\begin{abstract}
The development of the sixth generation (6G) of wireless networks is bound to streamline the transition of computation and learning towards the edge of the network. Hierarchical federated learning (HFL) becomes, therefore, a key paradigm to distribute learning across edge devices to reach global intelligence. In HFL, each edge device trains a local model using its respective data and transmits the updated model parameters to an edge server for local aggregation. The edge server, then, transmits the locally aggregated parameters to a central server for global model aggregation. The unreliability of communication channels at the edge and backhaul links, however, remains a bottleneck in assessing the true benefit of HFL-empowered systems. To this end, this paper proposes an unbiased HFL algorithm for unmanned aerial vehicle (UAV)-assisted wireless networks that counteracts the impact of unreliable channels by adjusting the update weights during local and global aggregations at UAVs and terrestrial base stations (BS), respectively. To best characterize the unreliability of the channels involved in HFL, we adopt tools from stochastic geometry to determine the success probabilities of the local and global model parameter transmissions. Accounting for such metrics in the proposed HFL algorithm aims at removing the bias towards devices with better channel conditions in the context of the considered UAV-assisted network.. The paper further examines the theoretical convergence guarantee of the proposed unbiased UAV-assisted HFL algorithm under adverse channel conditions. One of the developed approach's additional benefits is that it allows for optimizing and designing the system parameters, e.g., the number of UAVs and their corresponding heights. The paper results particularly highlight the effectiveness of the proposed unbiased HFL scheme as compared to conventional FL and HFL algorithms.

\end{abstract}
\begin{IEEEkeywords}
Federated learning, Hierarchical FL, HFL, stochastic geometry, unbiased aggregation, UAVs.
\end{IEEEkeywords}

%
\IEEEpeerreviewmaketitle


\section{Introduction}


The emergence of the sixth generation (6G) of wireless networks marks a significant leap in wireless communication technologies, introducing a new era of large-scale smart connectivity~\cite{asghar2022evolution}. Mobile edge computing (MEC) becomes a pivotal enabling technology, capitalizing on enhanced computational capabilities of edge devices and their access to vast data resources for the development of machine learning (ML)-based applications, including content caching~\cite{zhou2023edge}. Among the multiple ML paradigms, federated learning (FL) has emerged as a key solution for enabling edge devices to learn collaboratively while maintaining the local data privacy~\cite{nguyen2021federated}. FL fundamental operation relies on the principle of training models locally on devices such as smartphones, IoT devices, or edge servers and then aggregating the local updates at the central base station (BS) to train a global model~\cite{lim2020federated}. The main advantages of using FL in improving edge intelligence include 1) increased data privacy thanks to sharing the model weights/gradients instead of sensitive local data, 2) reduced latency and bandwidth requirements as a result of transmitting only the model weights that carry a substantially lower amount of information than locally collected data, 3) computation offloading by enabling devices to perform computational tasks and ML algorithms locally. This paper addresses the benefit of one particular FL variant, namely the hierarchical FL (HFL), as a means to counteract the impact of channel unreliability, often faced in the conventional FL operation.

More specifically, due to the nature of wireless communication, the communication quality between the edge devices and the server in conventional FL is subject to environmental impacts, losses, and fading~\cite{zhu2020toward}. Thus, it becomes challenging for edge devices to share their local models with the server. To address this issue, HFL emerge as an pertinent alternative that allows the local devices to communicate with nearby edge servers, which in turn deliver the locally aggregated updates to the main server~\cite{abad2020hierarchical}. In HFL, local devices are divided into multiple clusters. In each cluster, the device updates its local models independently. The edge server then aggregates the local models, and uploads them to the main server for global model aggregation. Compared to conventional FL, HFL provides enhanced accuracy and reduced latency by reducing the communication overhead and allowing more devices to participate in the learning~\cite{abad2020hierarchical}.

Nevertheless, due to the multipath fading and shadowing components immanent in the wireless channel, traditional ground BSs, acting as relays in HFL, may not guarantee good channel gains for all devices, thereby degrading communication performance~\cite{al2023edge}. A prevalent strategy for enhancing wireless network communication involves leveraging unmanned aerial vehicles (UAVs) due to their adaptability, straightforward installation, minimal maintenance costs, versatility, and relatively low operational expenses~\cite{almaameri2023overview}. In the context of HFL, UAVs can act as relays between edge devices and the central server to provide Line-of-Sight (LoS) connections and reduce the transmission distance to edge devices~\cite{bie2022deployment,li2021energy}. Furthermore, UAVs are exploited to provide computational offloading to BSs, given the sheer volume of edge devices and their limited resources~\cite{zhang2020enabling,zhan2020incentive}.

Despite the advantages provided by UAV-assisted HFL, its performance remains a strong function of several operational challenges~\cite{bakambekova2024interplay}. Specifically, the communication of the local and global models at the edge and backhaul channels between the devices, UAVs, and BS suffers from increased interference and path loss, which causes certain devices and UAVs to fail in successfully delivering the model updates~\cite{zhagypar2023characterization}. This creates a bias towards the devices and UAVs with better channel conditions during local and global aggregation, which drives the model farther from the global optimum. Although noticeable efforts have been devoted to developing efficient resource scheduling and allocation schemes to enhance the transmission of local and global models in UAV-assisted HFL~\cite{yang2021privacy}, the communication bottleneck persists, causing a global bias problem that requires further investigation to guarantee fair and reliable UAV-assisted HFL algorithms at the network edge~\cite{sirohi2023federated}. In this work, we aim to close such a performance gap by proposing a UAV-assisted unbiased HFL and studying its performance in terms of accuracy, convergence, and latency.

\subsection{Related Work}
The successful combination of FL and UAV-assisted wireless networks within the context of HFL raised great interest in the research community. In~\cite{wang2023uav}, an innovative two-tier HFL approach harnessing a swarm of UAVs is introduced. The authors in~\cite{wang2023uav} design a comprehensive joint optimization problem, encompassing device-UAV pairing, time constraints in HFL, and the number of local training iterations for devices. In~\cite{chen2023relay}, the authors study a federated edge learning (FEEL) network with relaying UAVs, addressing constraints related to latency and bandwidth. 
Despite providing convergence analysis, the conducted study~\cite{chen2023relay} does not account for the hierarchical structure intrinsic to the two-tier FL framework. The integration of UAVs within the multi-tier HFL (MT-HFL) framework, as explored in~\cite{farajzadeh2023self}, extends intelligent data processing from ground-level edge devices to aerial platforms, fostering decentralized decision-making and collaborative learning in Internet of things (IoT) networks. While the resultsin~\cite{farajzadeh2023self} demonstrate the performance enhancement of the proposed aerial HFL over terrestrial HFL, the challenges associated with channel unreliability are earmarked for future research directions. 

A UAV swarm architecture for personalized HFL is proposed in~\cite{wang2022uav}, and the UAVs act as network orchestrators that manages other UAVs within the swarm, which in turn perform local training, and coordinating relays. The authors in~\cite{wang2022uav} rely on reinforcement learning to optimize the swarm trajectory and learning duration design. A similar study on tuning the UAV trajectory and the global aggregation coefficient through mean squared error minimization is proposed in~\cite{zhong2022uav} for UAV-assisted over-the-air HFL. Furthermore, the authors in~\cite{masood2021content} investigate the performance of high altitude platform (HAP)-assisted multi-UAV networks supported with HFL for content caching. Hence, the significant potential of UAV-assisted wireless networks empowered by HFL becomes evident in terms of system performance, scalability, and the array of possible applications.



Most of the existing works in HFL, including studies in~\cite{you2023hierarchical,luo2020hfel,he2023three}, primarily focus on addressing challenges related to server resource scarcity and optimizing bandwidth allocation. The investigation conducted in~\cite{tursunboev2022hierarchical} delves into the performance of HFL with UAVs serving as local learning devices, operating under non-independent and identically distributed (non-IID) data conditions. However, the analysis in~\cite{tursunboev2022hierarchical} lacks the essential theoretical proof of HFL convergence and does not reflect on the impact of the wireless channel unreliability. 
In~\cite{chen2022federated}, the authors propose an HFL framework for multi-hop wireless networks by considering heterogeneous communication and computing resources. The considered strategy is able to accelerate the training process by designing a joint routing and spectrum allocation scheme. Authors of~\cite{tang2023multi} propose a novel algorithm for HFL to jointly decide the device sampling, CPU-cycle frequencies for local training, and bandwidths for intermediate result transmissions. This system promises to achieve optimal training latency and minimize energy consumption across diverse system settings. In order to achieve similar goals, authors of~\cite{hui2022quality} apply principles from game theory to improve the matching mechanisms within HFL, resulting in improved collaboration efficiency and more effective resource allocation. Another algorithm known as FogL is proposed in~\cite{hosseinalipour2022multi}, adopting a tree-based federated network structure. FogL utilizes multi-level and multi-stage tree aggregation techniques~\cite{cao2021layered,Yang_2021} to alleviate the possibility of system overload and enhance the efficiency of resource allocation. Such works, however, tend to overlook the issue of model bias, mainly caused by unreliable communication, at the edge and central servers. 

On the other hand, in conventional FL, notable efforts are spent on addressing the model bias during aggregation. The model bias, which skews the model from the global optimum, can be caused by the unreliability of the communication channels, data imbalance, and different computational capabilities of devices. Authors of~\cite{zhang2020fairfl} introduce FairFL, which mitigates bias due to data imbalances using reinforcement learning. A study in~\cite{wang2021federated} applies contract theory to fairly manage the involvement of UAVs in FL. Similar to others, these methods lack convergence guarantees. Some papers resolve FL communication bias by following an approach that relies on the use of stochastic geometry, firstly proposed in~\cite{salehi2021federated} and later extended to UAV-assisted networks in~\cite{zhagypar2023characterization}. The work on mitigating the bias for collaborative FL is presented in~\cite{yemini2022semi}, where the central server fetches the updates from edge devices through other devices, which in turn collaborate among themselves to ensure that redundant updates are mitigated. In~\cite{mou2021optimized}, the authors propose the FedBGVS algorithm that aims to reduce the severity of class bias by integrating a balanced global validation set. This approach refines the model aggregation algorithm through the utilization of the balanced global validation score. To sum up, although some papers aim to address the global bias problem in conventional FL schemes, there is no study that targets solving this issue in UAV-assisted HFL. This paper, therefore, aims to close this gap by proposing an unbiased UAV-assisted HFL and demonstrating its convergence using stochastic geometry tools.

\subsection{Contributions and Organization}
Different from existing studies, this paper proposes an HFL framework that is suited for UAV-assisted wireless networks. The network comprises global and local aggregators, with the BS serving as the global aggregators and UAVs serving as the local aggregators. Meanwhile, the edge devices undertake local training within this network architecture.
The core idea is to use stochastic geometry tools to characterize the edge and backhaul channels in order to derive upload and download success probabilities of local and global model transmissions for devices and UAVs. These probabilities are used to tune the update weights during the HFL local and global model aggregation. Furthermore, we study the performance of the proposed system and derive its theoretical convergence bound. The proposed unbiasing scheme for UAV-assisted HFL is independent of the resource scheduling techniques. Notably, this unbiasing method is adaptable and can seamlessly integrate with various scheduling approaches, as initially demonstrated in~\cite{salehi2021federated} for conventional FL. To the best of the authors' knowledge, this is the first work that addresses the problem of global biasing due to communication unreliability through the use of stochastic geometry tools in UAV-assisted HFL. In particular, the main contributions of the paper can be summarized as follows:


\begin{itemize}
    \item We develop a comprehensive analytical framework based on tools from stochastic geometry to characterize the edge and backhaul wireless channels of a UAV-assisted wireless network and derive success probabilities of transmissions over these channels in the context of UAV-assisted HFL. The derived expressions are used to tune the update weights in the HFL algorithm, allowing for the suppression of the bias during UAV aggregation, which encourages equal contributions among edge devices. The same unbiasing method is also used during the BS aggregation, leading to balanced updates from all UAVs.
        
    
    \item We derive the theoretical convergence bound of the proposed UAV-assisted unbiased HFL algorithm, reflecting the impact of the channel unreliability and highlighting that a reliable wireless channel holds equal significance to maintaining low upward and downward divergence criteria. The derived bound reflects the impact of the communication channels through the expressions of the derived transmission success probabilities. Furthermore, we study the latency of the proposed UAV-assisted unbiased HFL accounting for the times elapsed during both local and global aggregations.
    
    
    \item We compare the performance of the proposed algorithm with conventional FL and HFL algorithms using numerical simulations under different operational conditions and highlight the superiority of the proposed unbiased method. The developed framework allows for examining the impact of changing system parameters, such as UAV height and number, on the performance of the proposed HFL algorithm in terms of accuracy and latency. The obtained results provide several design insights on the potential of the proposed algorithm in real-world 6G applications.
    
\end{itemize}

The remainder of this paper is organized as follows. Section~\ref{sec:system_model} introduces the system model, where we discuss the network and channel models and association policies. In Section~\ref{sec:algorithm}, we present the details of the proposed algorithm. The analytical model used to derive the success probabilities for the proposed HFL is detailed in Section~\ref{sec:stochastic}. In Section~\ref{sec:performance}, we examine the performance of the proposed algorithm in terms of theoretical convergence and latency. We present numerical simulations to evaluate the performance of the proposed algorithm in Section~\ref{sec:results}. Finally, Section~\ref{sec:conclusions} provides the conclusion of the paper.

\begin{figure}
    \centering
    \includegraphics[viewport=100 200 500 400, trim=90 650 270 70, width=2.6in]{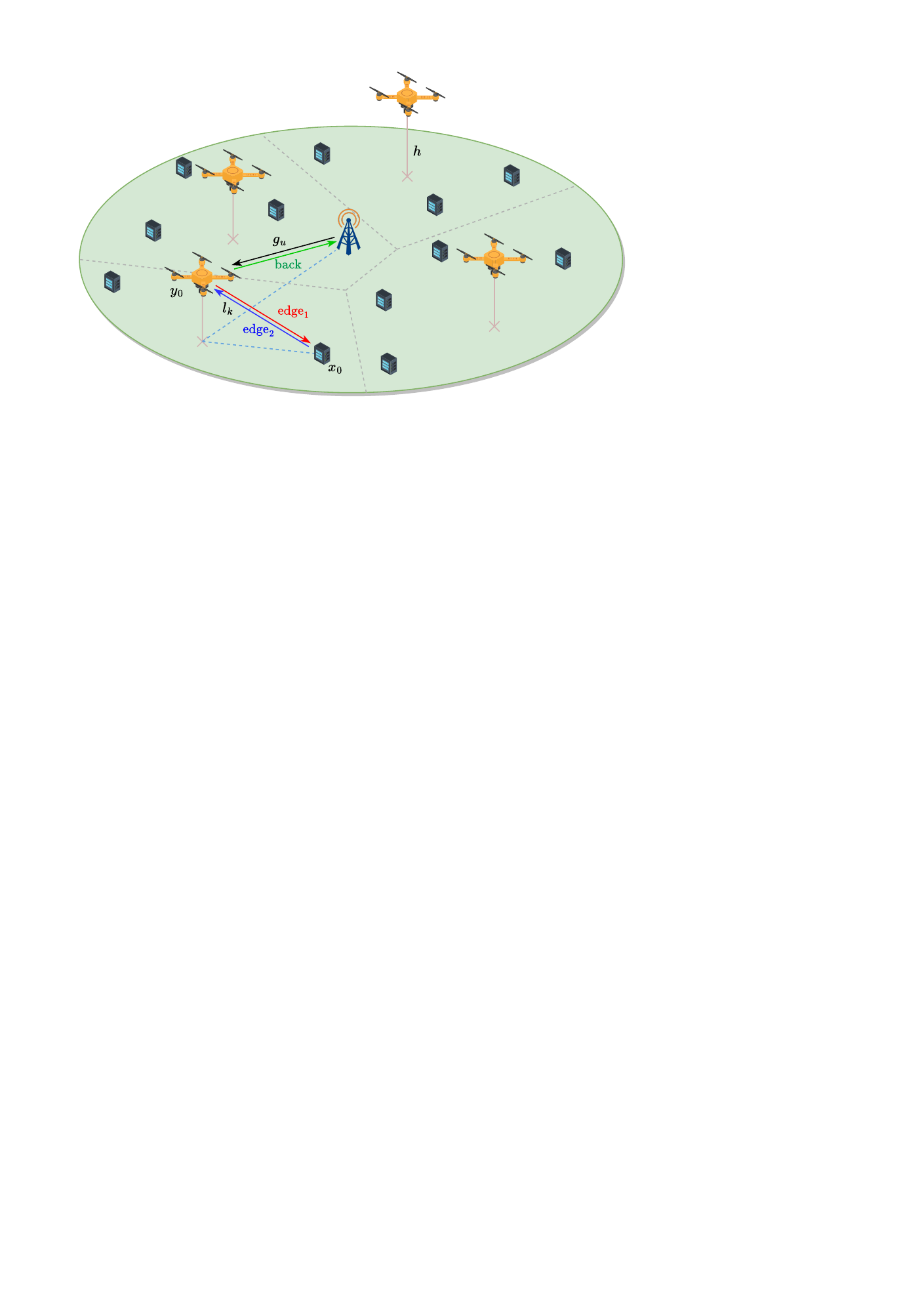}
    \caption{System model representation with $N_u=4$ and $N_d=12$.}
    \label{Systemmodel}
\end{figure}

\section{System Model}\label{sec:system_model}
\subsection{Network model}
A UAV-assisted HFL scenario is presented in Fig.~\ref{Systemmodel}, where $N_d$ devices train a global model assisted by one global server deployed within a terrestrial BS, and by $N_u$ UAVs that act as relays (edge servers). Besides assisting the data communication, UAVs can also perform model aggregation to reduce the computation at the BS. We consider that the locations of $N_d$ devices and $N_u$ UAVs are uniformly distributed in a finite area, forming separate instances of the Binomial Point Process (BPP), denoted as $\Phi_d$ and $\Phi_u$, respectively. The area in which the devices are distributed is considered as a disk $b(o',R)$ of radius $R$ centered at $o'=(0,0,0)$, where the BS with a global server is located. Although the devices can be strategically positioned to enhance network efficiency, in cases where precise traffic patterns are unknown, the assumption of BPP is justified using arguments similar to~\cite{chetlur2017downlink}. The locations of the devices are denoted by $\{x_k\}_{k=1:N_d} \equiv \Phi_d \subset \mathbb{R}^2$, whereas the UAVs are given by $\{y_u\}_{u=1:N_u} \equiv \Phi_u \subset \mathbb{R}^2$. The notations $x_0$ and $y_0$ are used to represent a reference device and a reference UAV, respectively\footnote{We use the notation of $x_0=||x_0||$ to represent both the reference device and its Euclidean distance in 3-D space. This also applies to the reference UAV located at $y_0=||y_0||$ and other devices and UAVs.}. For simplicity, we assume that all the UAVs are positioned at the same height $h$ and devices are located on the ground. We represent the 3-D distance of the link between transmitting UAV $u$ and the BS with $g_u$ and the link between a device $k$ and its serving UAV with $l_k$.

At the beginning of each global communication round of HFL, the BS broadcasts the global model parameters to all UAVs. In line with the existing literature, we assume that the downlink BS to UAV communication is reliable. This assumption is justified by the fact that the BS has sufficient power capabilities to deliver a high-quality signal to UAVs, rendering channel unreliability negligible in this scenario. In the second step, the UAVs further send the parameters in a downlink transmission with power $P_u$ to their linked devices through a UAV-to-device edge link denoted as `$\text{edge}_1$'. Then, the devices perform local training and transmit the resulting updates with power $P_d$ back to their UAVs using the device-to-UAV edge link denoted as `$\text{edge}_2$'. Unless specified otherwise, the term \textit{edge links} is used to collectively mean both links  $\text{edge}_1$ and $\text{edge}_2$. After performing local aggregations, each UAV sends the aggregated model parameters in an uplink transmission with power $P_u$ to the BS using the backhaul link denoted as `$\text{back}$'. This step completes a single global communication round in HFL. The mentioned links are also illustrated in Fig.~\ref{Systemmodel}.


\begin{figure}[t]
    \centering
    \resizebox{0.8\linewidth}{!}{\begin{tikzpicture}[thick,scale=1, every node/.style={scale=1.3},font=\Huge]
        \input{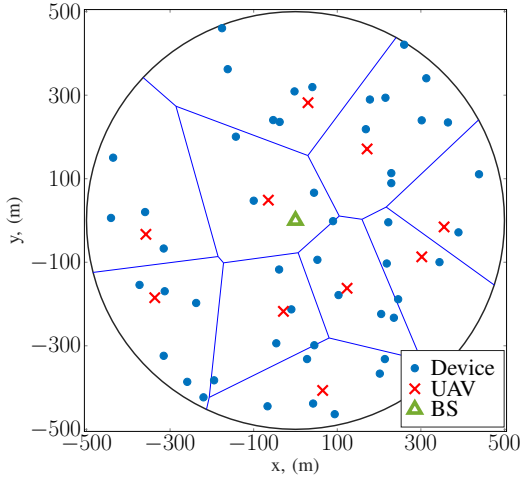}
        \end{tikzpicture}}
    \setlength{\belowcaptionskip}{-10pt}
    \caption{System model realization with $N_u=10$ and $N_d=50$.}
    \label{System_model_tex}
\end{figure}



\subsection{Channel Model}
One of the main advantages of introducing UAVs as relaying servers in HFL is their ability to establish LoS links with devices and the BS to increase signal strength due to their mobility and flexibility. Thus, the backhaul link between a UAV and the BS can be of LoS or non-LoS (NLoS) conditions with respect to the location of the UAV. Similarly, the edge links also experience LoS or NLoS conditions for model parameter transmissions between UAVs and devices. Due to the susceptibility of aerial links to environmental obstructions and the ground-level positioning of devices, we employ the LoS probability approximation proposed in~\cite{al2014optimal}. Specifically, the probability of establishing a LoS link with a 3-D distance of $r$, is expressed as follows:
\begin{equation}
P_\text{L}(r) = \frac{1}{1 + a \exp\left(-b \left[\frac{180}{\pi}\arctan\left(\frac{h}{\sqrt{r^2-h^2}}\right) - a\right]\right)},
\end{equation}
where the constants $a$ and $b$ are environment-dependent and can be found in~\cite[Table I]{bor2016efficient}, while $h$ represents the UAV's elevation height. The probability of NLoS communication, denoted as $P_\text{N}(r)$, is given by $P_\text{N}(r) = 1 - P_\text{L}(r)$.

We consider the effects of distance-dependent path loss with different path loss exponent for the LoS and NLoS edge and backhaul links denoted as $\alpha_{\text{L}}$ and $\alpha_{\text{N}}$, respectively. Furthermore, we assume that the communication channels are exposed to small-scale fading, which is modeled using the Nakagami-$m$ distribution with $m_{\text{L}}$ and $m_{\text{N}}$ being the corresponding parameters for the LoS and NLoS links. In our setup, there are a total of $M_b$ resource blocks (RBs) available at the BS and an additional $M_u$ RBs at each UAV. 
The BS uniformly selects $M_b$ devices out of $N_u$ UAVs without replacement, meaning that each UAV uses at most one resource block. Similarly, each UAV samples $M_u$ devices without replacement.
Universal frequency reuse is applied across all these RBs. To initiate the HFL process, the BS uses all $M_b$ RBs to establish connections with the UAVs, and each UAV makes use of $M_u$ RBs to communicate with its devices. For simplicity, we assume that RBs allocated to the BS and UAVs are distinct, ensuring that there is no interference between communication through backhaul and edge links.


\subsection{Association Policy and SINR}
During the downlink transmission of model parameters from UAVs to devices through the edge links, we adopt the {\color{black}maximum received average power selection policy~\cite{afshang2017fundamentals}.} This association policy assumes that the serving UAV provides the highest received energy out of all UAVs at the reference receiver device. The association rule implies the creation of exclusion regions on the locations of the interfering UAVs. Specifically, when a device $k$ associates with a LoS UAV with a distance of $l_k$, the exclusion region $E_{\text{LN}}(l_k)$ is created. When a device $k$ associates to a NLoS UAV with a distance of $l_k$, the corresponding exclusion region is given by $E_{\text{NL}}(l_k)$, where the expressions for these exclusion regions are given by
\begin{equation}
\begin{aligned}
    E_{\text{LN}}(l_k)= l_k^{\frac{\alpha_\text{L}}{\alpha_\text{N}}}, \\
    E_{\text{NL}}(l_k)= l_k^{\frac{\alpha_\text{N}}{\alpha_\text{L}}}.
    \label{exclusionRegion}
\end{aligned}
\end{equation}

During the uplink transmissions of the updates from UAVs to BS and devices to UAVs through backhaul and edge links, respectively, we adopt the uniform transmitter selection policy. This policy allows for the transmitting device/UAV to be chosen uniformly at random amongst other transmitters. Thanks to this policy, the proposed HFL algorithm is not constrained by any resource selection schemes and allows for all participants of HFL to upload their updates simultaneously. As a result, there are no exclusion regions created for the interfering UAVs or devices on the backhaul and uplink edge links.

We employ the signal-to-interference-plus-noise ratio (SINR) as a metric to describe and assess the quality of the backhaul and edge channels used for transmitting model parameters. 
Eventually, the SINR metric allows us to determine the success probability of transmissions over the link, which is used to implement the unbiasing idea. Hence, we can express the SINR of the backhaul link, denoted by $\text{SINR}^{\text{back}}_z$, used for UAV to BS communication as
\begin{equation}
\begin{aligned}
    \text{SINR}^{\text{back}}_z = \frac{P_u |h_{g_u}|^2_z g_u^{-\alpha_z}}{n_0^2 + I_{\text{back}}},
\end{aligned}
\end{equation}
where $z\in\{{\text{L}, \text{N}}\}$ denotes the LoS and NLoS links, $\alpha_z$ is the path-loss coefficient, $P_u$ is the transmission power of a UAV, $g_u$ is the 3-D distance between BS  and transmitting UAV $u$, $|h_{g_u}|^2_z$ is the power of the normalized small-scale Nakagami-$m$ fading for the main link $g_u$, $n_0^2$ is the noise power, and $I_{\text{back}}$ is the interference from all UAVs transmitting the aggregated model updates to the BS and is given as
\begin{equation}
\begin{aligned}
   I_{\text{back}}=\sum_{y_u\in \Phi_u\backslash y_0 } I_{y_0,y_u}(o') = \sum^{N_u^{\text{interf}}}_{u=1}I_{y_0,y_u}(o'),
\end{aligned}
\end{equation}
where $N_u^{\text{interf}}$ is the number of interfering UAVs that use the same RB as the main transmitting UAV, $I_{y_0,y_u}(o')$ represents the interference from UAV at $y_u$, which can have either LoS or NLoS links with the BS located at $o'$. Similarly, the SINR of the UAV to device downlink communication denoted by $\text{SINR}^{\text{edge}_1}_z$ is expressed as
\begin{equation}
\begin{aligned}
    \text{SINR}^{\text{edge}_1}_z = \frac{P_u |h_{l_k}|^2_z l_k^{-\alpha_z}}{n_0^2 + I_{\text{edge}_1}}, \ \ z\in\{\text{L},\text{N}\}
\end{aligned}
\end{equation}
where $l_k$ is the 3-D distance between the transmitting UAV and the receiving device $k$, $|h_{l_k}|^2_z$ is the power of the normalized small-scale Nakagami-$m$ fading for the main link $l_k$, and $I_{\text{edge}_1}$ is the interference from all UAVs transmitting to devices and is given as
\begin{equation}
\begin{aligned}
   I_{\text{edge}_1}=\sum_{y_u\in \Phi_u\backslash y_0 } I_{y_0,y_u}(x_0) = \sum^{N_u^{\text{interf}}}_{u=1}I_{y_0,y_u}(x_0),
\end{aligned}
\end{equation}
where $I_{y_0,y_u}(x_0)$ is the interference from UAV at $y_u$, which can have either LoS or NLoS links with the reference device at location $x_0$.

Finally, the SINR of the device-UAV uplink communication denoted by $\text{SINR}^{\text{edge}_2}_z$ is given by
\begin{equation}
\begin{aligned}
    \text{SINR}^{\text{edge}_2}_z = \frac{P_d |h_{l_k}|^2_z l_k^{-\alpha_z}}{n_0^2 + I_{\text{edge}_2}}, \ \ z\in\{\text{L},\text{N}\}
\end{aligned}
\end{equation}
where $P_d$ is the transmission power of a device, and $I_{\text{edge}_2}$ is the uplink interference from devices and is given as
\begin{equation}
\begin{aligned}
   I_{\text{edge}_2}=\sum_{x_k\in \Phi_d\backslash x_0 } I_{x_0,x_k}(y_0) = \sum^{N_d^{\text{interf}}}_{k=1}I_{x_0,x_k}(y_0),
\end{aligned}
\end{equation}
where $N_d^{\text{interf}}$ is the number of interfering devices that use the same RB as the main transmitting device, and $I_{x_0,x_k}(y_0)$ is the interference from the device at $x_k$, which can have either a LoS or a NLoS link with the UAV at location $y_0$.

\section{Proposed UAV-assisted Unbiased HFL}\label{sec:algorithm}
In this section, we provide the details of the proposed UAV-assisted unbiased HFL algorithm. 
The considered devices are partitioned into clusters of $N_d$ devices, each corresponding to a single UAV and denoted by $S_1, S_2, ..., S_{N_u}$. Thus, the objective of the central server at the terrestrial BS is to address the subsequent optimization challenge:
\begin{equation}
    \min_w \ f(w) = \sum^{N_u}_{u=1}\bar{p}_u f_u(w).
    \label{GlobalF}
\end{equation} 
Here, $w$ represents the learning model parameter, $\bar{p}_u=\bar{n}_u/\bar{n}$ denotes the weight of UAV $u$ with $\bar{n}_u = \sum_{k\in S_u}n_k$ data samples, $n_k$ is the number of data samples at device $k$ in its local dataset $\mathscr{D}_k$, and $\bar{n}$ signifies the total number of data samples. The term $f_u(\cdot)$ represents the averaged loss function of devices associated with UAV $u$ and is defined as follows
\begin{equation}\color{black}
    \min_w \ f_u(w) = \sum_{k\in S_u}p_k F_k(w),
    \label{UAV_F}
\end{equation} 
where $F_k(\cdot)$ denotes the local loss function at device $k$. During each local iteration $t$, each device $k$ updates its own model using stochastic gradient descent (SGD):
\begin{equation}
    w^{t+1}_{k} = w^{t}_{k} - \eta \nabla F_k(w^{t}_{k}),
\end{equation}
where $\eta$ is the learning rate, and $\nabla F_k(w^{t}_{k})$ is the gradient of the local loss function.

\begin{figure}
    \centering
    \includegraphics[viewport=800 200 300 400, trim=20 715 310 10, width=3.65in]{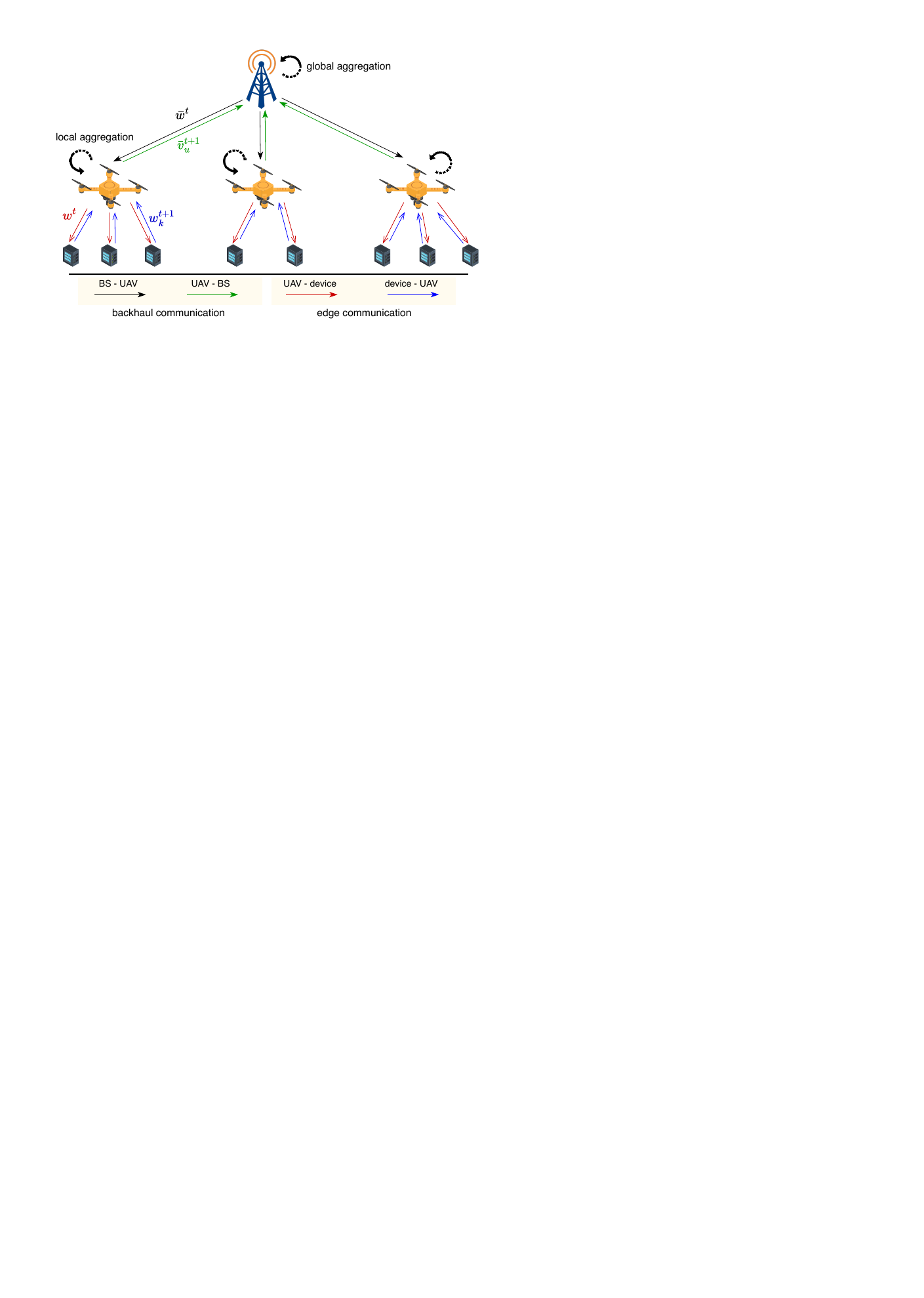}
    \caption{\color{black}Representation of HFL with a focus on the communication links.}
    \label{HFL}
\end{figure}

Fig.~\ref{HFL} represents a visual representation of the proposed UAV-assisted HFL framework, showcasing the backhaul and edge communication links among different participants. The update mechanisms in conventional HFL overlook the unreliability of communication channels, which can lead to certain updates not being successfully delivered. In such cases, the global model becomes skewed toward a UAV model benefiting from more favorable channel conditions. In turn, the UAV models themselves may be influenced by a bias toward devices with superior transmission capabilities. Hence, we propose aggregation rules that consider channel unreliability at both levels. In this approach, each update is assigned a weight based on the success probability of transmission defined as Definition~\ref{Def1}, effectively mitigating the impact of channel conditions.

\begin{definition}
Success probability of transmission over edge link in HFL is defined as the probability that both the SINR of a broadcast sent by a UAV at the reference device $k$ and the SINR of an update transmitted by the device at the same UAV jointly exceed the predefined threshold $\theta$ required to establish a successful connection. Mathematically, it can be considered as the conditional coverage probability, given the distance $l_k$ between the UAV-device pair.
\begin{equation}\small
\begin{aligned}
   \mathcal{P}^{\text{edge}}_k = {\color{black}\mathbb{E}_{z\in\{\text{L},\text{N}\}}\left[\mathbb{I}\left[\text{SINR}^{\text{edge}_1}_z>\theta,\text{SINR}^{\text{edge}_2}_z>\theta|l_k\right]\right]},
\end{aligned}
\end{equation}
where $\theta$ is the minimum SINR required to establish a successful connection.
\label{Def1}
\end{definition}

\begin{definition}
Success probability of transmission over backhaul link in HFL is defined as the probability that the SINR of an update transmitted by a UAV $u$ at the BS exceeds the predefined threshold $\theta$ required to establish a successful connection. Mathematically, it can be considered as the conditional coverage probability, given the distance $g_u$ between the transmitting UAV and receiving BS.
\begin{equation}
\mathcal{P}^{\text{back}}_u  = {\color{black}\mathbb{E}_{z\in\{\text{L},\text{N}\}}\left[\mathbb{I}\left[\text{SINR}^{\text{back}}_z>\theta|g_u\right]\right]}.
\end{equation}
\label{Def2}
\end{definition}

Note that a transmission of locally aggregated updates from a UAV to the BS through a backhaul link is received successfully if the SINR of the signal exceeds the threshold. Whereas, a UAV receives a local update from a device successfully if only both downlink and uplink transmissions from UAVs to devices and devices to UAVs through $\text{edge}_1$ and $\text{edge}_2$ links, respectively, jointly exceed a SINR threshold. This means that if a device successfully receives model parameters from its UAV, and yet the UAV fails to receive the update from its device, the device update does not contribute to the global model.

During a single communication round, local models are initially aggregated within {\color{black}UAV $u$ (where $u$ ranges from $1$ to $N_u$)} after every $E$ local iterations of stochastic gradient descent (SGD) using the proposed local aggregation rule. Particularly, at a local iteration $t$, we calculate $\bar{v}^{t}_u$ as the aggregation of models $w^{t}_k$ for all $k$ within the subset $S_u$:
\begin{equation}
\begin{aligned}
    \bar{v}_u^{t+1} = &\bar{v}_u^t +\sum_{k \in S_u} \frac{p_k / \bar{p}_u}{\mathcal{P}^{\text{edge}}_k} \times \mathbb{I}(\text{SINR}_k^{\text{edge}_1} > \theta, \text{SINR}_k^{\text{edge}_2} > \theta)\\
    &\times (w_k^{t+1} - \bar{v}_u^t),\ \ \ \ \ t = E, 2E, 3E,..
    \label{localAggregation1}
\end{aligned}
\end{equation}
where $\bar{v}^{t}_u$ represents the aggregated model for UAV $u$ at the given local iteration $t$. This UAV-level aggregation is performed at the corresponding UAV of the group of devices. The indicator function in (\ref{localAggregation1}) indicates that updates are successfully received only when the SINR values for both edge communications jointly exceed a predefined threshold value $\theta$. To cancel the effects of the unreliable channel, we divide the corresponding weights of the updates by edge success probability {\color{black}$P^{\text{edge}}_k$} for device $k$. The exact expression for $P^{\text{edge}}_k$ is derived later in section IV.

A global BS-level aggregation is performed every $G$ local iterations using the proposed global aggregation rule. At local iteration $t$, we update the global model $\bar{w}_t$ by aggregating the local models of each UAV using the following rule
\begin{equation}
\begin{aligned}
    \bar{w}^{t+1} = &\bar{w}^t + \sum_{u=1}^{N_u} \frac{\bar{p}_u}{\mathcal{P}^{\text{back}}_u} \times \mathbb{I}(\text{SINR}_u^{\text{back}} > \theta) \\
    &\times (\bar{v}_u^{t+1} - \bar{w}^t), \ \ \ \ \ t = G, 2G, 3G,..,
    \label{globalAggregation1}
    \end{aligned}
\end{equation}
It is important to note that we assume synchronous updates by all devices and that $G$ is chosen as a common multiple of $E$, i.e., $G=\gamma E, \gamma=1,2,...$. The parameter $\gamma$ is set by a user. Consequently, distributed SGD operates both at the local level within each UAV group and at the global level across all UAVs. Similar to the UAV-level aggregation, the updates from UAVs are included in the aggregation process in (\ref{globalAggregation1}) only when the SINR value for the backhaul link surpasses the predefined threshold. The impact of the unreliable channel is thus canceled by dividing the weights of the UAV updates by the backhaul success probability $P^{\text{back}}_u$ for UAV $u$. The exact expression for $P^{\text{back}}_u$ is derived in the following section.


The overall algorithm is depicted in Algorithm 1. Notably, the notation $G\ | \ t$ (or $G\ \nmid \ t$) in Algorithm 1 signifies that $t$ divides (or does not divide) $G$, meaning that $G$ is (or is not) an integer multiple of $t$. $E$ represents the number of local SGD steps at edge devices, which determines the period of UAV aggregation, while $G=\gamma E,\gamma =1,2,...$ is the period of global aggregation. 

\begin{algorithm*}\label{algo:HFL}
  \caption{Proposed UAV-assisted Unbiased Hierarchical FL (HFL)}
  \KwIn{Input parameters: $\bar{w}^0$, $\eta$, $\theta$, $G$, $S_u$ for $u \in \{1, 2, \ldots, N_u\}$}
  \KwOut{Global aggregated model $w^T$}
  \For{$t = 0$ to $T-1$}{
    \For{Each UAV $u \in \{1, 2, \ldots, N_u\}$, in parallel}{
      \For{Each device $k \in S_u$, in parallel}{
        Compute $\nabla F_k(w^{t}_k)$\;
        $w^{t+1}_{k} = w^{t}_{k} - \eta \nabla F_k(w^{t}_{k})$\;
        }
    \If{$E\ | \  t + 1$}{
      Local aggregation at UAV: $\bar{v}_u^{t+1} \leftarrow \bar{v}_u^t + \sum_{k \in S_u} \frac{p_k / \bar{p}_u}{\mathcal{P}^{\text{edge}}_k} \times \mathbb{I}(\text{SINR}_k^{\text{edge}_1} > \theta, \text{SINR}_k^{\text{edge}_2} > \theta) \times (w_k^{t+1} - \bar{v}_u^t)$ \;
      \If{$G \ \nmid \  t + 1$}{
      Distribute to devices: $w^{t+1}_k \leftarrow \bar{v}_u^{t+1} $ for all $k \in S_u$\;
    }
    }
    
    }
    \If{$G\ | \ t + 1$}{
      Global aggregation at BS: $\bar{w}^{t+1} \leftarrow \bar{w}^t + \sum_{u=1}^{N_u} \frac{\bar{p}_u}{\mathcal{P}^{\text{back}}_u} \times \mathbb{I}(\text{SINR}_u^{\text{back}} > \theta) \times (\bar{v}_u^{t+1} - \bar{w}^t)$\;
      Broadcast to UAVs: $\bar{v}^{t+1}_u \leftarrow \bar{w}^{t+1}$ for all $u \in N_u$\;
      UAVs distribute to devices: $w^{t+1}_k \leftarrow \bar{v}^{t+1}_u$ for all $k \in S_u$\;
    }
  }
  \Return{$w^T$}
\end{algorithm*}


\section{Success Probabilities Derivation}\label{sec:stochastic}
In this section, we derive the success probabilities of edge and backhaul transmissions involved during the training of HFL using stochastic geometry based on its prospects in analyzing and modeling wireless communication networks by means of realistically modeling network randomness and scalability. Stochastic geometry, in fact, serves as a crucial tool in the context of the proposed unbiased HFL algorithm, as it allows for quantitative estimation of the unreliability of the communication channels. The edge success probability in (\ref{localAggregation1}) is given in Theorem~\ref{Theorem1}, while the backhaul success probability in (\ref{globalAggregation1}) is given in Theorem~\ref{Theorem2}.

\begin{theorem}[Edge success probability]
In a UAV-assisted wireless network implementing HFL, the edge success probability for device {\color{black}$k, k\in S_u, u=1,2,...N_u$}, describing the reliability of local model parameters transmission between the device and its serving UAV acting as edge server is defined as the probability that the SINR values for both edge transmissions exceed predefined threshold $\theta$ and is expressed as
\begin{equation}\small
\begin{aligned}
   \mathcal{P}^{\text{edge}}_k = P_\text{L}(l_k) \ \mathcal{P}_k^{\text{L},\text{edge}_1} \  \mathcal{P}_k^{\text{L},\text{edge}_2} + P_\text{N}(l_k) \ \mathcal{P}_k^{\text{N},\text{edge}_1} \ \mathcal{P}_k^{\text{N},\text{edge}_2},
\end{aligned}
\end{equation}
where $l_k$ is the 3-D distance between a selected pair of device $k$ and a UAV. $\mathcal{P}_k^{z,\text{edge}_1}$ and $\mathcal{P}_k^{z,\text{edge}_2}$, $z\in\{\text{L},\text{N}\}$ correspond to the conditional probabilities of successful transmission from the UAV to the device, and conversely, from the device to the UAV when the serving link is a LoS or a NLoS link, respectively and are given as 
\begin{equation}\small
\begin{aligned}
   \mathcal{P}_k^{z,\text{edge}_1} =  &{\color{black}\sum_{j=1}^{m_z}} \binom{m_z}{j} (-1)^{j+1} \exp\left(\frac{-j \eta_z n_0^2 \theta}{P_u l_k^{-\alpha_z}}\right)\times \\
   &\mathcal{L}_{\text{edge}_1}^z\left(\exp\left(\frac{-j \eta_z \theta}{P_u l_k^{-\alpha_z}}\right)\right),
\end{aligned}
\end{equation}
\begin{equation}\small
\begin{aligned}
   \mathcal{P}_k^{z,\text{edge}_2} = 
   &\sum_{j=1}^{m_z} \binom{m_z}{j} (-1)^{j+1} \exp\left(\frac{-j \eta_z n_0^2 \theta}{P_d l_k^{-\alpha_z}}\right) \times \\ &\mathcal{L}_{\text{edge}_2}^z\left(\exp\left(\frac{-j \eta_z \theta}{P_d l_k^{-\alpha_z}}\right)\right),
\end{aligned}
\end{equation}
where $\eta_z = m_z(m_z!)^{-1/m_z}$, $m_z$ is the Nakagami fading parameter, $\mathcal{L}_{\text{edge}_1}^z(\cdot)$ and $\mathcal{L}_{\text{edge}_2}^z(\cdot)$ are the Laplace transforms of downlink edge interference $I_{\text{edge}_1}$ from interfering UAVs and uplink edge interference $I_{\text{edge}_2}$ from interfering devices given in Lemma~\ref{lemma1} and Lemma~\ref{lemma2}, respectively.
\begin{IEEEproof}
See Appendix~\ref{appendix:success_edge}.
\end{IEEEproof}
\label{Theorem1}
\end{theorem}

\begin{lemma}
\label{lemma1}
In a UAV-assisted wireless network modeled as a BPP, the Laplace transform $\mathcal{L}_{\text{edge}_1}^{z}(\cdot)$, $z\in\{\text{L, N}\}$ of the interference $I_{\text{edge}_1}$ on the UAV-to-device communication coming from other UAVs using the same RB take one of the following two forms depending on whether the serving link is LoS or NLoS and is given as
\begin{equation}\small
\begin{aligned}
   &\mathcal{L}_{\text{edge}_1}^z(s) = \left[\frac{1}{\int_{l_k}^{w_p} P_z(r) f_{R_1}(r|x_0)dr+\int_{E_{zz'}(l_k)}^{w_p}P_{z'}(r) f_{R_1}(r|x_0)dr}\right.\\
   &\left.\left(\int_{l_k}^{w_p} P_z(r) f_{R_1}(r|x_0)dr \right.\right.\\
   & \left.\left.\times \int_{l_k}^{w_p} \left(1+\frac{sP_u y_u^{-\alpha_z}}{m_z}\right)^{-m_z} f_{1,Y_z}(y_u|x_0)dy_u\right.\right.  \\
   & \left.\left.+ \int_{E_{zz'}(l_k)}^{w_p}P_{z'}(r) f_{R_1}(r|x_0)dr\right.\right.\\
   &\left.\left.\times \int_{E_{zz'}(l_k)}^{w_p} \left(1+\frac{sP_u y_u^{-\alpha_{z'}}}{m_{z'}}\right)^{-m_{z'}} f_{1,Y_{z'}}(y_u|x_0)dy_u\right)\right]^{N_u^{\text{interf}}},
   \label{Lemma1}
\end{aligned}
\end{equation}
where {\small$z\in\{\text{L, N}\}$} and {\small$z'\in\{\text{N, L}\}$}, $l_k$ is the serving distance from the UAV to device $k$, and {\small$N_u^{\text{interf}} = N_u\times P_{\text{active}}^{\text{edge}_1}-1$} represents the average number of interfering UAVs with {\small$P_{\text{active}}^{\text{edge}_1}=\frac{N_u/M_b}{N_u}$} is the probability of a UAV using the same RB as the serving UAV. The exclusion region $E_{zz'}(l_k)$ {\color{black}due to the maximum received average power policy} is defined in (\ref{exclusionRegion}). Also, according to \cite{chetlur2017downlink}, the distribution of the {distance \color{black}$R_1$} from the receiving device to the set of interfering UAVs, conditioned on the location of the device $x_0$ is defined as
\begin{equation}\small
\begin{aligned}
f_{R_1}(r|x_0) = 
 \begin{cases}
    \frac{2r}{R^2}, & \text{if } h \leq r \leq w_m\\
    \frac{2r}{\pi R^2} \arccos\left(\frac{r^2 + x_0^2 - d^2}{2x_0 \sqrt{r^2 - h^2}}\right),
& \text{if } w_m \leq r \leq w_p,
\end{cases},
\end{aligned}
\label{eq:fU}
\end{equation}
where $d = \sqrt{R^2+h^2},w_m = \sqrt{(R - x_0)^2 + h^2}$, $w_m = \sqrt{(R + x_0)^2 + h^2}$, and $R$ is the radius of the finite area. The distributions of the distance from the interfering LoS UAV and NLoS UAV to the reference receiving device~\cite{kouzayha2021analysis}, are given by $f_{1,Y_{z}}(y_u|x_0) = \frac{f_{R_1}(y_u|x_0) P_{z}(y_u)}{\int_{l_k}^{w_p} f_{R_1}(w|x_0) P_{z}(w)dw}$ and $f_{1,Y_{z'}}(y_u|x_0) = \frac{f_{R_1}(y_u|x_0) P_{z'}(y_u)}{\int_{E_{zz'}(l_k)}^{w_p} f_{R_1}(w|x_0) P_{z'}(w)dw}$, respectively.

\begin{IEEEproof}
See Appendix~\ref{appendix:laplace_edge1}.
\end{IEEEproof}
\end{lemma}

\begin{lemma}
In a UAV-assisted wireless network modeled as a BPP, the Laplace transform $\mathcal{L}_{\text{edge}_2}^{z}(\cdot)$ of the interference $I_{\text{edge}_2}$ on the device-to-UAV communication coming from other devices using the same RB can be of two forms depending on whether the desired link is with a LoS or NLoS UAV and is given as
\begin{equation}\small
\begin{aligned}
   &  \mathcal{L}_{\text{edge}_2}^{z}(s) = \left[\frac{1}{ \int_{h}^{w_p} P_{z}(r) f_{R_1}(r|y_0)du+\int_{h}^{w_p}P_{z'}(r) f_{R_1}(r|y_0)dr}\right.\\
   &\left.\left(\int_{h}^{w_p} P_{z}(r) f_{R_1}(r|y_0)dr \right.\right.\\
   &\left.\left. \times \int_{h}^{w_p} \left(1+\frac{sP_d x_k^{-\alpha_{z}}}{m_{z}}\right)^{-m_\text{L}} f_{1,X_{z}}(x_k|y_0)dx_k \right.\right. \\
   & \left.\left.+  \int_{h}^{w_p}P_{z'}(r) f_{R_1}(r|y_0)dr \right.\right.\\
   &\left. \left.\times \int_{h}^{w_p} \left(1+\frac{sP_d x_k^{-\alpha_{z'}}}{m_{z'}}\right)^{-m_{z'}} f_{1,X_{z'}}(x_k|y_0)dx_k\right) \right]^{N_d^{\text{interf}}},
\end{aligned}
\end{equation}
where {\small$z\in\{\text{L, N}\}$} and {\small$z'\in\{\text{N, L}\}$}, {\small$N_d^{\text{interf}}=N_d\times P_{\text{active}}^{\text{edge}_2}-1$} is the average number of interfering devices with {\small$P_{\text{active}}^{\text{edge}_2} =\frac{N_d/M_u}{N_d}$} represents the probability of a device using the same RB as the main transmitting device, $x_k$ is the distance of the UAV to the $k$-th interfering device. The distributions of the distances from the interfering device to the reference LoS UAV and NLoS UAV are given by  $f_{1,X_{z}}(x_k|y_0) = \frac{f_{R_1}(x_k|y_0) P_{z}(x_k)}{\int_{h}^{w_p} f_{R_1}(w|y_0) P_{z}(w)dw}$ and $f_{1,X_{z'}}(x_k|y_0) = \frac{f_{R_1}(x_k|y_0) P_{z'}(x_k)}{\int_{h}^{w_p} f_{R_1}(w|y_0) P_{z'}(w)dw}$, respectively.

\begin{IEEEproof}
The proof mirrors the steps used in the proof of Lemma 1, and thus is omitted here.
\end{IEEEproof}
\label{lemma2}
\end{lemma}

\begin{theorem}[Backhaul success probability]
In a UAV-assisted wireless network implementing HFL with a single BS in the center of the region, the backhaul success probability describing the
reliability of aggregated model parameters transmission from the UAV acting as edge server to the central server at the BS is defined as the probability that the SINR for transmissions from {\color{black}UAV $u, u=1,2,...N_u$} to BS exceeds the predefined threshold $\theta$ and can be expressed as
\begin{equation}
\begin{aligned}
   \mathcal{P}^{\text{back}}_u = P_\text{L}(g_u) \   \mathcal{P}^{\text{L},\text{back}}_u + P_\text{N}(g_u) \ \mathcal{P}^{\text{N},\text{back}}_u,
\end{aligned}
\end{equation}
where $g_u$ is the 3-D serving distance between a selected UAV and the BS,   $\mathcal{P}^{\text{L},\text{back}}_u$ and $\mathcal{P}^{\text{N},\text{back}}_u$ are the conditional success probabilities when the serving link is a LoS or a NLoS link and can be expressed as
\begin{equation}\small
\begin{aligned}
   &\mathcal{P}^{z,\text{back}}_u = \\
   &= \sum_{j=1}^{m_z} \binom{m_z}{j} (-1)^{j+1} \exp\left(\frac{-j \eta_z n_0^2 \theta}{P_u g_u^{-\alpha_z}}\right) \mathcal{L}_{\text{back}}\left(\exp\left(\frac{-j \eta_z \theta}{P_u g_u^{-\alpha_z}}\right)\right),
\end{aligned}
\label{eq:P_back}
\end{equation}
where $\mathcal{L}_{\text{back}}^z(\cdot)$ is the Laplace transform of the interference $I_{\text{back}}$ from all UAVs transmitting to the BS using the same RB and is given in Lemma~\ref{lemma:Laplace_back}.

\begin{IEEEproof}
The conditional backhaul success probability can be derived as
    \begin{equation}\small
    \begin{aligned}
       &\mathcal{P}^{z,\text{back}}_u = \mathbb{E}\left[\mathbb{I}\left[\text{SINR}^{\text{back}}_z>\theta|g_u\right]\right]\\
       &= \mathbb{E}\left[\mathbb{P}\left[ |h_0|^2 > \frac{\theta (n_0^2 + I_{\text{back}})}{P_u g_u^{-\alpha_z}}|g_u\right]\right]\\
       &\stackrel{\text{(a)}}{=} 1 - \mathbb{E}_{I_{\text{back}}}\left[\left(1-\exp\left( \frac{-\eta_z \theta (n_0^2 + I_{\text{back}})}{P_u g_u^{-\alpha_z}}\right)\right)^{m_z} \middle| g_u\right],
    \end{aligned}
    \end{equation}
    where (a) is derived from applying the cumulative distribution function (CDF) of the normalized gamma fading coefficient, while the final expression in~\eqref{eq:P_back} is obtained through the use of the binomial expansion.
\end{IEEEproof}
\label{Theorem2}
\end{theorem}

\begin{lemma}
In a UAV-assisted wireless network modeled as a BPP, the expression for the Laplace transform of the interference $I_{\text{back}}$ at the BS coming from other UAVs, which are using the same RB to transmit their own updates, can be derived as:
\begin{equation}\small
\begin{aligned}
   &  \mathcal{L}_{\text{back}}(s) = \left[\frac{1}{ \int_h^d P_\text{L}(r) f_{R_2}(r)dr+\int_h^d P_\text{N}(r) f_{R_2}(r)dr} \right.\\
   &\left.\left(\int_h^d P_\text{L}(r) f_{R_2}(r)dr\times \int_h^d \left(1+\frac{sP_u y_u^{-\alpha_{\text{L}}}}{m_{\text{L}}}\right)^{-m_{\text{L}}} f_{2,Y_{\text{L}}}(y_u)dy_u \right. \right. \\
   & \left.\left.+ \int_h^d P_\text{N}(r) f_{R_2}(r)dr\right. \right.\\
   &\left.\left.\times  \int_h^d \left(1+\frac{sP_u y_u^{-\alpha_{\text{N}}}}{m_{\text{N}}}\right)^{-m_{\text{N}}} f_{2,Y_{\text{L}}}(y_u)dy_u\right)\right]^{N_u^\text{interf}},
\end{aligned}
\end{equation}
where {\small$N_u^\text{interf}=N_u\times P_{\text{active}}^{\text{back}}-1$} is the average number of interfering UAVs with {\small$P_{\text{active}}^{\text{back}}=\frac{N_u/M_b}{N_u}$} is the probability of a UAV  using the same RB as the main transmitting UAV. The distance distribution of the interfering UAVs due to the uniform transmitter policy is given in~\cite{afshang2017fundamentals} by 
\begin{equation}
   f_{R_2}(r) =\begin{cases}
    \frac{2r}{R^2} & \text{if } h \leq r \leq d\\
    0 & \text{otherwise}, 
\end{cases}  
\end{equation}
where $d=\sqrt{R^2+h^2}$. The distribution of the distance from the interfering UAVs that have $z \in \{\text{L},\text{N}\}$ links with the BS is given by $f_{2,Y_z}(y_u) =\frac{f_{R_2}(y_u) P_z(y_u)}{\int_h^d f_{R_2}(w) P_z(w)dw}$.
\begin{IEEEproof}
See Appendix~\ref{appendix:laplace_back}.
\end{IEEEproof}
\label{lemma:Laplace_back}
\end{lemma}

\section{Performance analysis}\label{sec:performance}
\subsection{Proof of Convergence}
In this section, we derive a theoretical convergence bound for the proposed UAV-assisted unbiased HFL algorithm, offering a fundamental guarantee that the model converges to a steady state within a finite number of training iterations. The derived convergence bound serves as a critical tool, shedding light on the influential factors and system parameters that significantly impact the efficacy of the training process. In the pursuit of establishing the upper bound for the proposed unbiased UAV-assisted HFL, we propose an analytical methodology that involves an investigation on how the states of wireless channels would influence the algorithmic derivations. 


The main challenge of the analysis is that devices only perform local iterations before global aggregation in regular FL, whereas in HFL, devices not only perform local iterations but also do aggregations with other devices associated with the same UAV. If we directly apply the analysis of regular FL, the effect of local aggregations would be neglected, and the resulting bound would not be tight. To address this, the key idea is to consider the local parameter drift $||w_k-\bar{v}_u||, k\in S_u$ in the analysis of global parameter drift $||\bar{v}_u - \bar{w}||$ so that the analysis for the downward part can be incorporated in the analysis for the upward part. Such an approach allows us to reflect on the importance of the wireless channels during the aggregations on both UAV and BS separately. 

We next define a set of analytical assumptions used in the convergence analysis:
\begin{assumption}
\textit{The loss functions at the devices and UAVs have L-Lipschitz continuous gradient for some L$>$0}
\begin{equation}
    ||\nabla F_k(w) -\nabla F_k(w') || \le L ||w-w'||, \forall j,w,w', \text{and}
\end{equation}
\begin{equation}
    ||\nabla f_u(w) -\nabla f_u(w') || \le L ||w-w'||, \forall j,w,w'.
\end{equation}
\label{assumption1}
\end{assumption}

\begin{assumption}
\textit{The upward divergence between the gradients of the UAVs' and the BS's loss functions are upper bounded as}
\begin{equation}
    \sum_u^{N_u} \bar{p}_u||\nabla f_u(w) - \nabla f(w)||^2 \le \epsilon^2, \forall w.
\end{equation}
\label{assumption2}
\end{assumption}
\begin{assumption}
\textit{The downward divergence between the gradients of the devices' and the corresponding UAV's loss functions are upper bounded as}
\begin{equation}
    \sum_{k\in S_u} \frac{p_k}{\bar{p}_u}||\nabla F_k(w) - \nabla f_u(w)||^2 \le \hat{\epsilon}^2, \forall w.
\end{equation}
\label{assumption3}
\end{assumption}
\begin{assumption}
\textit{The expected squared norms of the stochastic gradients of the devices' and the corresponding UAV's loss functions are uniformly bounded as}
\begin{equation}
    \mathbb{E}||\nabla F_k (w)||^2 \le A^2_2, \forall k, w,
\end{equation}
and
\begin{equation}
    \mathbb{E}||\nabla f_u (w)||^2 \le A^2_1, \forall u, w.
\end{equation}
\label{assumption4}
\end{assumption}

The presented assumptions are meant to bound the gradients of the loss functions in a controlled manner, which allows to establish the convergence pattern.  
As a result of applying these assumptions, we formulate the convergence bound in the following theorem.
\begin{theorem} 
Considering the UAV-assisted unbiased HFL in Algorithm 1 with non-IID data and non-convex objective function, and that satisfies assumptions~\ref{assumption1}-\ref{assumption4}, the learning rate of $\eta<\frac{1}{4\sqrt{3}GL}$ and $T\geq1$, we have
\begin{equation}\small
    \begin{aligned}
        &\frac{1}{T}\sum^{T-1}_{t=0}\mathbb{E}\left\| \nabla f(\bar{w}^t)\right\|^2\le  \frac{2}{\eta T}\left[\mathbb{E} f(\Bar{w}^{0}) - f^* \right]+ 4B_1^2A_1^2+ 4B_1^2A_2^2 \\
        &  \ \ \ \ \ \ \ \  + 2C\eta^2 G^2 A_2^2 L^2 (B_3+B_2) + 5C\eta^2G^2\epsilon^2L^2  \\
        & \ \ \ \ \ \ \ \ + C\eta^2 E^2 A^2_2 L^2 B_3 + C\eta^2 L^2\sum_{u=1}^{N_u}\bar{p}_u E^2 \hat{\epsilon}^2,
    \end{aligned}
\end{equation}
where \small{$C=112/5$, $B_1 =\sum^{N_u}_{u=1}\bar{p}_u \left(\frac{1}{\mathcal{P}^{\text{back}}_u}-1\right)$, $B_2 =\sum_{u=1}^{N_u}\sum_{k\in S_u}p_k\left(\frac{1}{\mathcal{P}^{\text{back}}_u \mathcal{P}^{\text{edge}}_k}-1\right)$}, and  \small{$B_3 =\sum^{N_u}_{u=1}\sum_{k\in S_u} p_k\left(\frac{1}{\mathcal{P}^{\text{edge}}_k}-1\right)$}.

\begin{IEEEproof}
See Appendix~\ref{appendix:convergence}.
\end{IEEEproof}
\label{Theorem3}
\end{theorem}

The upper bound expression is dependent on various system parameters. Specifically, it incorporates terms such as $B_1$, $B_2$, and $B_3$, which reflect the condition of wireless channels between Unmanned Aerial Vehicles (UAVs) and devices through $\mathcal{P}^{\text{edge}}_k$, as well as between UAVs and Base Stations (BS) through $\mathcal{P}^{\text{back}}_u$. Typically, this bound is heavily influenced by the divergence criteria, as determined by assumptions about the behavior of loss functions. However, our analysis indicates that the reliability of wireless channels also plays a crucial role along with the divergence criteria ${\epsilon}^2$ and $\hat{\epsilon}^2$ since all these terms are scaled by the same terms $G^2$ and $E^2$. Consequently, alterations in the periods of local and global aggregations similarly impact the convergence of HFL.
Comparing with the upper bound derived in~\cite{wang2022demystifying}, where the effect of noise due to training on mini-batches is taken into account, we draw the conclusion that the importance of having a reliable wireless channel surpasses the significance of reducing mini-batch noise. This inference arises from the observation that the term related to mini-batch noise is only scaled by $G$ and $E$. Additionally, we note that the combined influence of $B_2$ and $B_3$ outweighs their individual impacts, primarily due to the relationship $G>E$. In practical terms, this suggests that enhancing solely the edge channels may not yield the anticipated improvements, unlike scenarios where both edge and backhaul channels witness simultaneous enhancements.

Following a similar methodology, an upper bound on the convergence of the UAV-assisted unbiased HFL algorithm when a random number of devices are linked to each UAV at specific intervals can be derived. In such a scenario, the number of devices associated with a UAV is random at each global communication round. 
To derive the upper bound, the following assumption on the bound of the global divergence must be introduced.
\begin{assumption}
    \textit{The global divergence between the gradients of devices' and the BS's loss functions are upper bounded as}
    \begin{equation}
        \sum_{k=1}^{N_d} p_k||\nabla F_k(w) - \nabla f(w) ||^2\le \Tilde{\epsilon}^2, \forall w.
    \end{equation}
\end{assumption}
Following similar steps as in~\cite{wang2022demystifying}, we derive the following convergence upper bound for the random grouping case.
\begin{corollary}
\label{corollary:bound_uniform}
\textit{The convergence bound of the UAV-assisted HFL algorithm presented in Algorithm~\ref{algo:HFL} when using the uniform random grouping strategy for devices association with UAVs with $\eta<\frac{1}{4\sqrt{3}GL}$ is given by}
\begin{equation}\small
    \begin{aligned}
        \frac{1}{T}\sum^{T-1}_{t=0}&\mathbb{E}\left\| \nabla f(\bar{w}^t)\right\|^2\le  \frac{2}{\eta T}\left[\mathbb{E} f(\Bar{w}^{0}) - f^* \right]+ 4B_1^2A_1^2+ 4B_1^2A_2^2 \\
        & +5C\eta^2L^2\left[\left(\frac{N_u-1}{\bar{n}-1}\right)G^2 +\left(1-\frac{N_u-1}{\bar{n}-1}\right)E^2  \right]\Tilde{\epsilon}^2 \\
        & + 2C\eta^2L^2A^2_2\left[\left(B_3+B_2\right)G^2 +B_3 E^2\right].
    \end{aligned}
\end{equation}
\end{corollary}

The outcome of corollary~\ref{corollary:bound_uniform} highlights the important role of favorable wireless channel conditions on both edge and backhaul links. Note that a tighter convergence bound can be obtained by increasing the value of global iterations {\small$G=\gamma E, \gamma =1,2,...$} to {\small$G'=lG, 1<l<\sqrt{\frac{1}{m^2}\frac{\bar{n}-N_u}{N_u}+1}$} while decreasing the value of local iterations {\small$E$} to {\small$E'=qE, q \le \sqrt{1-m^2(l^2-1)\frac{N_u}{\bar{n}-N_u}}$} when the channel conditions are robust. Thus, the previous claim can only be achieved when $B_2$ and $B_3$ satisfy the following condition
\begin{equation}\small
    \begin{aligned}
        B_3 \le \frac{-\frac{5}{2}\frac{N_u\Tilde{\epsilon}^2}{A^2_2}\left[1+\frac{N_u-1}{\bar{n}-1}-\frac{N_u-1}{N_u}\right]+B_2(\bar{n}-N_u)}{2N_u-\bar{n}}.
    \end{aligned}
\end{equation}

\begin{figure}[t]
     \centering
      {\includegraphics[width=0.95\linewidth]{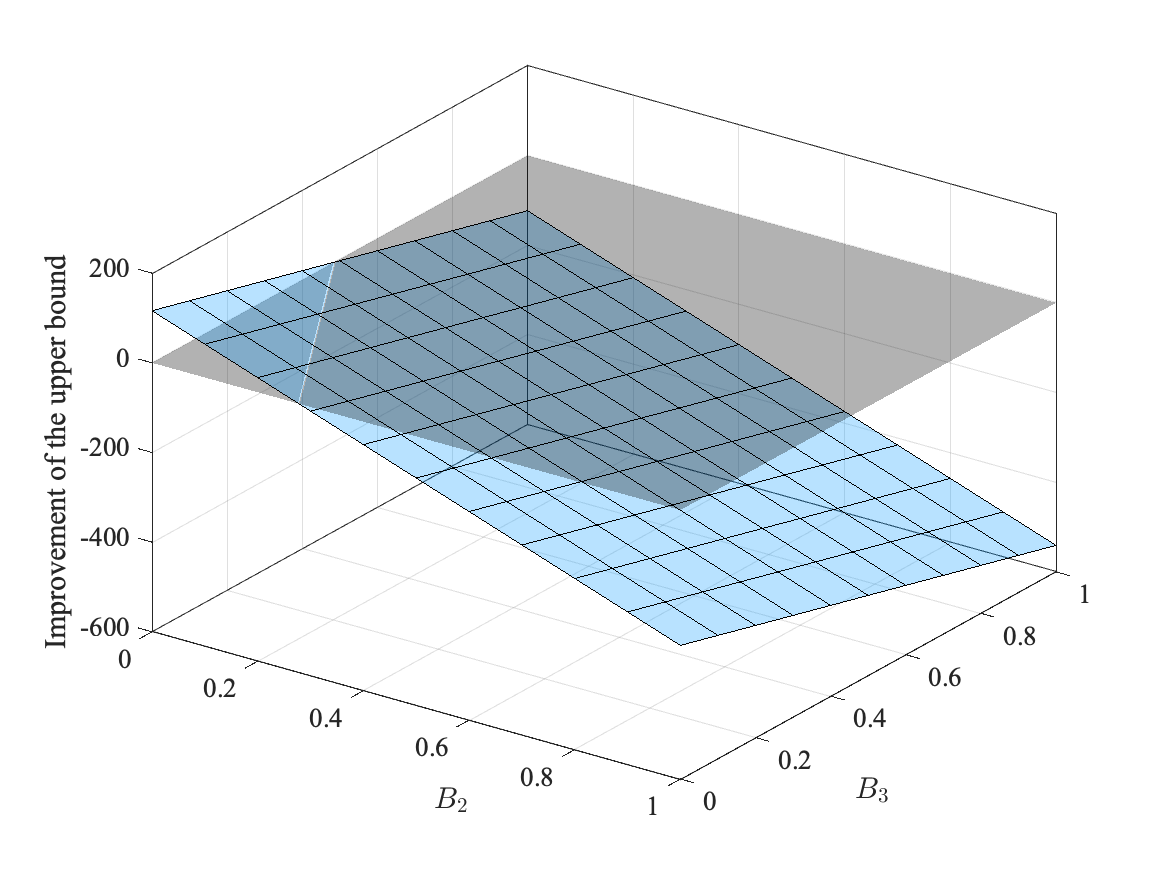}
      }
     \setlength{\belowcaptionskip}{-12pt}
     \vspace{-0.5cm}
     \caption{Illustration of the improvement of upper bound as a function of $B_2$ and $B_3$. The blue plane represents attainable bound improvement, while the gray plane is for the reference and means no improvement. }
    \label{fig:BoundImprov}
\end{figure}

A visual representation of the constraint can be seen in Fig.~\ref{fig:BoundImprov}, which illustrates the achievable decrease in the bound as a function of $B_2$ and $B_3$. The gray plane in Fig.~\ref{fig:BoundImprov} represents the zero value, which means that it points out to the case where no improvement in the upper bound is obtained. The blue plane, on the other hand, contains the values of the improvement when we increase $G$ and decrease $E$. We observe that the positive values are attainable only when $B_2$ and $B_3$ have lower values, meaning that the states of the wireless channel should be favorable (high values of $\mathcal{P}^{\text{back}}_u, u=1,2...,N_u$ and $\mathcal{P}^{\text{edge}}_k, k\in S_u$).  

\subsection{Latency Analysis}
In this section, we describe the components of the global communication latency, which is the required time to complete one round of global training. It consists of the computational latency at devices and the  communication latency between BS, UAVs, and devices. The computation time at a device $k$ is characterized by $c_{k}$, representing the number of CPU cycles required for the device $k$ to execute the training of one data sample, and $\sigma_{k}$, representing the CPU-cycle frequency of the device $k$. Consequently, the computation time for the device $k$ per a single local SGD iteration $T_{\text{cmp}}^k$ can be expressed as~\cite{you2023hierarchical}
\begin{equation}
T_{\text{cmp}}^k = c_k\frac{n_k}{\sigma_{k}},
\label{compTime}
\end{equation}
where $n_k$ is the dataset size available at device $k$.

The communication latency is usually larger than the computation delay due to the frequent and data-intensive nature of communication between the BS and distributed nodes, compounded by factors such as network constraints and synchronization requirements. The round trip communication for the model transfer between a UAV $u$ and device $k$ depends on the rates of the edge link $r^{u,k}$ and $r^{k,u}$, which depends on the bandwidth $b^{k}$ allocated to the device $k$ as well as the SINRs $\text{SINR}^{k,u}$ and $\text{SINR}^{u,k}$ of the transmissions between UAV $u$ and device $k$. The expression of $\text{SINR}^{k,u}$ and $\text{SINR}^{u,k}$ are the forms defined in (5) and (7), respectively. Whether the links are LoS or NLoS depend on the which UAVs the devices are connected to.
Thus, the respective expressions of communication latencies of the device-to-UAV communication and the UAV-to-device communication are given by
\begin{equation}
T_{\text{com}}^{u,k} = \frac{Z}{r^{u,k}} = \frac{Z}{b^{k}\log{(1+\text{SINR}^{u,k}})},
\end{equation}
and
\begin{equation}
T_{\text{com}}^{k,u} = \frac{Z}{r^{k,u}} = \frac{Z}{b^{k}\log{(1+\text{SINR}^{k,u}})},
\end{equation}
where $Z$ is the size of the model parameters transmitted between the participants of the HFL algorithm. Similarly, the latency of the UAV-to-BS communication can be expressed as
\begin{equation}
T_{\text{com}}^{u,\text{BS}} = \frac{Z}{r^{u,\text{BS}}} = \frac{Z}{b^{u}\log{(1+\text{SINR}^{u,\text{BS}})}},
\end{equation}
where $r^{u,\text{BS}}$ is the rate of the backhaul link between UAV $u$ and BS, $b^{u}$ is the bandwidth allocated to the UAV $u$, and $\text{SINR}^{u,\text{BS}}$ represents the SINR value of the received signal at the BS. The expression for $\text{SINR}^{u,\text{BS}}$ follows the format outlined in Equation (3). The selection of a particular UAV determines whether the connection with BS is LoS or NLoS. Thus, the total time elapsed for UAV  $u$ during one global communication can be expressed as:
\begin{equation}\small
\begin{aligned}
&T^u = T_{\text{com}}^{u} + T_{\text{cmp}}^u \\
&= \max_{k\in S_u}(\frac{G}{E}T_{\text{com}}^{u,k}) + \max_{k\in S_u}(\frac{G}{E} T_{\text{com}}^{k,u}) + T_{\text{com}}^{u,\text{BS}} + G \max_{k \in S_u}(T_{\text{cmp}}^k),  
\label{Tudelay}
\end{aligned}
\end{equation}
where $\max(\cdot)$ is due to the synchronous nature of the HFL algorithm, meaning that a UAV has to wait for the update from the slowest device to perform local aggregation. Thus, the total latency for one global update is
\begin{equation}
\begin{aligned}
 T = \max_{u\in N_u}(T^u).
\end{aligned}
\end{equation}
Similar to (\ref{Tudelay}), $\max(\cdot)$ is used because of the synchronous nature of the HFL algorithm, where the BS waits for the update from the slowest UAV to perform global aggregation.

\section{Numerical results}\label{sec:results}
\begin{figure}
    \centering
    \includegraphics[viewport=100 200 300 300, trim=-30 670 270 40, width=3.8in]{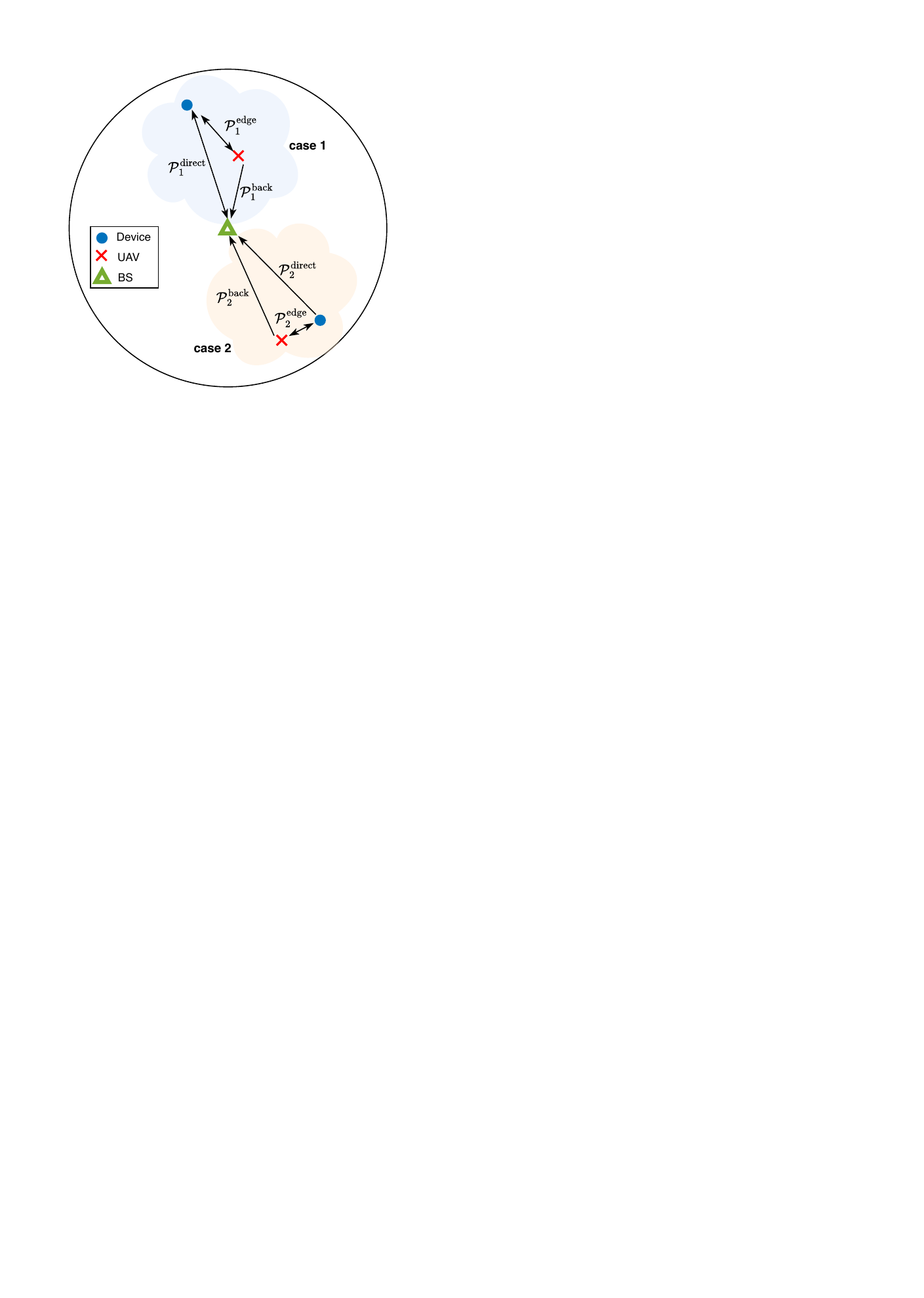}
    \caption{Illustration of two pairs of devices and UAVs used to represent the success probabilities of the edge and backhaul links. The placements are taken from the system realization in Fig.~\ref{System_model_tex}.}
    \label{Systemmodel_v2}
\end{figure}

In this section, we present analytical results and Monte-Carlo simulations to corroborate the accuracy of the success probabilities outlined in Theorems~\ref{Theorem1} and~\ref{Theorem2}. Furthermore, we offer simulation findings that highlight the accuracy and latency of the proposed unbiased HFL algorithm in UAV-assisted networks. Unless mentioned otherwise, the simulation parameters can be found in Table~\ref{tab:my-table}. In Fig.~\ref{Systemmodel_v2}, we consider two pairs of UAVs and devices with the goal of examining the success probabilities involved in HFL transmissions. The benchmark scenario considers a conventional FL training, where the devices directly communicate with the server at the BS. In order to consider an unbiased FL algorithm, we derive the success probability of transmissions from devices to the BS in Lemma 4. 
\begin{lemma}
In a wireless network modeled as a BPP, the upload success probability of a device $k$, $k=1,..,N_d$, to participate in FL training can be expressed as
\begin{equation}\small
\begin{aligned}
    \mathcal{P}^{\text{direct}}_k = &\sum_{j=1}^{m} \binom{m}{j} (-1)^{j+1} \exp\left(-\frac{j \eta_d \theta n_0^2}{P_d q_k^{-\alpha}}\right) \\
    &\times \mathcal{L}_{\text{direct}}\left(\exp\left(\frac{j \eta_d \theta}{P_d q_k^{-\alpha}}\right)\right),
\end{aligned}
\end{equation}
where $q_k$ is the 3-D distance between a device $k$ and the BS, $\eta_d=m(m\!)^{-1/m}$, $m$ and $\alpha$ are the Nakagami fading parameter and path-loss exponent for the device-to-BS link, and $\mathcal{L}_{\text{direct}}(\cdot)$ represents the Laplace transform of interference sourcing from other devices that upload their local models to the same BS, and can be expressed as
\begin{equation}\small
\begin{aligned}
   \mathcal{L}_{\text{direct}}(s)  &= \left( \int_0^R \left(1+\frac{sP_d q'^{-\alpha}}{m}\right)^{-m} \frac{2q'}{R^2}\, dq' \right)^{N_d\times P^{\text{direct}}_{\text{active}}-1},
\end{aligned}
\end{equation}
where $q'$ represents the interfering distances from the BS to other devices, and {\small$P^{\text{direct}}_{\text{active}}=\frac{N_d/M_d}{N_d}$} is the probability of a device using the same RB as the main transmitting device, where $M_d$ denotes the total number of RBs available for device to BS communication.
\end{lemma}
Thus, we denote these success probabilities in Fig.~\ref{Systemmodel_v2} as $\mathcal{P}^{\text{direct}}_1$ and $\mathcal{P}^{\text{direct}}_2$ for the first and second cases, respectively. The system parameter values considered for the device-to-BS communication are assumed as $m=2, \alpha=2.5$, and $M_d=20$.  
We plot the edge, backhaul, and direct success probabilities of the two considered device/UAV pairs as a function of the SINR threshold in Fig.~\ref{fig:fig1}. In the first case, the UAV is located in between the BS and the device, whereas in the second case, it is almost at the same distance from the BS as the device. 
Fig.~\ref{fig:fig1} shows that, for the first case, both backhaul and edge success probabilities significantly surpass the success probability of device-to-BS transmissions. This is the main advantage of integrating the UAVs as intermediate aggregators, meaning that it increases the chance of successful communication. This trend remains consistent in the second case, primarily attributable to the UAV's amplified transmission power and enhanced LoS communication.

\begin{table}[t]
    \centering
    \caption{System parameter values}
    \label{tab:my-table}
    \resizebox{\columnwidth}{!}{%
    \begin{tabular}{|c|c|l|}
    \hline
    \textbf{Parameters} & \textbf{Values} & \textbf{Descriptions} \\
    \hline
    $N_d, N_u$ & $50$, $10$ & Number of devices, Number of UAVs \\
    $R,h$ & $500$, $120$ m & Radius, UAV altitude \\
    $P_d, P_u$ & $0.75$, $1.5$~W & Device transmit power, UAV power \\
    $\alpha_L, \alpha_N$ & $2.0$, $3.5$ & Path loss exponents of LoS and NLoS links \\
    $n_0^2$ & $4.14 \times 10^{-6}$ W & Noise power \\
    $a, b$ & $9.61$, $0.16$ & Environmental parameters \\
    $m_L, m_N$ & $4$, $1$ & Nakagami-m fading parameters \\
    $\theta$ & $-5$~dB & SINR threshold \\
    $M_b, M_u$ & $5$, $15$ & RBs at BS, RBs at each UAV\\
    $b^k=b^u$ & $1$~MHz & Bandwidth at a device and a UAV\\
    $\sigma_k$ & $2$~GHz & CPU cycle frequency \\
    $c_k$ & $20$ cycles/bit & Number of CPU cycles per training \\
    $\eta$ & $0.01$ & Learning rate \\
    $\xi$ & $64$ & Local batch size \\
    $E, G$ & $2$, $1E$ & Local and global aggregation periods \\
    \hline
    \end{tabular}%
    }
\end{table}

\begin{figure}[t]
    \centering
    \resizebox{0.85\linewidth}{!}{\begin{tikzpicture}[thick,scale=1, every node/.style={scale=1.3},font=\Huge]
%
%
\definecolor{mycolor1}{rgb}{0.00000,0.44706,0.74118}%
\definecolor{mycolor2}{rgb}{0.85098,0.32549,0.09804}%
\definecolor{mycolor3}{rgb}{0.49412,0.18431,0.55686}%
\definecolor{mycolor4}{rgb}{0.85000,0.32500,0.09800}%
%

\begin{axis}[%
width=8in,
height=6in,
at={(0.78in,0.556in)},
scale only axis,
xmin=-20,
xmax=10,
ytick={0,0.1,...,1},
xtick={-20,-15,...,10},
ticklabel style={font=\huge},
xlabel style={font=\color{white!15!black}},
xlabel={\huge{$\text{SINR threshold, } \theta$}},
ymin=0,
ymax=1,
ylabel style={font=\color{white!15!black}},
ylabel={\huge{$\text{Success probability} $}},
axis background/.style={fill=white},
xmajorgrids,
ymajorgrids,
legend style={at={(0.85,0.55)}, anchor=south west, legend cell align=left, align=left, draw=white!15!black,font=\LARGE,row sep=0.5em}
]
]
\addplot [color=mycolor1, line width=2pt]
  table[row sep=crcr]{%
-20	0.730396\\
-18	0.730375\\
-16	0.73027\\
-14	0.72979\\
-12	0.727902\\
-10	0.721727\\
-8	0.705402\\
-6	0.670778\\
-4	0.610672\\
-2	0.522419\\
0	0.410375\\
2	0.284879\\
4	0.159193\\
6	0.0580767\\
8	0.00969978\\
10	0.0004536\\
};
\addlegendentry{$\mathcal{P}^{\text{edge}}_1$}

\addplot [color=mycolor1, line width=2.5pt, only marks, mark size=8.0pt, mark=x, mark options={solid, mycolor1},forget plot]
  table[row sep=crcr]{%
-20	0.732529265697056\\
-18	0.732529265697056\\
-16	0.731465058531394\\
-14	0.731110322809507\\
-12	0.727562965590635\\
-10	0.712577496452643\\
-8	0.691317599907991\\
-6	0.655276418183532\\
-4	0.600596925073164\\
-2	0.510356283827377\\
0	0.397717347990201\\
2	0.271191556929541\\
4	0.146236065498625\\
6	0.048855449444617\\
8	0.00212841433132316\\
10	0\\
};

\addplot [color=mycolor2, line width=2pt]
  table[row sep=crcr]{%
-20	0.467356\\
-18	0.467336\\
-16	0.467243\\
-14	0.466861\\
-12	0.465538\\
-10	0.461778\\
-8	0.453137\\
-6	0.436953\\
-4	0.411626\\
-2	0.377417\\
0	0.335911\\
2	0.286558\\
4	0.221268\\
6	0.134082\\
8	0.051153\\
10	0.00944879\\
};
\addlegendentry{$\mathcal{P}^{\text{back}}_1$}

\addplot [color=mycolor2, line width=2.5pt, only marks, mark size=8.0pt, mark=x, mark options={solid, mycolor2},forget plot]
  table[row sep=crcr]{%
-20	0.467044444444444\\
-18	0.467044444444444\\
-16	0.466966666666667\\
-14	0.466677777777778\\
-12	0.4654\\
-10	0.4606\\
-8	0.451166666666667\\
-6	0.432344444444444\\
-4	0.404888888888889\\
-2	0.367822222222222\\
0	0.324355555555556\\
2	0.275111111111111\\
4	0.205166666666667\\
6	0.109877777777778\\
8	0.038144444444444\\
10	0.0021\\
};

\addplot [color=mycolor3, line width=2pt]
  table[row sep=crcr]{%
-20	0.784509\\
-18	0.690011\\
-16	0.5663\\
-14	0.417488\\
-12	0.260549\\
-10	0.125475\\
-8	0.0403404\\
-6	0.00690147\\
-4	0.000440273\\
-2	5.9926e-06\\
0	7.23257e-09\\
2	1.93987e-13\\
4	1.29924e-20\\
6	6.82817e-32\\
8	1.16917e-49\\
10	1.09049e-77\\
};
\addlegendentry{$\mathcal{P}^{\text{direct}}_1$}

\addplot [color=mycolor3, line width=2.5pt, only marks, mark size=8.0pt, mark=x, mark options={solid, mycolor3},forget plot]
  table[row sep=crcr]{%
-20	0.78397008\\
-18	0.681178037\\
-16	0.549195176\\
-14	0.395261985\\
-12	0.238509581\\
-10	0.111447339\\
-8	0.036056798\\
-6	0.006127367\\
-4	0.000426523\\
-2	1.0403e-05\\
0	0\\
2	0\\
4	0\\
6	0\\
8	0\\
10	0\\
};

\addplot [color=mycolor1, dashed,    dash pattern=on 8pt off 4pt, line width=2pt]
  table[row sep=crcr]{%
-20	0.931816\\
-18	0.931791\\
-16	0.931772\\
-14	0.931675\\
-12	0.931225\\
-10	0.929419\\
-8	0.923311\\
-6	0.906399\\
-4	0.868424\\
-2	0.798872\\
0	0.693526\\
2	0.558099\\
4	0.402565\\
6	0.237456\\
8	0.0933435\\
10	0.0173552\\
};
\addlegendentry{$\mathcal{P}^{\text{edge}}_2$}

\addplot [color=mycolor1, line width=2.5pt, only marks, mark size=8.0pt, mark=x, mark options={solid, mycolor1},forget plot]
  table[row sep=crcr]{%
-20	0.932152805411858\\
-18	0.932152805411858\\
-16	0.932152805411858\\
-14	0.932152805411858\\
-12	0.930760047751691\\
-10	0.92658177477119\\
-8	0.915240748109829\\
-6	0.8949980103462\\
-4	0.855483700756068\\
-2	0.784632076004775\\
0	0.677188287306009\\
2	0.540798738559491\\
4	0.387116756028651\\
6	0.221343479108635\\
8	0.08044022284123\\
10	0.00166661161957818\\
};

\addplot [color=mycolor4,dashed,    dash pattern=on 8pt off 4pt, line width=2pt]
  table[row sep=crcr]{%
-20	0.230939\\
-18	0.230851\\
-16	0.230517\\
-14	0.229452\\
-12	0.226696\\
-10	0.220922\\
-8	0.210972\\
-6	0.196443\\
-4	0.177845\\
-2	0.155828\\
0	0.128672\\
2	0.091343\\
4	0.046488\\
6	0.0133248\\
8	0.00161561\\
10	5.4495e-05\\
};
\addlegendentry{$\mathcal{P}^{\text{back}}_2$}

\addplot [color=mycolor2, line width=2.5pt, only marks, mark size=8.0pt, mark=x, mark options={solid, mycolor2},forget plot]
  table[row sep=crcr]{%
-20	0.229955555555556\\
-18	0.229855555555556\\
-16	0.229555555555556\\
-14	0.228455555555556\\
-12	0.225744444444444\\
-10	0.218633333333333\\
-8	0.211353385555556\\
-6	0.195383153333333\\
-4	0.175003822222222\\
-2	0.151416843333333\\
0	0.123352758888889\\
2	0.0835688588888889\\
4	0.042177777777778\\
6	0.011344444444444\\
8	0.000166666666666667\\
10	0\\
};

\addplot [color=mycolor3,dashed,    dash pattern=on 8pt off 4pt, line width=2pt]
  table[row sep=crcr]{%
-20	0.70701762\\
-18	0.61142664\\
-16	0.48883683\\
-14	0.34603347\\
-12	0.20273628\\
-10	0.088484664\\
-8	0.024399294\\
-6	0.00328383\\
-4	0.0001439082\\
-2	1.0873653e-06\\
0	5.20893e-10\\
2	3.2656671e-15\\
4	2.2125816e-23\\
6	3.1206522e-36\\
8	1.7490882e-56\\
10	1.8990135e-88\\
};
\addlegendentry{$\mathcal{P}^{\text{direct}}_2$}

\addplot [color=mycolor3, line width=2.5pt, only marks, mark size=8.0pt, mark=x, mark options={solid, mycolor3},forget plot]
  table[row sep=crcr]{%
-20	0.70394\\
-18	0.61417\\
-16	0.48064\\
-14	0.33133\\
-12	0.18777\\
-10	0.07945\\
-8	0.02122\\
-6	0.00287\\
-4	0.00017\\
-2	0\\
0	0\\
2	0\\
4	0\\
6	0\\
8	0\\
10	0\\
};

\end{axis}

\begin{axis}[%
width=7.778in,
height=5.833in,
at={(0in,0in)},
scale only axis,
xmin=0,
xmax=1,
ymin=0,
ymax=1,
axis line style={draw=none},
ticks=none,
axis x line*=bottom,
axis y line*=left
]
\end{axis}
        \end{tikzpicture}}
    \setlength{\belowcaptionskip}{-10pt}
    \caption{Success probabilities as a function of the SINR threshold for the devices and UAVs pairs illustrated in Fig.~\ref{Systemmodel_v2}.}
    \label{fig:fig1}
\end{figure}
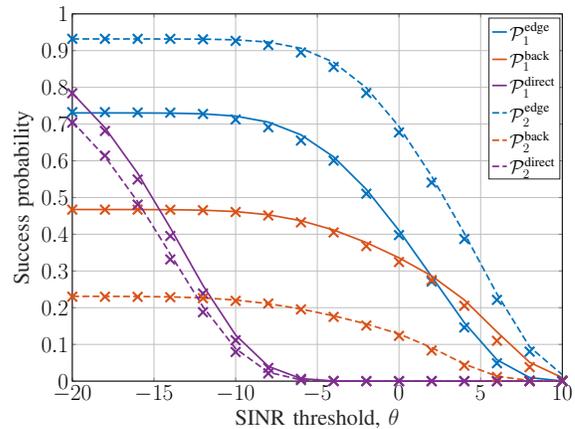
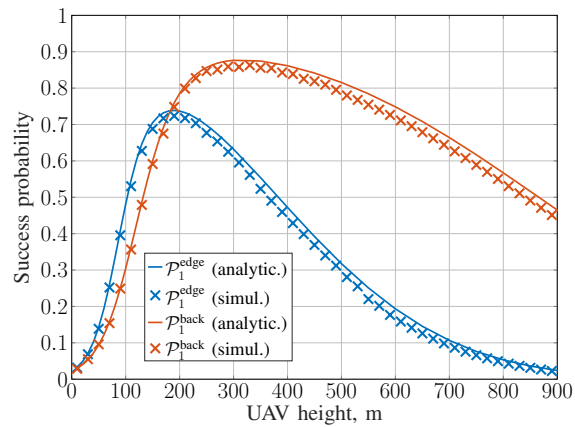
\begin{figure}[t]
    \centering
    \resizebox{0.85\linewidth}{!}{\begin{tikzpicture}[thick,scale=1, every node/.style={scale=1.3},font=\Huge]
%
%
\definecolor{mycolor1}{rgb}{0.00000,0.44706,0.74118}%
\definecolor{mycolor2}{rgb}{0.85098,0.32549,0.09804}%
\definecolor{mycolor3}{rgb}{0.63500,0.07800,0.18400}%
%

\begin{axis}[%
width=8in,
height=6in,
at={(0.78in,0.556in)},
scale only axis,
xmin=0,
xmax=900,
ytick={0,0.1,...,1},
xtick={0,100,...,900},
ticklabel style={font=\huge},
xlabel style={font=\color{white!15!black}},
xlabel={\huge{UAV height, m}},
ymin=0,
ymax=1,
ylabel style={font=\color{white!15!black}},
ylabel={\huge{$\text{Success probability} $}},
axis background/.style={fill=white},
xmajorgrids,
ymajorgrids,
legend style={at={(0.15,0.05)}, anchor=south west, legend cell align=left, align=left, draw=white!15!black,font=\LARGE,row sep=0.5em}
]
\addplot [color=mycolor1, line width=2pt]
  table[row sep=crcr]{%
10	0.0328158\\
20	0.0485832\\
30	0.0707774\\
40	0.101511\\
50	0.142808\\
60	0.195828\\
70	0.259957\\
80	0.332253\\
90	0.407782\\
100	0.480881\\
110	0.546729\\
120	0.602365\\
130	0.646818\\
140	0.680597\\
150	0.705014\\
160	0.721636\\
170	0.731965\\
180	0.737288\\
190	0.738644\\
200	0.736845\\
210	0.732516\\
220	0.726131\\
230	0.718056\\
240	0.708573\\
250	0.697902\\
260	0.686221\\
270	0.673674\\
280	0.660385\\
290	0.646456\\
300	0.631981\\
350	0.554019\\
400	0.472187\\
450	0.392077\\
500	0.317503\\
550	0.250848\\
600	0.193361\\
650	0.145396\\
700	0.106629\\
750	0.0762527\\
800	0.0531671\\
850	0.0361423\\
900	0.0239541\\
950	0.0154797\\
1000	0.00975463\\
};
\addlegendentry{$\mathcal{P}^{\text{edge}}_1$ (analytic.)}

\addplot [color=mycolor1, line width=2.5pt, only marks, mark size=8.0pt, mark=x, mark options={solid, mycolor1}]
  table[row sep=crcr]{%
10	0.031831326\\
30	0.068654078\\
50	0.13852376\\
70	0.25215829\\
90	0.39554854\\
110	0.53032713\\
130	0.62741346\\
150	0.68786358\\
170	0.71700605\\
190	0.723648468\\
210	0.71754052\\
230	0.70351432\\
250	0.67696494\\
270	0.65346378\\
290	0.62506232\\
310	0.5955288\\
330	0.5614163268\\
350	0.5232764771\\
370	0.4905786737\\
390	0.4596917494\\
410	0.4289393738\\
430	0.3985981715\\
450	0.3689052493\\
470	0.3400610189\\
490	0.3122301378\\
510	0.2805471547\\
530	0.2551183913\\
550	0.220228832\\
570	0.20173170844\\
590	0.176672077916\\
610	0.158416849024\\
630	0.141484299716\\
650	0.125858848688\\
670	0.111511930216\\
690	0.09840459104\\
710	0.0864886156804\\
730	0.075709556136\\
750	0.0660064721956\\
770	0.0573142686336\\
790	0.0495653399032\\
810	0.0426904357644\\
830	0.0366201328516\\
850	0.0312857868644\\
870	0.0266203116328\\
890	0.0225588716196\\
910	0.019039920674\\
930	0.0160049423432\\
950	0.0133996617516\\
970	0.0111732665356\\
990	0.00927953216\\
};
\addlegendentry{$\mathcal{P}^{\text{edge}}_1$ (simul.)}

\addplot [color=mycolor2, line width=2pt]
  table[row sep=crcr]{%
10	0.0296776\\
20	0.0400866\\
30	0.0537171\\
40	0.0714057\\
50	0.0940756\\
60	0.122645\\
70	0.157872\\
80	0.200149\\
90	0.249295\\
100	0.304408\\
110	0.363859\\
120	0.425462\\
130	0.486803\\
140	0.545616\\
150	0.600089\\
160	0.649032\\
170	0.691882\\
180	0.728601\\
190	0.759519\\
200	0.785184\\
210	0.806241\\
220	0.823341\\
230	0.837095\\
240	0.848048\\
250	0.856668\\
260	0.863348\\
270	0.868415\\
280	0.872136\\
290	0.874727\\
300	0.876367\\
350	0.874399\\
400	0.861494\\
450	0.841402\\
500	0.815509\\
550	0.784415\\
600	0.748512\\
650	0.70822\\
700	0.664082\\
750	0.616798\\
800	0.56721\\
850	0.516267\\
900	0.46497\\
950	0.41431\\
1000	0.36521\\
};
\addlegendentry{$\mathcal{P}^{\text{back}}_1$ (analytic.)}

\addplot [color=mycolor2, line width=2.5pt, only marks, mark size=8.0pt, mark=x, mark options={solid, mycolor2}]
  table[row sep=crcr]{%
10	0.0286\\
30	0.0551\\
50	0.0959\\
70	0.1541\\
90	0.249\\
110	0.3564\\
130	0.479\\
150	0.5914\\
170	0.675\\
190	0.7482\\
210	0.7986\\
230	0.8274\\
250	0.8469\\
270	0.8511\\
290	0.8592\\
310	0.8581\\
330	0.863\\
350	0.856\\
370	0.8549\\
390	0.8429\\
410	0.8387\\
430	0.8255\\
450	0.8188\\
470	0.8095\\
490	0.7953\\
510	0.779803108227\\
530	0.767536588269\\
550	0.754511294493\\
570	0.740751484374\\
590	0.726280445088\\
610	0.711127285605\\
630	0.695322085194\\
650	0.67889880432\\
670	0.661896254943\\
690	0.644358100518\\
710	0.626329945098\\
730	0.607861273932\\
750	0.589005453465\\
770	0.569819731338\\
790	0.55036329579\\
810	0.53069533506\\
830	0.510878918583\\
850	0.490975175196\\
870	0.471048144633\\
890	0.45115992603\\
910	0.431371648224\\
930	0.411742499454\\
950	0.392328757062\\
970	0.373184757792\\
990	0.35436289779\\
};
\addlegendentry{$\mathcal{P}^{\text{back}}_1$ (simul.)}

\end{axis}

\begin{axis}[%
width=7.778in,
height=5.833in,
at={(0in,0in)},
scale only axis,
xmin=0,
xmax=1,
ymin=0,
ymax=1,
axis line style={draw=none},
ticks=none,
axis x line*=bottom,
axis y line*=left
]
\end{axis}
        \end{tikzpicture}}
    \setlength{\belowcaptionskip}{-10pt}
    \caption{Edge and backhaul success probabilities of the first device/UAV pair as a function of the UAV height.}
    \label{DifferentHeightPlot_v1}
\end{figure}
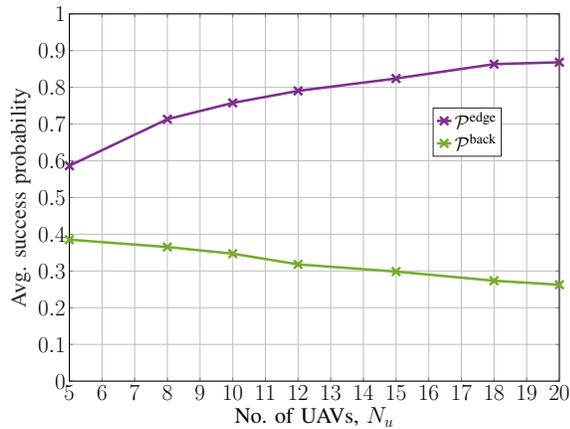
\begin{figure}[t]
    \centering
    \resizebox{0.85\linewidth}{!}{\begin{tikzpicture}[thick,scale=1, every node/.style={scale=1.3},font=\Huge]
%
%

\definecolor{mycolor1}{rgb}{0.46667,0.67451,0.18824}%
\definecolor{mycolor2}{rgb}{0.49400,0.18400,0.55600}%

\begin{axis}[%
width=8in,
height=6in,
at={(0.18in,0.556in)},
scale only axis,
bar shift auto,
xmin=5,
xmax=20,
xticklabel style={font=\fontsize{20pt}{18pt}\selectfont },
xlabel style={font=\fontsize{18pt}{18pt}\selectfont },
xlabel={\huge{$\text{No. of UAVs, $N_u$} $}},
ymin=0,
ymax=1,
yticklabel style={font=\fontsize{20pt}{18pt}\selectfont },
ylabel style={font=\color{white!15!black}},
ylabel={\huge{$\text{Avg. success probability} $}},
axis background/.style={fill=white},
xmajorgrids,
ymajorgrids,
legend style={at={(0.742,0.612)}, anchor=south west, legend cell align=left, align=left, draw=white!15!black,font=\fontsize{18pt}{20pt}\selectfont,}
]
\addplot[color=mycolor2, line width=3.0pt,mark repeat={1}, mark size=7.0pt, mark=x, mark options={solid, mycolor2}] table[row sep=crcr] {%
5	0.5866\\
8	0.7133\\
10	0.7576\\
12	0.7904\\
15	0.8241\\
18	0.8632\\
20	0.8683\\
};\addlegendentry{$\mathcal{P}^{\text{edge}}$}

\addplot[color=mycolor1, line width=3.0pt,mark repeat={1}, mark size=7.0pt, mark=x, mark options={solid, mycolor1}] table[row sep=crcr] {%
5	0.3853\\
8	0.3652\\
10	0.3471\\
12	0.3180\\
15	0.2984\\
18	0.2734\\
20	0.2625\\
};\addlegendentry{$\mathcal{P}^{\text{back}}$}
\end{axis}

\begin{axis}[%
width=7.778in,
height=5.833in,
at={(0in,0in)},
scale only axis,
xmin=0,
xmax=1,
ymin=0,
ymax=1,
axis line style={draw=none},
ticks=none,
axis x line*=bottom,
axis y line*=left
]
\end{axis}
        \end{tikzpicture}}
    \setlength{\belowcaptionskip}{-10pt}
    \caption{\color{black}Average edge and backhaul success probabilities as a function of the number of UAVs.}
    \label{DifferentUAVs_v2}
\end{figure}

Using the developed framework, we investigate the impact of different system parameters, including the height of UAVs. Fig.~\ref{DifferentHeightPlot_v1} illustrates how the edge and backhaul success probabilities for the first pair of UAVs and devices in Fig.~\ref{Systemmodel_v2} varies with increasing the UAVs height $h$.
From Fig.~\ref{DifferentHeightPlot_v1}, it is evident that both backhaul and edge success probabilities initially increase before gradually declining as UAV heights are raised. This pattern arises due to the initial elevation contributing to an enhanced LoS probability. However, at a certain height, the LoS probability reaches saturation, achieving a value of $1$. Beyond this point, further height increments result in a steady decrease due to the path-loss factor starting to dominate. 

To assess the effectiveness of the proposed HFL algorithm, we rely on the training and testing accuracies and the latency of the global model as the main performance metrics. The considered model is based on a multi-layer perceptron (MLP) consisting of a single hidden layer with $300$ neurons followed by the rectified linear unit (ReLU) activation function. The simulations are based on the MNIST dataset~\cite{deng2012mnist}, which has become a baseline for evaluating FL models. The dataset is distributed in a non-IID fashion. Thus, the input of the model is a flattened $784$-dimensional image. Although our convergence analysis assumes full batch training, the simulations are done with a local mini-batch of size $\xi=64$. It should be noted that training in batches can be easily integrated into the analysis, as shown in~\cite{wang2022demystifying}.

\begin{figure}[t]
    \centering
    \resizebox{0.85\linewidth}{!}{\begin{tikzpicture}[thick,scale=1, every node/.style={scale=1.3},font=\Huge]
        \input{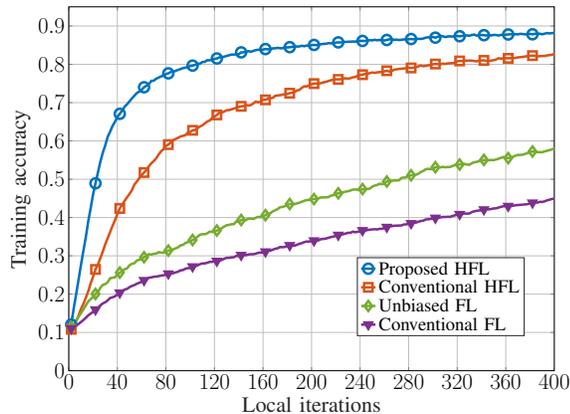}
        \end{tikzpicture}}
    \setlength{\belowcaptionskip}{-10pt}
    \caption{Training accuracy of the global model for different FL frameworks.}
    \label{fig:fig2}
\end{figure}
\begin{figure}[t]
    \centering
    \resizebox{0.85\linewidth}{!}{\begin{tikzpicture}[thick,scale=1, every node/.style={scale=1.3},font=\Huge]
        \input{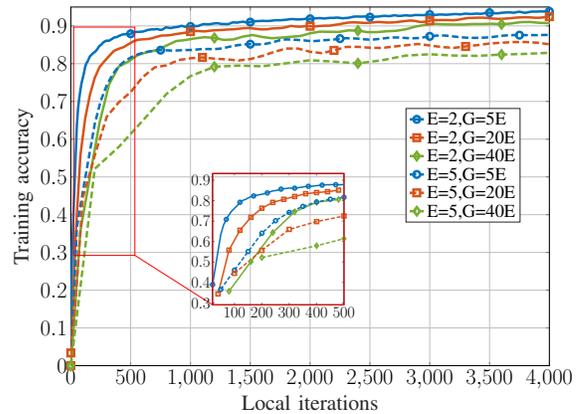}
        \end{tikzpicture}}
    \setlength{\belowcaptionskip}{-10pt}
    \caption{Accuracy plots of the proposed HFL for different values of local and global aggregation periods.}
    \label{fig:fig3}
\end{figure}
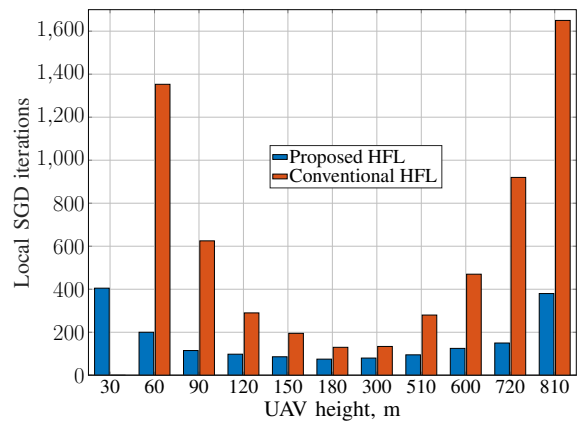
\begin{figure}[t]
    \centering
    \resizebox{0.85\linewidth}{!}{\begin{tikzpicture}[thick,scale=1, every node/.style={scale=1.3},font=\Huge]
%
%
\definecolor{mycolor2}{rgb}{0.00000,0.44700,0.74100}%
\definecolor{mycolor1}{rgb}{0.85000,0.32500,0.09800}%
%

\begin{axis}[%
width=8in,
height=6in,
at={(0.18in,0.556in)},
scale only axis,
bar shift auto,
xmin=0.514285714285714,
xmax=11.4857142857143,
xtick={1,2,3,4,5,6,7,8,9,10,11},
xticklabels={{30},{60},{90},{120},{150},{180},{300},{510},{600},{720},{810}},
xticklabel style={
        font=\fontsize{18pt}{18pt}\selectfont 
    },
xlabel={\huge{$\text{UAV height, m} $}},
ymin=0,
ymax=1700,
ytick={0,200,...,1700},
yticklabel style={font=\fontsize{20pt}{18pt}\selectfont },
ylabel style={font=\color{white!15!black}},
ylabel={\huge{$\text{Local SGD iterations} $}},
axis background/.style={fill=white},
xmajorgrids,
ymajorgrids,
legend style={at={(0.372,0.512)}, anchor=south west, legend cell align=left, align=left, draw=white!15!black,font=\fontsize{18pt}{20pt}\selectfont}
]

\addplot[ybar, bar width=0.329, fill=mycolor2, draw=black, area legend] table[row sep=crcr] {%
1	405\\
2	200\\
3	115\\
4	98\\
5	86\\
6	75\\
7	80\\
8	95\\
9	125\\
10	150\\
11	380\\
};
\addplot[forget plot, color=white!15!black] table[row sep=crcr] {%
0.514285714285714	0\\
11.4857142857143	0\\
};
\addlegendentry{Proposed HFL}

\addplot[ybar, bar width=0.329, fill=mycolor1, draw=black, area legend] table[row sep=crcr] {%
1	0\\
2	1353\\
3	625\\
4	290\\
5	195\\
6	130\\
7	134\\
8	280\\
9	470\\
10	920\\
11	1650\\
};
\addplot[forget plot, color=white!15!black] table[row sep=crcr] {%
0.514285714285714	0\\
11.4857142857143	0\\
};
\addlegendentry{Conventional HFL}

\end{axis}

\begin{axis}[%
width=7.778in,
height=5.833in,
at={(0in,0in)},
scale only axis,
xmin=0,
xmax=1,
ymin=0,
ymax=1,
axis line style={draw=none},
ticks=none,
axis x line*=bottom,
axis y line*=left
]
\end{axis}
        \end{tikzpicture}}
    \setlength{\belowcaptionskip}{-10pt}
    \caption{Number of iterations to reach $80\%$ of testing accuracy across different UAV heights.}
    \label{DifferentHeightPlot_v2}
\end{figure}

Fig.~\ref{DifferentUAVs_v2} depicts the average backhaul and edge success probabilities as function of the number of UAVs deployed as edge servers for the proposed HFL algorithm. When only $5$ UAVs are available, no interference from other UAVs occurs since $M_b=5$ RBs are available at the BS for backhaul communication with UAVs. This corresponds to the highest backhaul success probability as seen in Fig.~\ref{DifferentUAVs_v2} and yet the lowest edge success probability due to the increased distances between a UAV and its devices. On the other side, when the number of available UAVs is $20$, the backhaul success probability is reduced due to increased interference levels caused by the low number of backhaul RBs. However, the edge success probability is maximized due to smaller device clusters and shorter edge links between UAVs and their associated devices. 

Fig.~\ref{fig:fig2} illustrates the training accuracy of the global model at the BS for different learning paradigms. Namely, we plot the training accuracy of the proposed HFL algorithm in addition to the conventional HFL when no unbiasing is considered. Furthermore, we plot the training accuracy of the conventional FL algorithm, known as FedAvg, as well as its unbiased counterpart studied in~\cite{zhagypar2023characterization}. In this algorithm, the transmissions occur between the BS and devices directly. As expected, the unbiased FedAvg performs better than the conventional one by integrating the success probabilities into the aggregation rule. However, the use of UAVs in the conventional HFL improves the performance of the system. Yet, the proposed unbiased HFL brings a significant leap in performance, reaching the highest accuracy in fewer local iterations.

Given the hierarchical structure of the proposed algorithm, it becomes intriguing to investigate how the number of local aggregations at UAVs before the global aggregation impacts the system. Fig.~\ref{fig:fig3} showcases the results of simulations, displaying accuracy plots of the proposed HFL algorithm under various values of the local and global aggregations $E$ and $G$. Notably, the key observation is that as we increase both the local and global aggregation periods, the training accuracy of the proposed algorithm demonstrates a decline. This outcome is intuitive, suggesting that extended training periods without aggregations lead to individual updates deviating from global optimality. This aligns with the findings from Theorem~\ref{Theorem3}, wherein increased values of $E$ and $G$ result in looser upper bounds, reinforcing the observed trend in the simulations.


The learning performance exhibits a strong dependence on the UAVs' altitude, as illustrated in Fig.~\ref{DifferentHeightPlot_v2}. The presented bar plot shows how many local iterations are needed to reach an $80\%$ testing accuracy on the global model for different values of the UAV height for the proposed and the conventional HFL algorithms. We can notice that at around $180$~meters, there is an optimal UAV height requiring the fewest iterations to hit the desired accuracy. The lowest number of iterations corresponds to the UAV height that optimizes the edge and backhaul success probabilities, as shown in Fig.~\ref{DifferentHeightPlot_v1}. Furthermore, the conventional HFL presents a higher number of iterations to achieve the required accuracy and a larger variation in the number of iterations compared with the proposed HFL algorithm. Such behavior is mainly due to the impact of the edge and backhaul channels, which is mitigated thanks to the unbiasing method applied in the proposed HFL algorithm. Fig.~\ref{DifferentHeightPlot_v2} also shows that, at the highest and lowest ends of the height range, conventional HFL exhibits a rapid increase in required iterations, unlike the proposed HFL algorithm. This discrepancy occurs because, at extreme UAV heights, the BS receives fewer updates from devices, mainly those with robust channel conditions. This unequal input of updates heavily biases the global model. However, the proposed method tackles this bias by adjusting updates based on success probabilities, effectively mitigating the bias typical in conventional methods. Note that there is no bar for the conventional HFL at $30$ meters, indicating its inefficiency in achieving the expected accuracy level even provided with an unlimited number of iterations.


\begin{figure}[t]
    \centering
    \resizebox{0.85\linewidth}{!}{\begin{tikzpicture}[thick,scale=1, every node/.style={scale=1.3},font=\Huge]
%
%
\definecolor{mycolor1}{rgb}{0.00000,0.44700,0.74100}%
\definecolor{mycolor2}{rgb}{0.85000,0.32500,0.09800}%


\begin{axis}[%
width=8in,
height=6in,
at={(0.18in,0.556in)},
scale only axis,
bar shift auto,
xtick={1,2,3,4,5,6,7},
xticklabel style={font=\fontsize{20pt}{18pt}\selectfont },
xticklabels={{5},{8},{10},{12},{15},{18},{20}},
xlabel style={font=\fontsize{18pt}{18pt}\selectfont },
xlabel={\huge{$\text{No. of UAVs, $N_u$} $}},
ymin=0,
ymax=500,
yticklabel style={font=\fontsize{20pt}{18pt}\selectfont },
ylabel style={font=\color{white!15!black}},
ylabel={\huge{$\text{Iterations} $}},
axis background/.style={fill=white},
xmajorgrids,
ymajorgrids,
legend style={at={(0.342,0.712)}, anchor=south west, legend cell align=left, align=left, draw=white!15!black,font=\fontsize{18pt}{20pt}\selectfont}
]
\addplot[ybar, bar width=0.25, fill=mycolor1, draw=black, area legend] table[row sep=crcr] {%
1	180\\
2	200\\
3	132\\
4	190\\
5	210\\
6	160\\
7	150\\
};\addlegendentry{Proposed HFL}


\addplot[ybar, bar width=0.25, fill=mycolor2, draw=black, area legend] table[row sep=crcr] {%
1	190\\
2	270\\
3	285\\
4	295\\
5	300\\
6	365\\
7	480\\
};\addlegendentry{Conventional HFL}

\end{axis}

\begin{axis}[%
width=7.778in,
height=5.833in,
at={(0in,0in)},
scale only axis,
xmin=0,
xmax=1,
ymin=0,
ymax=1,
axis line style={draw=none},
ticks=none,
axis x line*=bottom,
axis y line*=left
]
\end{axis}
        \end{tikzpicture}}
    \setlength{\belowcaptionskip}{-10pt}
    \caption{Number of iterations to reach $80\%$ of testing accuracy across different numbers of UAVs.}
    \label{DifferentUAVs}
\end{figure}
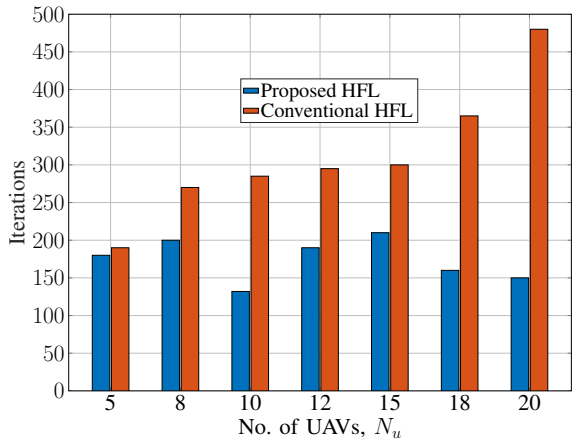
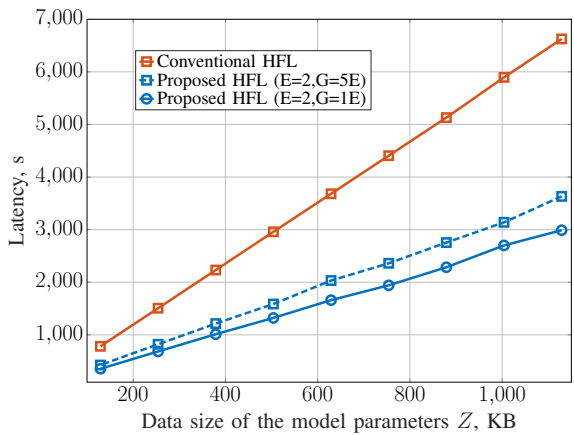
\begin{figure}[t]
    \centering
    \resizebox{0.85\linewidth}{!}{\begin{tikzpicture}[thick,scale=1, every node/.style={scale=1.3},font=\Huge]
%
%
\definecolor{mycolor1}{rgb}{0.00000,0.44700,0.74100}%
\definecolor{mycolor2}{rgb}{0.85000,0.32500,0.09800}%
\definecolor{mycolor3}{rgb}{0.46667,0.67451,0.18824}%
\definecolor{mycolor4}{rgb}{0.49400,0.18400,0.55600}%
%

\begin{axis}[%
width=8in,
height=6in,
at={(0.78in,0.556in)},
scale only axis,
xmin=100,
xmax=1150,
ticklabel style={font=\huge},
xlabel style={font=\color{white!15!black}},
xlabel={\huge{$\text{Data size of the model parameters $Z$, KB}$}},
ymin=100,
ymax=7000,
ytick={0,1000,2000,...,7000},
xtick={0,200,400,...,1150},
ylabel style={font=\color{white!15!black}},
ylabel={\huge{$\text{Latency, s} $}},
axis background/.style={fill=white},
xmajorgrids,
ymajorgrids,
legend style={at={(0.1,0.75)}, anchor=south west, legend cell align=left, align=left, draw=white!15!black,font=\LARGE}
]
\addplot [color=mycolor2, line width=3.0pt, mark repeat={1}, mark size=5.0pt, mark=square, mark options={solid, mycolor2}]
  table[row sep=crcr]{%
1129	6625\\
1004	5895\\
879	5130\\
754	4406\\
629	3682\\
504	2959\\
379	2232\\
254	1506\\
129	782\\
};
\addlegendentry{ Conventional HFL}

\addplot [color=mycolor1, line width=3.0pt,  dashed,    dash pattern=on 8pt off 4pt, mark repeat={1}, mark size=5.0pt, mark=square, mark options={solid, mycolor1}]
  table[row sep=crcr]{%
1129	3632.4\\
1004	3139.2\\
879	2754\\
754	2360.4\\
629	2032.8\\
504	1586.4\\
379	1214.4\\
254	820.8\\
129	426\\
};
\addlegendentry{ Proposed HFL (E=2,G=5E)}

\addplot [color=mycolor1, line width=3.0pt, mark repeat={1}, mark size=5.5pt, mark=o, mark options={solid, mycolor1}]
  table[row sep=crcr]{%
1129	2990\\
1004	2700\\
879	2285\\
754	1942\\
629	1660\\
504	1322\\
379	1012\\
254	684\\
129	355\\
};
\addlegendentry{ Proposed HFL (E=2,G=1E)}

\end{axis}

\begin{axis}[%
width=7.764in,
height=5.819in,
at={(0in,0in)},
scale only axis,
xmin=0,
xmax=1,
ymin=0,
ymax=1,
axis line style={draw=none},
ticks=none,
axis x line*=bottom,
axis y line*=left
]
\end{axis}
        \end{tikzpicture}}
    \setlength{\belowcaptionskip}{-10pt}
    \caption{Latency as a function of the model parameters' data size for the conventional and the proposed HFL algorithms.}
    \label{fig:fig6}
\end{figure}
A bar plot in Fig.~\ref{DifferentUAVs} illustrates the number of iterations needed to reach $80\%$ of accuracy as a function of the number of UAVs deployed as edge servers for the proposed and the conventional HFL algorithms. As the number of UAVs increases, more iterations are required for the conventional HFL to reach target accuracy. This is mainly due to the fact that, as more UAVs are added as edge servers, higher interference levels are expected at the backhaul connection as more UAVs will compete over a limited number of resource blocks ($M_b=5$). This can be clearly noticed in Fig~\ref{DifferentUAVs_v2} which shows the decrease of the backhaul probability with the number of UAVs. On the other side, although the edge success probability improves when more UAVs are added to the network, the impact of the backhaul channel in more prominent causing the increase in the number of iterations. The reason for such behavior is that aggregated updates from UAVs are more appreciated than local updates from single devices. Furthermore, as the number of UAVs decreases, each UAV will manage more devices and data samples, resulting in their aggregated updates being less susceptible to a strong bias and thus being closer to a global optimum. Conversely, when the number of UAVs increases, each UAV update represents only a small dataset fraction, which puts it further from the global optimum. On the other side, a different behavior is noticed for the proposed HFL algorithm as the impact of edge and backhaul channels is mitigated through the applied unbiasing technique. At the two extreme values of the UAVs number ($N_u=5$ and $N_u=20$), a lower number of iterations is required due to the excellence of one of the success probabilities compared to other cases. Yet, the fewest number of iterations is achieved when $N_u=10$ UAVs are added. Here, both backhaul and edge success probabilities attain relatively average values, enhancing the overall success of update transmissions from devices to the BS via UAVs. The importance of UAV updates has a slighter impact compared to conventional HFL since the unbiasing technique mitigates biases at both UAV and BS levels. Note that although we unbias the aggregation rules, the lower values of the success probabilities still result in fewer device participation, which leads to slower convergence. Overall, the proposed HFL algorithm achieves the expected accuracy faster than conventional HFL in all cases, showcasing its efficacy.

The final simulation result displayed in Fig.~\ref{fig:fig6} depicts the total elapsed latency to reach $80\%$ of accuracy for the global model plotted against different data sizes of the model parameters. The main observation is that it takes less time for the proposed HFL algorithm to complete the training compared to the conventional HFL algorithm. Note that the conventional and unbiased FL algorithms take way longer to reach the same accuracy value, thus, not plotted in Fig~\ref{fig:fig6}. Additionally, the figure illustrates latency plots for two different periods of global aggregation. The dashed blue plot highlights that extending the global period leads to a slower convergence, aligning with the conclusions drawn from Fig.~\ref{fig:fig3}.


    

\section{Conclusion}\label{sec:conclusions}
HFL is expected to be a key enabler of large-scale distributed learning in future 6G networks. In the paper, we propose a robust HFL algorithm designed for UAV-assisted wireless networks. The inherent unreliability of wireless channels often introduces biases favoring devices and UAVs with stronger channel conditions. To this end, our paper proposes an HFL algorithm that leverages stochastic geometry tools to characterize the reliability of edge and backhaul links, so as to effectively mitigate biases during local and global data aggregations. The convergence of the proposed algorithm is proven theoretically and via simulations. The derived theoretical convergence bound, which is a strong function of the system parameters, accurately reflects the impact of channel unreliability. Through extensive simulations, our algorithm demonstrates superior performance compared to conventional FL, conventional HFL, and unbiased FL algorithms. Notably, our framework allows for the optimization of critical system parameters such as the number and altitude of UAVs, highlighting its numerical prospects in the context of future HFL-empowered networks.

\bibliographystyle{IEEEtran}
\bibliography{IEEEabrv,Bibliography}


\appendix
\subsection{Proof of Theorem 1}
\label{appendix:success_edge}
In the UAV-assisted wireless network where UAVs and devices are modeled as independent BPPs, the expression of the edge success probability $\mathcal{P}^{\text{edge}}_k$ of device $k$ can be defined as
\begin{equation}\small
\begin{aligned}
   \mathcal{P}^{\text{edge}}_k &= \mathbb{E}_{z\in\{\text{L},\text{N}\}}\left[\mathbb{I}\left[\text{SINR}^{\text{edge}_1}_z>\theta,\text{SINR}^{\text{edge}_2}_z>\theta|l_k\right]\right]\\
   &\stackrel{\text{(a)}}{=} P_\text{L}(l_k) \mathcal{P}_k^{\text{L},\text{edge}_1} \mathcal{P}_k^{\text{L},\text{edge}_2} + P_\text{N}(l_k) \mathcal{P}_k^{\text{N},\text{edge}_1} \mathcal{P}_k^{\text{N},\text{edge}_2},
\end{aligned}
\end{equation}
where (a) is obtained due to the independence of the downlink and uplink transmissions when conditioned on the serving distance $l_k$ between the device and its serving UAV. $\mathcal{P}_k^{z,\text{edge}_1}$ and $\mathcal{P}_k^{z,\text{edge}_2}$, $z \in \{\text{L},\text{N}\}$ represent the UAV-to-device and device-to-UAV success probabilities for device $k$ when associated with a UAV through a LoS or a NLoS link. The expressions of these edge success probabilities can be obtained as
\begin{equation}\small
\begin{aligned}
   &\mathcal{P}_k^{z,\text{edge}_1} = \mathbb{E}\left[\mathbb{I}\left[\text{SINR}^{\text{edge}_1}_z>\theta|l_k\right]\right]\\
   &= \mathbb{E}\left[\mathbb{P}\left[ |h_{l_k}|^2 > \frac{\theta (n_0^2 + I_{\text{edge}_1})}{P_u l_k^{-\alpha_z}}|l_k\right]\right]\\
   &\stackrel{\text{(a)}}{=} 1 - \mathbb{E}_{I_{\text{edge}_1}}\left[\left(1-\exp\left( \frac{-\eta_z \theta (n_0^2 + I_{\text{edge}_1})}{P_u l_k^{-\alpha_z}}\right)\right)^{m_z} \middle| l_k\right]\\
   &\stackrel{\text{(b)}}{=} \sum_{j=1}^{m_z} \!\binom{m_z}{j} (-1)^{j+1} \!\!\exp\left(\frac{-j \eta_z n_0^2 \theta}{P_u l_k^{-\alpha_z}}\right) \!\mathcal{L}_{\text{edge}_1}^z\!\!\left(\exp\left(\frac{-j \eta_z \theta}{P_u l_k^{-\alpha_z}}\right)\right),
\end{aligned}
\end{equation}
where $\eta_z = m_z(m_z!)^{-1/m_z}$, $m_z$ is the Nakagami fading parameter, (a) follows from applying the cumulative distribution function (CDF) of the normalized gamma fading coefficient, and (b) follows from the use of binomial expansion. 

Following similar steps, we can obtain the expression of the success probability of update transmissions from a device to UAV $\mathcal{P}_k^{z,\text{edge}_1}$ as follows
\begin{equation}\small
\begin{aligned}
   \mathcal{P}_k^{z,\text{edge}_2}&= \sum_{j=1}^{m_z} \binom{m_z}{j} (-1)^{j+1} \exp\left(\frac{-j \eta_z n_0^2 \theta}{P_d l_k^{-\alpha_z}}\right)\\
   &\times \mathcal{L}_{\text{edge}_2}^z\left(\exp\left(\frac{-j \eta_z \theta}{P_d l_k^{-\alpha_z}}\right)\right).
\end{aligned}
\end{equation}
\subsection{Proof of Lemma 1}
\label{appendix:laplace_edge1}
The total interference $I_{\text{edge}_1}$ on the UAV-to-device communication has signals coming from both LoS and NLoS interfering UAVs. The Laplace transform of the interference coming from LoS UAVs $\mathcal{L}_{\text{edge}_1}^\text{L}(s)$ is derived as
\begin{equation}\small
\begin{aligned}
   \mathcal{L}_{\text{edge}_1}^\text{L}(s) &=  \mathbb{E}\left[\exp{(-sI_{\text{edge}_1})}\right]\\
   &=   \mathbb{E}\left[\exp{(-s \sum^{N_u^\text{interf}}_{u=1}I_{y_0,y_u})}\right]\\
   &\stackrel{\text{(a)}}{=} \prod^{N_u^\text{interf}}_{u=1} \mathbb{E}_{I_{y_0,y_u}}\left[\exp{(-sI_{y_0,y_u})}\right]\\
   &\stackrel{\text{(b)}}{=}  \left( \mathbb{E}_{I_{y_0,y_u}}\left[\exp{(-sI_{y_0,y_u})}\right]\right)^{N_u^\text{interf}}
\end{aligned}
\end{equation}
where (a) results from the IID distribution of the fading gains and from their independence of the BPP process, (b) follows from the IID distribution of the interferers distance, {\small$N_u^{\text{interf}} = N_u\times P_{\text{active}}^{\text{edge}_1}-1$} represents the average number of interfering UAVs with {\small$P_{\text{active}}^{\text{edge}_1}=\frac{N_u/M_b}{N_u}$} is the probability of a UAV using the same RB as the serving UAV, and $\mathbb{E}_{I_{y_0,y_u}}\left[\exp{(-sI_{y_0,y_u})}\right]$ can be calculated as

\begin{equation}\footnotesize
\begin{aligned}
   &\mathbb{E}_{I_{y_0,y_u}}\left[\exp{(-sI_{y_0,y_u})}\right] \\
   &= \frac{ \int_{l_k}^{w_p} P_\text{L}(r) f_{R_1}(r|x_0)dr \times \mathbb{E}_{|h_{y_u}|^2_\text{L},y_u}\left[\exp\left( -sP_u |h_{y_u}|^2_\text{L} y_u^{-\alpha_\text{L}} \right) \right]}{ \int_{l_k}^{w_p} P_\text{L}(r) f_{R_1}(r|x_0)dr+\int_{E_{\text{LN}}(l_k)}^{w_p}P_\text{N}(r) f_{R_1}(r|x_0)dr}  \\
   & + \frac{ \int_{E_{\text{LN}}(l_k)}^{w_p}P_\text{N}(r) f_{R_1}(r|x_0)dr\times \mathbb{E}_{|h_{y_u}|^2_\text{N},y_u}\left[\exp\left( -sP_u |h_{y_u}|^2_\text{N} y_u^{-\alpha_\text{N}} \right) \right]}{ \int_{l_k}^{w_p} P_\text{L}(r) f_{R_1}(r|x_0)dr+\int_{E_{\text{LN}}(l_k)}^{w_p}P_\text{N}(r) f_{R_1}(r|x_0)dr} ,
\end{aligned}
\end{equation}

where $y_u$ is the distance of the device to the $u$-th interfering UAV. The lower limit of the integral in the numerator indicates that for any of the $N_u^\text{interf}$ interfering UAVs, we have NLoS UAVs located further than the distance $E_{\text{LN}}(l_k)= l_k^{\frac{\alpha_\text{L}}{\alpha_\text{N}}}$ due to {\color{black}maximum received average power association policy}, where $l_k$ is the distance to the serving UAV.
The expectation term in the numerator can further be expressed as
\begin{equation}\small
\begin{aligned}
&\mathbb{E}_{|h_{y_u}|^2_\text{L},y_u}\left[\exp\left( -sP_u |h_{y_u}|^2_\text{L} y_u^{-\alpha_\text{L}} \right) \right] \\
&= \mathbb{E}_{y_u}\left[ \left(1+\frac{sP_u y_u^{-\alpha_\text{L}}}{m_\text{L}}\right)^{-m_\text{L}} \right]\\
&= \int_{l_k}^{w_p} \left(1+\frac{sP_u y_u^{-\alpha_\text{L}}}{m_\text{L}}\right)^{-m_\text{L}} f_{1,Y_\text{L}}(y_u|x_0)dy_u,
\end{aligned}
\end{equation}
where  $f_{Y_\text{L}}(y_u|x_0)$ is the distribution of the distance from the interfering LoS UAV to the reference receiving device~\cite{kouzayha2021analysis}, given by $f_{1,Y_\text{L}}(y_u|x_0) = \frac{f_{R_1}(y_u|x_0) P_\text{L}(y_u)}{\int_{l_k}^{w_p} f_{R_1}(w|x_0) P_\text{L}(w)dw}$.

Similarly,
\begin{equation}\small
\begin{aligned}
&\mathbb{E}_{|h_{y_u}|^2_\text{N},y_u}\left[\exp\left( -sP_u |h_{y_u}|^2_\text{N} y_u^{-\alpha_\text{N}} \right) \right]\\
&= \int_{E_{\text{LN}}(l_k)}^{w_p} \left(1+\frac{sP_u y_u^{-\alpha_\text{N}}}{m_\text{N}}\right)^{-m_\text{N}} f_{1,Y_\text{N}}(y_u|x_0)dy_u,
\end{aligned}
\end{equation}
where $f_{1,Y_\text{N}}(y_u|x_0) = \frac{f_{R_1}(y_u|x_0) P_\text{N}(y_u)}{\int_{E_{\text{LN}}(l_k)}^{w_p} f_{R_1}(w|x_0) P_\text{N}(w)dw}$.

Following similar steps, we can also obtain the expression of the Laplace transform of interference on the UAV-device link from existing NLoS UAVs. The final expression of $\mathcal{L}_{\text{edge}_1}^\text{N}(s)$ is given in (\ref{Lemma1}) for $z'=\text{N}$.

\subsection{Proof of Lemma 3}
\label{appendix:laplace_back}
The total interference $I_{\text{back}}$ has signals coming from both LoS and NLoS interfering UAVs. We can derive the expression of the Laplace transform of the interference coming from UAVs with LoS link as:
\begin{equation}\small
\begin{aligned}
   \mathcal{L}_{\text{back}}(s) &=  \mathbb{E}\left[\exp{(-sI_{\text{back}})}\right]\\
   &=   \mathbb{E}\left[\exp{(-s \sum^{N_u^\text{interf}}_{u=1}I_{y_0,y_u})}\right]\\
   &\stackrel{\text{(a)}}{=} \prod^{N_u^\text{interf}}_{u=1} \mathbb{E}_{I_{y_0,y_u}}\left[\exp{(-sI_{y_0,y_u})}\right]\\
   &\stackrel{\text{(b)}}{=} \left( \mathbb{E}_{I_{y_0,y_u}}\left[\exp{(-sI_{y_0,y_u})}\right]\right)^{N_u^\text{interf}}
\end{aligned}
\end{equation}
where (a) results from the IID distribution of the fading gains and from their independence of the BPP process, (b) follows from the IID distribution of the interferers distance, and $\mathbb{E}_{I_{y_0,y_u}}\left[\exp{(-sI_{y_0,y_u})}\right]$ can be calculated as

\begin{equation}\footnotesize
\begin{aligned}
   &\mathbb{E}_{I_{y_0,y_u}}\left[\exp{(-sI_{y_0,y_u})}\right] \\
   &= \frac{ \int_h^d P_\text{L}(u) f_U(u)du\times \mathbb{E}_{|h_{y_u}|^2_\text{L},y_u}\left[\exp\left( -sP_u |h_{y_u}|^2_\text{L} y_u^{-\alpha_\text{L}} \right) \right] }{ \int_h^d P_\text{L}(u) f_U(u)du+\int_h^d P_\text{N}(u) f_U(u)du} \\
   & + \frac{ \int_h^d P_\text{N}(u) f_U(u)du\times  \mathbb{E}_{|h_{y_u}|^2_\text{N},y_u}\left[\exp\left( -sP_u |h_{y_u}|^2_\text{N} y_u^{-\alpha_\text{N}} \right) \right]}{ \int_h^d P_\text{L}(u) f_U(u)du+\int_h^d P_\text{N}(u) f_U(u)du},
\end{aligned}
\end{equation}
where the expression of $\mathbb{E}_{|h_{y_u}|^2_z,y_u}\left[\exp\left( -sP_u |h_{y_u}|^2_z y_u^{-\alpha_z} \right) \right]$ is given by

\begin{equation}\small
\begin{aligned}
&\mathbb{E}_{|h_{y_u}|^2_z,y_u}\left[\exp\left( -sP_u |h_{y_u}|^2_z y_u^{-\alpha_z} \right) \right] \\
&= \mathbb{E}_{y_u}\left[ \left(1+\frac{sP_u y_u^{-\alpha_z}}{m_z}\right)^{-m_z} \right]\\
&= \int_h^d \left(1+\frac{sP_u y_u^{-\alpha_z}}{m_z}\right)^{-m_z} f_{Y_z}(y_u)dy_u,
\end{aligned}
\end{equation}
where $f_{Y_z}(y_u) =\frac{f_U(y_u) P_z(y_u)}{\int_h^d f_U(w) P_z(w)dw}$.

\subsection{Proof of Theorem 3}
\label{appendix:convergence}
To establish the convergence bound for the proposed HFL algorithm, we utilize the following inequalities consistently throughout our derivations, serving as fundamental elements in our analysis.
\begin{equation}\small
    \begin{aligned}
        \left\|\sum_k^{N_d} p_kx\right\|^2 \le  \sum_k^{N_d} p_k\left\|x\right\|^2,
    \end{aligned}
    \label{a1}
\end{equation}
\begin{equation}\small
    \begin{aligned}
        \mathbb{E}\left\|\sum^{N_d}_k x\right\|^2 \le N_d \mathbb{E}\sum^{N_d}_k \left\|x\right\|^2,
    \end{aligned}
    \label{a2}
\end{equation}
and 
\begin{equation}\small
    \begin{aligned}
        \sum_k^{N_d} p_k\left\|x-\bar{x}\right\|^2 =  \sum_k^{N_d} p_k\left\|x\right\|^2 -\left\|\bar{x}\right\|^2  \le  \sum_k^{N_d} p_k\left\|x§\right\|^2.
        \label{a3}
    \end{aligned}
\end{equation}

We start the proof of convergence by showing that global model update can be written in the form of local SGD update:
\begin{equation}\small
    \begin{aligned}
        \bar{w}^{t+1}-\bar{w}^t = \sum^{N_u}_{u=1} \frac{\bar{p}_u}{\mathcal{P}^{\text{back}}_u} \mathbb{I}_1(\cdot) (\bar{v}_u^{t+1}- \bar{w}^t)
        \label{eq1}
    \end{aligned}
\end{equation}

\begin{equation}\small
    \begin{aligned}
       \bar{v}_u^{t+1}-\bar{v}_u^t =  \sum_{k \in S_u} \frac{p_k / \bar{p}_u}{\mathcal{P}^{\text{edge}}_k}  \mathbb{I}_2(\cdot) (w_k^{t+1} - \bar{v}_u^t)
       \label{eq2}
    \end{aligned}
\end{equation}

\begin{equation}\small
    \begin{aligned}
      w^{t+1}_{k} = w^{t}_{k} - \eta \nabla F_k(w^{t}_{k})
      \label{eq3}
    \end{aligned}
\end{equation}

Combining (\ref{eq1}), (\ref{eq2}), and  (\ref{eq3}), we can write:
\begin{equation}\footnotesize
    \begin{aligned}
      &\bar{w}^{t+1} - \bar{w}^{t}  \\
      &= - \eta \sum^{N_u}_{u=1} \frac{\bar{p}_u}{\mathcal{P}^{\text{back}}_u} \mathbb{I}(\text{SINR}_u^{\text{back}} > \theta) \sum_{k \in S_u} \frac{p_k / \bar{p}_u}{\mathcal{P}^{\text{edge}}_k} \cdot\mathbb{I}_2(\cdot)  (\nabla F_k(w^{t}_{k}))
    \end{aligned}
\end{equation}
For convenience, \small{$\mathbb{I}_1(\cdot) = \mathbb{I}(\text{SINR}_u^{\text{back}} > \theta)$ and  $\mathbb{I}_2(\cdot) =  \mathbb{I}(\text{SINR}_k^{\text{edge}_1} > \theta, \text{SINR}_k^{\text{edge}_2} > \theta) $}.

\begin{equation}\footnotesize
    \begin{aligned}
      &\mathbb{E}f(\Bar{w}^{t+1}) \\
      &= \mathbb{E}f(\Bar{w}^{t}- \eta \sum^{N_u}_{u=1} \frac{\bar{p}_u}{\mathcal{P}^{\text{back}}_u} \mathbb{I}_1(\cdot) \sum_{k \in S_u} \frac{p_k / \bar{p}_u}{\mathcal{P}^{\text{edge}}_k} \cdot \mathbb{I}_2(\cdot)\nabla F_k(w^{t}_{k}))\\
      &\stackrel{\text{(a)}}{\le} \mathbb{E}f(\bar{w}^t) - \textcolor{black}{\eta\mathbb{E} \left\langle \nabla f(\bar{w}^t), \sum^{N_u}_{u=1} \frac{\bar{p}_u \mathbb{I}_1(\cdot)}{\mathcal{P}^{\text{back}}_u}  \sum_{k \in S_u} \frac{p_k / \bar{p}_u \mathbb{I}_2(\cdot)}{\mathcal{P}^{\text{edge}}_k} \nabla F_k(w^{t}_{k})\right\rangle }\\
      &+ \frac{\eta^2L}{2} \textcolor{black}{\mathbb{E}\left|\left| \sum^{N_u}_{u=1} \frac{\bar{p}_u}{\mathcal{P}^{\text{back}}_u} \mathbb{I}_1(\cdot) \sum_{k \in S_u} \frac{p_k / \bar{p}_u}{\mathcal{P}^{\text{edge}}_k} \cdot \mathbb{I}_2(\cdot)\nabla F_k(w^{t}_{k})\right|\right|^2},
    \end{aligned}
    \label{eq4}
\end{equation}
where (a) is due to the Lipschitz smoothness of the loss function. The inner term can be expanded as follows 

\begin{equation}\footnotesize
    \begin{aligned}
        &\textcolor{black}{-\eta\mathbb{E} \left\langle \nabla f(\bar{w}^t), \sum^{N_u}_{u=1} \frac{\bar{p}_u}{\mathcal{P}^{\text{back}}_u} \mathbb{I}_1(\cdot) \sum_{k \in S_u} \frac{p_k / \bar{p}_u}{\mathcal{P}^{\text{edge}}_k} \cdot \mathbb{I}_2(\cdot)\nabla F_k(w^{t}_{k})\right\rangle }=\\
        &-\frac{\eta}{2}\left(\mathbb{E} \left|\left| \nabla f(\bar{w}^t) -  \sum^{N_u}_{u=1} \frac{\bar{p}_u}{\mathcal{P}^{\text{back}}_u} \mathbb{I}_1(\cdot) \sum_{k \in S_u} \frac{p_k / \bar{p}_u}{\mathcal{P}^{\text{edge}}_k} \cdot \mathbb{I}_2(\cdot)\nabla F_k(w^{t}_{k})\right|\right|^2 \right.\\
        &\left.- \mathbb{E} \left|\left| \nabla f(\bar{w}^t)\right|\right|^2 \right. \\
        &\left.- \mathbb{E} \left|\left| \sum^{N_u}_{u=1} \frac{\bar{p}_u}{\mathcal{P}^{\text{back}}_u} \mathbb{I}_1(\cdot) \sum_{k \in S_u} \frac{p_k / \bar{p}_u}{\mathcal{P}^{\text{edge}}_k} \cdot \mathbb{I}_2(\cdot)\nabla F_k(w^{t}_{k})\right|\right|^2 \right).
    \end{aligned}
\end{equation}

Substituting the result to (\ref{eq4}) and combining with the last term, we have:
\begin{equation}\footnotesize
    \begin{aligned}
        &\mathbb{E}f(\Bar{w}^{t+1})\\
        &\le \mathbb{E}f(\Bar{w}^{t}) -\frac{\eta}{2}\mathbb{E}\left|\left| \nabla f(\bar{w}^t)\right|\right|^2 \\
        &- (\frac{\eta}{2}-\frac{\eta^2 L}{2}) \mathbb{E}\left\| \sum^{N_u}_{u=1} \frac{\bar{p}_u}{\mathcal{P}^{\text{back}}_u} \mathbb{I}_1(\cdot) \sum_{k \in S_u}\frac{p_k/\bar{p}_u}{\mathcal{P}^{\text{edge}}_k} \mathbb{I}_2(\cdot)\nabla F_k(w^t_k) \right\|^2 +\\
        & \frac{\eta}{2} \mathbb{E}\left\|  f(\bar{w}^t) -  \sum^{N_u}_{u=1} \frac{\bar{p}_u}{\mathcal{P}^{\text{back}}_u} \mathbb{I}_1(\cdot) \sum_{k \in S_u}\frac{p_k/\bar{p}_u}{\mathcal{P}^{\text{edge}}_k} \mathbb{I}_2(\cdot)\nabla F_k(w^t_k) \right\|^2.
    \end{aligned}
\end{equation}

Suppose $ (\frac{\eta}{2}-\frac{\eta^2 L}{2}) \ge 0$, then we have
\begin{equation}\footnotesize
    \begin{aligned}
        &\mathbb{E}f(\Bar{w}^{t+1}) \le \mathbb{E}f(\Bar{w}^{t}) -\frac{\eta}{2}\mathbb{E}\left|\left| \nabla f(\bar{w}^t)\right|\right|^2\\
        &+ \frac{\eta}{2} \textcolor{black}{\mathbb{E}\left\|  f(\bar{w}^t) -  \sum^{N_u}_{u=1} \frac{\bar{p}_u}{\mathcal{P}^{\text{back}}_u} \mathbb{I}_1(\cdot) \sum_{k \in S_u}\frac{p_k/\bar{p}_u}{\mathcal{P}^{\text{edge}}_k} \mathbb{I}_2(\cdot)\nabla F_k(w^t_k) \right\|^2},
    \end{aligned}
    \label{eq7}
\end{equation}
where the last term be upper bounded as
\begin{equation}\footnotesize
    \begin{aligned}
        &\textcolor{black}{\mathbb{E}\left\|  f(\bar{w}^t) -  \sum^{N_u}_{u=1} \frac{\bar{p}_u}{\mathcal{P}^{\text{back}}_u} \mathbb{I}_1(\cdot) \sum_{k \in S_u}\frac{p_k/\bar{p}_u}{\mathcal{P}^{\text{edge}}_k} \mathbb{I}_2(\cdot)\nabla F_k(w^t_k) \right\|^2} \\
        &= \mathbb{E}\left\|  f(\bar{w}^t) - \sum^{N_u}_{u=1}\bar{p}_u\nabla f_u(\bar{v}_u^t)+ \sum^{N_u}_{u=1}\bar{p}_u\nabla f_u(\bar{v}_u^t)\right.\\
        &\left.- \sum^{N_u}_{u=1} \frac{\bar{p}_u}{\mathcal{P}^{\text{back}}_u} \mathbb{I}_1(\cdot) \sum_{k \in S_u}\frac{p_k/\bar{p}_u}{\mathcal{P}^{\text{edge}}_k} \mathbb{I}_2(\cdot)\nabla F_k(w^t_k) \right\|^2\\
        &\stackrel{\text{(a)}}{\le} \textcolor{black}{2\mathbb{E}\left\| \nabla f(\bar{w}^t) - \sum^{N_u}_{u=1}\bar{p}_u\nabla f_u(\bar{v}_u^t)\right\|^2 }\\
        &+  \textcolor{black}{2\mathbb{E}\left\|\sum^{N_u}_{u=1}\bar{p}_u\nabla f_u(\bar{v}_u^t) - \sum^{N_u}_{u=1} \frac{\bar{p}_u}{\mathcal{P}^{\text{back}}_u} \mathbb{I}_1(\cdot) \sum_{k \in S_u}\frac{p_k/\bar{p}_u}{\mathcal{P}^{\text{edge}}_k} \mathbb{I}_2(\cdot)\nabla F_k(w^t_k)\right\|^2},
    \end{aligned}
    \label{eq5}
\end{equation}
where (a) is due to the inequality in (\ref{a2}). The first term is further bounded using $ \nabla f(\bar{w}^t)=\sum_{u=1}^{N_u}\frac{\bar{p}_u}{\mathcal{P}^{\text{back}}_u}\nabla f_u(\bar{w}^t)$ as
\begin{equation}\footnotesize
    \begin{aligned}
        &\textcolor{black}{2\mathbb{E}\left\| \nabla f(\bar{w}^t) - \sum^{N_u}_{u=1}\bar{p}_u\nabla f_u(\bar{v}_u^t)\right\|^2 }  \\
        &=  2\mathbb{E}\left\| \nabla f(\bar{w}^t) - \sum^{N_u}_{u=1}\bar{p}_u\nabla f_u(\bar{w}^t)+ \sum^{N_u}_{u=1}\bar{p}_u\nabla f_u(\bar{w}^t) - \sum^{N_u}_{u=1}\bar{p}_u\nabla f_u(\bar{v}_u^t)\right\|^2\\
        &\stackrel{\text{(a)}}{\le} 4\mathbb{E}\sum^{N_u}_{u=1}\bar{p}_u \left\|\frac{\mathbb{I}_1(\cdot)-\mathcal{P}^{\text{back}}_u}{\mathcal{P}^{\text{back}}_u}\nabla f_u(\bar{w}^t)\right\|^2 +4\mathbb{E} \sum^{N_u}_{u=1}\bar{p}_u\left\|f_u(\bar{w}^t) -f_u(\bar{v}^t_u)\right\|^2\\
         &\stackrel{\text{(b)}}{\le}4\sum^{N_u}_{u=1}\bar{p}_u \frac{\mathbb{E}[\mathbb{I}_1(\cdot)^2]-{\mathcal{P}^{\text{back}}_u}^2}{{\mathcal{P}^{\text{back}}_u}^2} \mathbb{E}\left\|\nabla f_u(\bar{w}^t)\right\|^2 +4 L^2 \sum^{N_u}_{u=1}\bar{p}_u\mathbb{E} \left\|\bar{w}^t -\bar{v}^t_u\right\|^2\\
        &\stackrel{\text{(c)}}{\le}4\sum^{N_u}_{u=1}\bar{p}_u (\frac{1}{\mathcal{P}^{\text{back}}_u}-1)A_1^2 + 4 L^2 \sum^{N_u}_{u=1}\bar{p}_u\mathbb{E} \left\|\bar{w}^t -\bar{v}^t_u\right\|^2\\
        &= 4 B_1 A_1^2 + 4 L^2 \sum^{N_u}_{u=1}\bar{p}_u\mathbb{E} \left\|\bar{w}^t -\bar{v}^t_u\right\|^2,
    \end{aligned}
\end{equation}
where (a) is derived by applying (\ref{a2}) and (\ref{a1}), (b) is due to assumption 2, and (c) is the result of using assumption 4 and the fact that {\footnotesize$\mathbb{E}[\mathbb{I}_1(\cdot)^2]=\mathcal{P}^{\text{back}}_u$} due to the random resource allocation without replacement.

The second term in (\ref{eq5}) can be rewritten as
\begin{equation}\footnotesize
    \begin{aligned}
        &\textcolor{black}{2\mathbb{E}\left\|\sum^{N_u}_{u=1}\bar{p}_u\nabla f_u(\bar{v}_u^t) - \sum^{N_u}_{u=1} \frac{\bar{p}_u}{\mathcal{P}^{\text{back}}_u} \mathbb{I}_1(\cdot) \sum_{k \in S_u}\frac{p_k/\bar{p}_u}{\mathcal{P}^{\text{edge}}_k} \mathbb{I}_2(\cdot)\nabla F_k(w^t_k)\right\|^2}\\
        &=2\mathbb{E}\left\|\sum^{N_u}_{u=1}\bar{p}_u\nabla f_u(\bar{v}_u^t) - \sum^{N_u}_{u=1}\bar{p}_u\sum_{k\in S_u} \frac{p_k}{\bar{p}_u} \nabla F_k(w_k^t)\right.\\
        &\left.+\sum^{N_u}_{u=1}\bar{p}_u\sum_{k\in S_u} \frac{p_k}{\bar{p}_u} \nabla F_k(w_k^t)- \sum^{N_u}_{u=1} \frac{\bar{p}_u \mathbb{I}_1(\cdot)}{\mathcal{P}^{\text{back}}_u}  \sum_{k \in S_u}\frac{p_k/\bar{p}_u}{\mathcal{P}^{\text{edge}}_k} \mathbb{I}_2(\cdot)\nabla F_k(w^t_k)\right\|^2\\
       &\stackrel{\text{(a)}}{\le} \textcolor{black}{4\mathbb{E}\left\|\sum^{N_u}_{u=1}\bar{p}_u\nabla f_u(\bar{v}_u^t) - \sum^{N_u}_{u=1}\bar{p}_u\sum_{k\in S_u} \frac{p_k}{\bar{p}_u} \nabla F_k(w_k^t)\right\|^2} \\
        &+ \textcolor{black}{ 4\mathbb{E}\left\|\sum^{N_u}_{u=1}\bar{p}_u\sum_{k\in S_u} \frac{p_k}{\bar{p}_u} \nabla F_k(w_k^t)\right.}\\
        &\textcolor{black}{ \left.- \sum^{N_u}_{u=1} \frac{\bar{p}_u}{\mathcal{P}^{\text{back}}_u} \mathbb{I}_1(\cdot) \sum_{k \in S_u}\frac{p_k/\bar{p}_u}{\mathcal{P}^{\text{edge}}_k} \mathbb{I}_2(\cdot)\nabla F_k(w^t_k)\right\|^2},
    \end{aligned}
    \label{eq6}
\end{equation}
where (a) is derived by applying (\ref{a2}) and the first term is bounded using assumption 1 as
\begin{equation}\footnotesize
    \begin{aligned}
      &\textcolor{black}{4\mathbb{E}\left\|\sum^{N_u}_{u=1}\bar{p}_u\nabla f_u(\bar{v}_u^t) - \sum^{N_u}_{u=1}\bar{p}_u\sum_{k\in S_u} \frac{p_k}{\bar{p}_u} \nabla F_k(w_k^t)\right\|^2 }\\
      &\le 4L^2\sum^{N_u}_{u=1}\bar{p}_u\sum_{k\in S_u}\frac{p_k}{\bar{p}_u}\mathbb{E}\left\|\bar{v}_u^t -w_k^t \right\|^2,
    \end{aligned}
\end{equation}
and the second term can be further derived as
\begin{equation}\footnotesize
    \begin{aligned}
        &4\mathbb{E}\left\|\sum^{N_u}_{u=1}\bar{p}_u\sum_{k\in S_u} \frac{p_k}{\bar{p}_u} \nabla F_k(w_k^t)\right.\\
        &\left. - \sum^{N_u}_{u=1} \frac{\bar{p}_u}{\mathcal{P}^{\text{back}}_u} \mathbb{I}_1(\cdot) \sum_{k \in S_u}\frac{p_k/\bar{p}_u}{\mathcal{P}^{\text{edge}}_k} \mathbb{I}_2(\cdot)\nabla F_k(w^t_k)\right\|^2\\
        &=4\mathbb{E}\left\|\left(\sum^{N_u}_{u=1}\sum_{k\in S_u} p_k \frac{\mathcal{P}^{\text{edge}}_k \mathcal{P}^{\text{back}}_u- \mathbb{I}_2(\cdot) \mathbb{I}_1(\cdot)}{\mathcal{P}^{\text{edge}}_k \mathcal{P}^{\text{back}}_u} \right)\nabla F_k(w^t_k)\right\|^2\\
         &\stackrel{\text{(a)}}{\le} 4\mathbb{E}\sum^{N_u}_{u=1}\sum_{k\in S_u} p_k \left\|\frac{\mathcal{P}^{\text{edge}}_k \mathcal{P}^{\text{back}}_u- \mathbb{I}_2(\cdot) \mathbb{I}_1(\cdot)}{\mathcal{P}^{\text{edge}}_k \mathcal{P}^{\text{back}}_u} \nabla F_k(w^t_k)\right\|^2\\
        &= 4\sum^{N_u}_{u=1}\sum_{k\in S_u}p_k\frac{\mathbb{E}[\mathbb{I}_2^2(\cdot) \mathbb{I}_1^2(\cdot)] - {\mathcal{P}^{\text{edge}}_k}^2{\mathcal{P}^{\text{back}}_u}^2}{{\mathcal{P}^{\text{edge}}_k}^2 {\mathcal{P}^{\text{back}}_u}^2}\mathbb{E} \left\|\nabla F_k(w^t_k)\right\|^2\\
         &\stackrel{\text{(b)}}{\le}4\sum^{N_u}_{u=1}\sum_{k\in S_u}p_k\left( \frac{1}{\mathcal{P}^{\text{edge}}_k \mathcal{P}^{\text{back}}_u}-1\right)A^2_2 \\
        &= \textcolor{black}{4B_2A^2_2},
    \end{aligned}
\end{equation}
where (a) is obtained due to (\ref{a1}), and (b) is the result of applying the assumption 4 and the fact that {\footnotesize$\mathbb{E}[\mathbb{I}_2^2(\cdot) \mathbb{I}_1^2(\cdot)]={\mathcal{P}^{\text{edge}}_k \mathcal{P}^{\text{back}}_u}$} due to the random resource allocation without replacement.

The results are substituted back into (\ref{eq7}) to obtain the following expression
\begin{equation}\footnotesize
    \begin{aligned}
        &\frac{\eta}{2}\mathbb{E}\left|\left| \nabla f(\bar{w}^t)\right|\right|^2 \le \mathbb{E}f(\Bar{w}^{t}) - \mathbb{E}f(\Bar{w}^{t+1}) \\
        &+ \frac{\eta}{2}\left[ \textcolor{black}{4 B_1 A_1^2 + 4 L^2 \sum^{N_u}_{u=1}\bar{p}_u\mathbb{E} \left\|\bar{w}^t -\bar{v}^t_u\right\|^2} \right.\\
        &\left.+ \textcolor{black}{4B_2A^2_2} +  \textcolor{black}{4L^2\sum^{N_u}_{u=1}\bar{p}_u\sum_{k\in S_u}\frac{p_k}{\bar{p}_u}\mathbb{E}\left\|\bar{v}_u^t -w_k^t \right\|^2}\right].
    \end{aligned}
\end{equation}

Dividing both sides of (\ref{eq7})  by $\frac{\eta}{2}$ and taking average over time, we get
\begin{equation}\footnotesize
    \begin{aligned}
        &\frac{1}{T}\sum^{T-1}_{t=0}\mathbb{E}\left|\left| \nabla f(\bar{w}^t)\right|\right|^2 \\
        &\le \frac{2}{\eta T}\left[ f(\Bar{w}^{0}) - \mathbb{E}f(\Bar{w}^{T})\right] + \frac{4L^2}{T}\sum^{T-1}_{t=0}\sum^{N_u}_{u=1}\bar{p}_u \mathbb{E}\left\|\bar{w}^t -\bar{v}^t_u \right\|^2+\\
        &+ \frac{4L^2}{T}\sum^{T-1}_{t=0}\sum^{N_u}_{u=1} \sum_{k\in S_u} p_k \mathbb{E}\left\|\bar{v}^t_u -w^t_k \right\|^2 + 4B_1^2A_1^2+ 4B_2^2A_2^2 \\
        &\le \frac{2}{\eta T}\left[ f(\Bar{w}^{0}) - f^* \right] + \frac{4L^2}{T}\sum^{T-1}_{t=0}\textcolor{black}{\sum^{N_u}_{u=1}\bar{p}_u \mathbb{E}\left\|\bar{w}^t -\bar{v}^t_u \right\|^2}+\\
        &+ \frac{4L^2}{T}\sum^{T-1}_{t=0}\sum^{N_u}_{u=1} \sum_{k\in S_u} p_k \mathbb{E}\left\|\bar{v}^t_u -w^t_k \right\|^2 + 4B_1^2A_1^2+ 4B_2^2A_2^2.
    \end{aligned}
    \label{eq8}
\end{equation}
Thus, the result depends on both upward and downward parameter mean squared errors (MSE). 

\textit{E. 1: Bounding upward parameter MSE}

We continue with bounding the upward parameter MSE found in the second term of (\ref{eq8}). For that, we remember that when $t=aG,\ \  a=0,1,2...$, we have $\bar{w}^t = \bar{v}^t_u$, $u=1,2,...,N_u$. Suppose $t=aG+c,\ \  c=0,1,2...$, so when $1\le c\le G-1$, we have
\begin{equation}\footnotesize
    \begin{aligned}
        &\textcolor{black}{\sum^{N_u}_{u=1}\bar{p}_u \mathbb{E}\left\|\bar{w}^t -\bar{v}^t_u \right\|^2} = \\
        &= \sum^{N_u}_{u=1}\bar{p}_u \mathbb{E}\left\|   \left(\bar{w}^{t-1} -\eta \sum^{N_u}_{u'=1}\frac{\bar{p}_{u'} \mathbb{I}_1(\cdot)}{\mathcal{P}^{\text{back}}_{u'}}\sum_{k\in S_{u'}}\frac{p_k \mathbb{I}_2(\cdot)}{\bar{p}_{u'} \mathcal{P}^{\text{edge}}_k}\nabla F_k(w_k^{t-1})\right)\right. \\
        &\left.-\left(\bar{v}^{t-1}_u  - \eta \sum_{{k'}\in S_u}\frac{p_{k'}}{\bar{p}_u \mathcal{P}^{\text{edge}}_{k'}}\mathbb{I}_2(\cdot)\nabla F_{k'}(w_{k'}^{t-1}) \right) \right\|^2 \\
        &= \sum^{N_u}_{u=1}\bar{p}_u \mathbb{E}\left\|  \left(\cancel{\bar{w}^{aG}} -\sum^{aG+c-1}_{\tau=aG} \eta \sum^{N_u}_{{u'}=1}\frac{\bar{p}_{u'}\mathbb{I}_1(\cdot)}{\mathcal{P}^{\text{back}}_{u'}}\sum_{k\in S_{u'}}\frac{p_k \mathbb{I}_2(\cdot)}{\bar{p}_{u'} \mathcal{P}^{\text{edge}}_k}\nabla F_k(w_k^{\tau})\right)\right.\\
        &\left.- \left(\cancel{\bar{v}^{aG}_u} -  \sum^{aG+c-1}_{\tau=aG}\eta \sum_{{k'}\in S_u}\frac{p_{k'} \mathbb{I}_2(\cdot)}{\bar{p}_u \mathcal{P}^{\text{edge}}_{k'}}\nabla F_{k'}(w_{k'}^{\tau}) \right) \right\|^2 \\
        &= \eta^2 \sum^{N_u}_{u=1}\bar{p}_u \mathbb{E} \left\| \sum^{aG+c-1}_{\tau=aG} \left(\sum_{{k'}\in S_u} \frac{p_{k'}}{\bar{p}_u \mathcal{P}^{\text{edge}}_{k'}}\mathbb{I}_2(\cdot)\nabla F_{k'}(w_{k'}^{\tau})\right.\right. \\
        &\left.\left.- \sum^{N_u}_{{u'}=1}\frac{\bar{p}_{u'}}{\mathcal{P}^{\text{back}}_{u'}}\mathbb{I}_1(\cdot)\sum_{k\in S_{u'}}\frac{p_k}{\bar{p}_{u'} \mathcal{P}^{\text{edge}}_k}\mathbb{I}_2(\cdot)\nabla F_k(w_k^{\tau}) \right) \right\|^2\\
        &= \eta^2 \sum^{N_u}_{u=1}\bar{p}_u \mathbb{E} \left\| \sum^{aG+c-1}_{\tau=aG} \left(\sum_{{k'}\in S_u} \frac{p_{k'} \mathbb{I}_2(\cdot)}{\bar{p}_u \mathcal{P}^{\text{edge}}_{k'}}\nabla F_{k'}(w_{k'}^{\tau}) -  \nabla f_u(\bar{v}_u^\tau)  \right.\right. \\
        &\left.\left.+ \nabla f_u(\bar{v}_u^\tau) - 
        \sum_{{u'}=1}^{N_u}\bar{p}_{u'}\nabla f_{u'}(\bar{v}_{u'}^\tau) \right.\right.\\
        &\left.\left.+ \sum_{{u'}=1}^{N_u}\bar{p}_{u'}\nabla f_{u'}(\bar{v}_{u'}^\tau) - \sum^{N_u}_{{u'}=1}\frac{\bar{p}_{u'} \mathbb{I}_1(\cdot)}{\mathcal{P}^{\text{back}}_{u'}}\sum_{k\in S_{u'}}\frac{p_k \mathbb{I}_2(\cdot)}{\bar{p}_{u'} \mathcal{P}^{\text{edge}}_k}\nabla F_k(w_k^{\tau}))    \right) \right\|^2\\
        &=\eta^2 \sum^{N_u}_{u=1}\bar{p}_u  \mathbb{E} \left\| \sum^{aG+c-1}_{\tau=aG} \left(\sum_{{k'}\in S_u} \frac{p_{k'}}{\bar{p}_u \mathcal{P}^{\text{edge}}_{k'}}\mathbb{I}_2(\cdot)\nabla F_{k'}(w_{k'}^{\tau})\right.\right. \\
        &\left.\left.-  \sum_{{k'}\in S_u}\frac{p_{k'}}{\bar{p}_u}\nabla F_{k'}(w_{k'}^{\tau})   +   \sum_{{k'}\in S_u}\frac{p_{k'}}{\bar{p}_u}\nabla F_{k'}(w_{k'}^{\tau})    \right. \right. \\
        &\left. \left. - \sum_{{k'}\in S_u}\frac{p_{k'}}{\bar{p}_u}\nabla F_{k'}(\bar{v}_u^{\tau})-   \sum^{N_u}_{{u'}=1}\frac{\bar{p}_{u'} \mathbb{I}_1(\cdot)}{\mathcal{P}^{\text{back}}_{u'}}\sum_{k\in S_{u'}}\frac{p_k \mathbb{I}_2(\cdot)}{\bar{p}_{u'} \mathcal{P}^{\text{edge}}_k}\nabla F_k(w_k^{\tau})  \right.\right. \\
        &\left.\left.+ \sum_{{u'}=1}^{N_u}\bar{p}_{u'}\sum_{k\in S_{u'}} \frac{p_k}{\bar{p}_{u'}}\nabla F_k(w_k^{\tau})-   \sum_{{u'}=1}^{N_u}\bar{p}_{u'}\sum_{k\in S_{u'}} \frac{p_k}{\bar{p}_{u'}}\nabla F_k(w_k^{\tau})   \right. \right.  \\
        & \left. \left.  + \sum_{{u'}=1}^{N_u}\bar{p}_{u'}\sum_{k\in S_{u'}} \frac{p_k}{\bar{p}_{u'}}\nabla F_k(\bar{v}_{u'}^{\tau})+\nabla f_u(\bar{v}_u^\tau)    -  \sum_{{u'}=1}^{N_u}\bar{p}_{u'}\nabla f_{u'}(\bar{v}_{u'}^\tau)   \right) \right\|^2\\
        &\stackrel{\text{(a)}}{\le} \overbrace{4\eta^2 \sum^{N_u}_{u=1}\bar{p}_u  \mathbb{E} \left\| \sum^{aG+c-1}_{\tau=aG} \left(\sum_{{k'}\in S_u} \frac{p_{k'}}{\bar{p}_u \mathcal{P}^{\text{edge}}_{k'}}\mathbb{I}_2(\cdot)\nabla F_{k'}(w_{k'}^{\tau})\right.\right.}^{\textcircled{1}}  \\
        &\left.\left.-  \sum_{{k'}\in S_u}\frac{p_{k'}}{\bar{p}_u}\nabla F_{k'}(w_{k'}^{\tau})\right) \right\|^2\\
        &+ \overbrace{4\eta^2 \sum^{N_u}_{u=1}\bar{p}_u  \mathbb{E} \left\| \sum^{aG+c-1}_{\tau=aG} \left(    \sum_{{k'}\in S_u}\frac{p_{k'}}{\bar{p}_u}\nabla F_{k'}(w_{k'}^{\tau})   - \sum_{{k'}\in S_u}\frac{p_{k'}}{\bar{p}_u}\nabla F_{k'}(\bar{v}_u^{\tau}) \right.\right.}^{\textcircled{2}}\\
        &\left.\left.-   \sum_{{u'}=1}^{N_u}\bar{p}_{u'}\sum_{k\in S_{u'}} \frac{p_k}{\bar{p}_{u'}}\nabla F_k(w_k^{\tau})  + \sum_{{u'}=1}^{N_u}\bar{p}_{u'}\sum_{k\in S_{u'}} \frac{p_k}{\bar{p}_{u'}}\nabla F_k(\bar{v}_{u'}^{\tau}) \right) \right\|^2+ \\
        \label{eq9}
    \end{aligned}
    \nonumber
\end{equation}

\begin{equation}\footnotesize
    \begin{aligned}
        &+ \overbrace{4\eta^2 \sum^{N_u}_{u=1}\bar{p}_u  \mathbb{E} \left\| \sum^{aG+c-1}_{\tau=aG} \left( \sum_{{u'}=1}^{N_u}\bar{p}_{u'}\sum_{k\in S_{u'}} \frac{p_k}{\bar{p}_{u'}}\nabla F_k(w_k^{\tau}) \right.\right.}^{\textcircled{3}}\\
        &\left.\left.-     \sum^{N_u}_{{u'}=1}\frac{\bar{p}_{u'}}{\mathcal{P}^{\text{back}}_{u'}}\mathbb{I}_1(\cdot)\sum_{k\in S_{u'}}\frac{p_k}{\bar{p}_{u'} \mathcal{P}^{\text{edge}}_k}\mathbb{I}_2(\cdot)\nabla F_k(w_k^{\tau})   \right) \right\|^2 \\
        &+\overbrace{4\eta^2 \sum^{N_u}_{u=1}\bar{p}_u  \mathbb{E} \left\| \sum^{aG+c-1}_{\tau=aG} \left(     \nabla f_u(\bar{v}_u^\tau)    -  \sum_{{u'}=1}^{N_u}\bar{p}_{u'}\nabla f_{u'}(\bar{v}_{u'}^\tau) \right) \right\|^2 }^{\textcircled{4}},
    \end{aligned}
\end{equation}
where (a) is the result of applying (\ref{a2}). Hence, the upward parameter MSE can be bounded by the expressions of \textcircled{1} -- \textcircled{4}:

\begin{equation}\footnotesize
    \begin{aligned}
    &\textcircled{1}:\\
      &4\eta^2 \sum^{N_u}_{u=1}\bar{p}_u  \mathbb{E} \left\| \sum^{aG+c-1}_{\tau=aG} \left(\sum_{{k'}\in S_u} \frac{p_{k'}}{\bar{p}_u \mathcal{P}^{\text{edge}}_{k'}}\mathbb{I}_2(\cdot)\nabla F_{k'}(w_{k'}^{\tau}) \right.\right.\\
      &\left.\left.-  \sum_{{k'}\in S_u}\frac{p_{k'}}{\bar{p}_u}\nabla F_{k'}(w_{k'}^{\tau})\right) \right\|^2\\
    &\stackrel{\text{(a)}}{\le} 4\eta^2 G\sum^{N_u}_{u=1}\bar{p}_u  \sum^{aG+c-1}_{\tau=aG} \mathbb{E} \left\| \left(\sum_{{k'}\in S_u} \frac{p_{k'}}{\bar{p}_u \mathcal{P}^{\text{edge}}_{k'}}\mathbb{I}_2(\cdot)\nabla F_{k'}(w_{k'}^{\tau}) \right.\right.\\
    &\left.\left.-  \sum_{{k'}\in S_u}\frac{p_{k'}}{\bar{p}_u}\nabla F_{k'}(w_{k'}^{\tau})\right) \right\|^2\\
    &\stackrel{\text{(b)}}{\le} 4\eta^2 G\sum^{N_u}_{u=1}\bar{p}_u  \sum^{aG+c-1}_{\tau=aG} \mathbb{E} \sum_{{k'}\in S_u} \frac{p_{k'}}{\bar{p}_u}\left\|\frac{\mathbb{I}_2(\cdot)-\mathcal{P}^{\text{edge}}_{k'}}{\mathcal{P}^{\text{edge}}_{k'}}\nabla F_{k'}(w_{k'}^{\tau})\right\|^2\\
    &\stackrel{\text{(c)}}{\le} 4\eta^2 G\sum^{N_u}_{u=1} \sum^{aG+c-1}_{\tau=aG}\sum_{{k'}\in S_u} p_{k'}\left(\frac{1}{\mathcal{P}^{\text{edge}}_{k'}}-1\right) A_2^2\\
    &\le4\eta^2 G^2 A_2^2 \overbrace{\sum^{N_u}_{u=1}\sum_{{k'}\in S_u} p_{k'}\left(\frac{1}{\mathcal{P}^{\text{edge}}_{k'}}-1\right)}^{B_3}\\
    &=4\eta^2 G^2 A_2^2  B_3,
    \end{aligned}
\end{equation}
where (a) and (b) are due to the use of (\ref{a2}), and (c) is obtained by applying assumption 4.

\begin{equation}\footnotesize
    \begin{aligned}
    &\textcircled{2}:\\ 
      &4\eta^2 \sum^{N_u}_{u=1}\bar{p}_u  \mathbb{E} \left\| \sum^{aG+c-1}_{\tau=aG} \left(    \sum_{{k'}\in S_u}\frac{p_{k'}}{\bar{p}_u}\nabla F_{k'}(w_{k'}^{\tau})\right.\right.   \\
      &\left.\left.- \sum_{{k'}\in S_u}\frac{p_{k'}}{\bar{p}_u}\nabla F_{k'}(\bar{v}_u^{\tau})\right)   -  \left( \sum_{{u'}=1}^{N_u}\bar{p}_{u'}\sum_{k\in S_{u'}} \frac{p_k}{\bar{p}_{u'}}\nabla F_k(w_k^{\tau})  \right.\right. \\
      &\left.\left.- \sum_{{u'}=1}^{N_u}\bar{p}_{u'}\sum_{k\in S_{u'}} \frac{p_k}{\bar{p}_{u'}}\nabla F_k(\bar{v}_u^{\tau}) \right) \right\|^2 \\
     &\stackrel{\text{(a)}}{\le}  4\eta^2 \sum^{N_u}_{u=1}\bar{p}_u  \mathbb{E} \left\| \sum^{aG+c-1}_{\tau=aG} \left(       \sum_{{k'}\in S_u}\frac{p_{k'}}{\bar{p}_u}\nabla F_{k'}(w_{k'}^{\tau}) \right.\right.\\
    &\left.\left.- \sum_{{k'}\in S_u}\frac{p_{k'}}{\bar{p}_u}\nabla F_{k'}(\bar{v}_u^{\tau})    \right) \right\|^2\\
    &\stackrel{\text{(b)}}{\le}4\eta^2GL^2\sum^{aG+c-1}_{\tau=aG} \sum^{N_u}_{u=1}\bar{p}_u \sum_{{k'}\in S_u}\frac{p_{k'}}{\bar{p}_u} \mathbb{E}\left\| w_{k'}^\tau - \bar{v}_u^\tau\right\|^2,
    \end{aligned}
\end{equation}
where (a) is due to the use of (\ref{a3}), and (b) results from the application of (\ref{a2}) and assumption 1.

\begin{equation}\footnotesize
    \begin{aligned}
    &\textcircled{3}:\\  
    &4\eta^2 \sum^{N_u}_{u=1}\bar{p}_u  \mathbb{E} \left\| \sum^{aG+c-1}_{\tau=aG} \left( \sum_{{u'}=1}^{N_u}\bar{p}_{u'}\sum_{k\in S_{u'}} \frac{p_k}{\bar{p}_{u'}}\nabla F_k(w_k^{\tau})  \right.\right.\\
    &\left.\left.-     \sum^{N_u}_{{u'}=1}\frac{\bar{p}_{u'}}{\mathcal{P}^{\text{back}}_{u'}}\mathbb{I}_1(\cdot)\sum_{k\in S_{u'}}\frac{p_k}{\bar{p}_{u'} \mathcal{P}^{\text{edge}}_k}\mathbb{I}_2(\cdot)\nabla F_k(w_k^{\tau})   \right) \right\|^2 \\
    &\stackrel{\text{(a)}}{\le}  4\eta^2G \sum^{N_u}_{u=1}\bar{p}_u \sum^{aG+c-1}_{\tau=aG}  \mathbb{E} \left\|  \sum_{{u'}=1}^{N_u}\sum_{k\in S_{u'}}  \left(p_k - \frac{p_k\mathbb{I}_1(\cdot)\mathbb{I}_2(\cdot)}{\mathcal{P}^{\text{back}}_{u'} \mathcal{P}^{\text{edge}}_k}\right)\nabla F_k(w_k^{\tau})    \right\|^2 \\
    &\stackrel{\text{(b)}}{\le}  4\eta^2G \sum^{N_u}_{u=1}\bar{p}_u \sum^{aG+c-1}_{\tau=aG}   \overbrace{\sum_{{u'}=1}^{N_u}\sum_{k\in S_{u'}}p_k\left(\frac{1}{\mathcal{P}^{\text{back}}_{u'} \mathcal{P}^{\text{edge}}_k}-1\right)}^{B_2} A_2^2\\
    &= 4\eta^2G^2 \sum^{N_u}_{u=1}\bar{p}_u B_2A_2^2\\
    &= 4\eta^2G^2 A_2^2 B_2,
    \end{aligned}
\end{equation}
where (a) is the result of applying (\ref{a2}), and (b) is due to assumption 4.

\begin{equation}\footnotesize
    \begin{aligned}
    &\textcircled{4}:\\ 
    &4\eta^2 \sum^{N_u}_{u=1}\bar{p}_u  \mathbb{E} \left\| \sum^{aG+c-1}_{\tau=aG} \left(     \nabla f_u(\bar{v}_u^\tau)    -  \sum_{{u'}=1}^{N_u}\bar{p}_{u'}\nabla f_{u'}(\bar{v}_{u'}^\tau) \right) \right\|^2 \\
    &= 4\eta^2 \sum^{N_u}_{u=1}\bar{p}_u  \mathbb{E} \left\| \sum^{aG+c-1}_{\tau=aG} \left(     \nabla f_u(\bar{v}_u^\tau)    -  \nabla f(\bar{v}_u^\tau) +  \nabla f(\bar{v}_u^\tau) \right.\right.\\
    &\left.\left.-\nabla f(\bar{w}^\tau) +\nabla f(\bar{w}^\tau) - \sum_{{u'}=1}^{N_u}\bar{p}_{u'}\nabla f_{u'}(\bar{v}_{u'}^\tau) \right) \right\|^2 \\
    &\stackrel{\text{(a)}}{\le}  12\eta^2 \sum^{N_u}_{u=1}\bar{p}_u \left(\mathbb{E}\left\| \sum^{aG+c-1}_{\tau=aG}    \nabla f_u(\bar{v}_u^\tau)    -  \nabla f(\bar{v}_u^\tau)  \right\|^2   \right.  \\
    &\left.+     \mathbb{E}\left\|\sum^{aG+c-1}_{\tau=aG} \nabla f(\bar{v}_u^\tau) -\nabla f(\bar{w}^\tau)  \right\|^2   \right.  \\
    &\left. +     \mathbb{E}\left\| \sum^{aG+c-1}_{\tau=aG} \nabla f(\bar{w}^\tau) - \sum_{{u'}=1}^{N_u}\bar{p}_{u'}\nabla f_{u'}(\bar{v}_{u'}^\tau)  \right\|^2      \right)\\
     &\stackrel{\text{(b)}}{\le} 12\eta^2 \sum^{N_u}_{u=1}\bar{p}_u \left(G \sum^{aG+c-1}_{\tau=aG}    \mathbb{E}\left\|\nabla f_u(\bar{v}_u^\tau)    -  \nabla f(\bar{v}_u^\tau)  \right\|^2   \right. \\
    & \left. +    G\sum^{aG+c-1}_{\tau=aG} \mathbb{E}\left\| \nabla f(\bar{v}_u^\tau) -\nabla f(\bar{w}^\tau)  \right\|^2   \right. \\
    & \left.+      G \sum^{aG+c-1}_{\tau=aG}\mathbb{E}\left\| \nabla f(\bar{w}^\tau) - \sum_{{u'}=1}^{N_u}\bar{p}_{u'}\nabla f_{u'}(\bar{v}_{u'}^\tau )  \right\|^2      \right)\\
     &\stackrel{\text{(c)}}{\le} 12\eta^2G^2\epsilon^2 +  12\eta^2GL^2  \sum^{aG+c-1}_{\tau=aG}\sum^{N_u}_{u=1}\bar{p}_u \mathbb{E}\left\|\bar{v}^\tau_u-\bar{w}^\tau \right\|^2   \\
    &+  12\eta^2GL^2 \sum^{aG+c-1}_{\tau=aG}\sum^{N_u}_{{u'}=1}\bar{p}_{u'}\mathbb{E}\left\|\bar{w}^\tau-\bar{v}^\tau_{u'} \right\|^2 \\
    &=12\eta^2G^2\epsilon^2 + 24\eta^2L^2G \sum^{aG+c-1}_{\tau=aG}\sum^{N_u}_{{u'}=1}\bar{p}_{u'}\mathbb{E}\left\|\bar{w}^\tau-\bar{v}^\tau_{u'} \right\|^2,
    \end{aligned}
\end{equation}
where (a) is due to (\ref{a1}), (b) is obtained by applying (\ref{a2}), and (c) is the result of applying the assumption 1 and 2.

Substituting the results of \textcircled{1} -- \textcircled{4} back into back to the inequality (\ref{eq9}), we can obtain:
\begin{equation}\footnotesize
    \begin{aligned}
        &\textcolor{black}{\sum^{N_u}_{u=1}\bar{p}_u \mathbb{E}\left\|\bar{w}^t -\bar{v}^t_u \right\|^2} = \\
        &\le 4\eta^2 G^2 A_2^2  B_3 + 4\eta^2G^2 A_2^2 B_2  \\
        & +  4\eta^2GL^2\sum^{aG+c-1}_{\tau=aG} \sum^{N_u}_{u=1}\bar{p}_u \sum_{{k'}\in S_u}\frac{p_{k'}}{\bar{p}_u} \mathbb{E}\left\| w_{k'}^\tau - \bar{v}_u^\tau\right\|^2 \\
        & + 12\eta^2G^2\epsilon^2 + 24\eta^2L^2G \sum^{aG+c-1}_{\tau=aG}\sum^{N_u}_{{u'}=1}\bar{p}_{u'}\mathbb{E}\left\|\bar{w}^\tau-\bar{v}^\tau_{u'} \right\|^2 \\
        &\stackrel{\text{(a)}}{\le} 4\eta^2 G^2 A_2^2  B_3 + 4\eta^2G^2 A_2^2 B_2  \\
        & +  4\eta^2GL^2\sum^{aG+G-1}_{\tau=aG} \sum^{N_u}_{u=1}\bar{p}_u \sum_{{k'}\in S_u}\frac{p_{k'}}{\bar{p}_u} \mathbb{E}\left\| w_{k'}^\tau - \bar{v}_u^\tau\right\|^2 \\
        &+ 12\eta^2G^2\epsilon^2 + 24\eta^2L^2G \sum^{aG+G-1}_{\tau=aG}\sum^{N_u}_{{u'}=1}\bar{p}_{u'}\mathbb{E}\left\|\bar{w}^\tau-\bar{v}^\tau_{u'} \right\|^2,
    \end{aligned}
\end{equation}
where (a) is because of $1\le c\le G-1$.

Assuming $T-1 = dG,\ d=0,1,2,...$, we take the average over time on both sides to obtain:
\begin{equation}\footnotesize
    \begin{aligned}
        &\frac{1}{T}\sum^{T-1}_{t=0}\sum^{N_u}_{u=1}\bar{p}_u  \mathbb{E}\left\|\bar{w}^t -\bar{v}^t_u \right\|^2\\
         &\stackrel{\text{(a)}}{=}\frac{1}{T}\sum^{T-2}_{t=0}\sum^{N_u}_{u=1}\bar{p}_u  \mathbb{E}\left\|\bar{w}^t -\bar{v}^t_u \right\|^2\\
        &\le 4\eta^2 G^2 A_2^2  B_3 + 4\eta^2G^2 A_2^2 B_2  +  12\eta^2G^2\epsilon^2 \\
        & + 4\eta^2GL^2\frac{1}{T}\sum^{d-1}_{a=0}\left(G\sum^{aG+G-1}_{\tau=aG} \sum^{N_u}_{u=1}\bar{p}_u \sum_{{k'}\in S_u}\frac{p_{k'}}{\bar{p}_u} \mathbb{E}\left\| w_{k'}^\tau - \bar{v}_u^\tau\right\|^2\right)   \\
        &+ 24\eta^2GL^2\frac{1}{T}\sum^{d-1}_{a=0}\left(G \sum^{aG+G-1}_{\tau=aG}\sum^{N_u}_{{u'}=1}\bar{p}_{u'}\mathbb{E}\left\|\bar{w}^\tau-\bar{v}^\tau_{u'} \right\|^2 \right) \\
         &\stackrel{\text{(b)}}{\le} 4\eta^2 G^2 A_2^2  B_3 + 4\eta^2G^2 A_2^2 B_2  +  12\eta^2G^2\epsilon^2 \\
        & + 4\eta^2G^2L^2\frac{1}{T}\sum^{T-2}_{t=0} \sum^{N_u}_{u=1}\bar{p}_u \sum_{{k'}\in S_u}\frac{p_{k'}}{\bar{p}_u} \mathbb{E}\left\| w_{k'}^\tau - \bar{v}_u^\tau\right\|^2\\
        &+  24\eta^2G^2L^2\frac{1}{T}\sum^{T-1}_{t=0}\sum^{N_u}_{{u'}=1}\bar{p}_{u'}\mathbb{E}\left\|\bar{w}^\tau-\bar{v}^\tau_{u'} \right\|^2,
    \end{aligned}
\end{equation}
where (a) and (b) hold due to $\left\|\bar{w}^{T-1} - \bar{v}^{T-1}_u \right\|^2 = 0$ and $\left\|{w}^{T-1}_{k'} - \bar{v}^{T-1}_u \right\|^2 = 0$.

Moving the last term to the left side and dividing both sides by $(1-24\eta^2G^2L^2)$, we have

\begin{equation}\footnotesize
    \begin{aligned}
        &\frac{1}{T}\sum^{T-1}_{t=0}\sum^{N_u}_{u=1}\mathbb{E}\left\|\bar{w}^t -\bar{v}^t_u \right\|^2\\
         &\le \frac{4\eta^2 G^2 A_2^2  B_3}{(1-24\eta^2G^2L^2)} + \frac{4\eta^2G^2 A_2^2 B_2}{(1-24\eta^2G^2L^2)}  +  \frac{12\eta^2G^2\epsilon^2}{(1-24\eta^2G^2L^2)} \\
         & + \frac{4\eta^2G^2L^2}{(1-24\eta^2G^2L^2)}\frac{1}{T}\sum^{T-2}_{t=0} \sum^{N_u}_{u=1}\bar{p}_u \sum_{k'\in S_u}\frac{p_{k'}}{\bar{p}_u} \mathbb{E}\left\| w_{k'}^\tau - \bar{v}_u^\tau\right\|^2.
    \end{aligned}
    \label{eq13}
\end{equation}

\textit{E. 2: Bounding downward parameter MSE}

We begin with bounding the downward parameter MSE, which corresponds to the last term in (\ref{eq8}), $\frac{1}{T}\sum^{T-1}_{t=0} \sum^{N_u}_{u=1}\sum_{k\in S_u}p_k\mathbb{E}\left\|\bar{v}^t_u -{w}^t_k \right\|^2$. This aggregation occurs at the UAV level. For that, we assume $t=bE+c$, where $b = 0,1,2...\frac{G}{E}-1$ and $c=0,1,2,...,E-1$. We have $\bar{v}_{u'}^t = w_k^t$ $\forall k\in S_{u'}, {u'}=1,2...N_u$ when $c=0$. For cases when  $c\geq 1$, we have
\begin{equation}\footnotesize
    \begin{aligned}
        & \sum^{N_u}_{{u'}=1}\sum_{k\in S_{u'}}p_k\mathbb{E}\left\|\bar{v}^t_{u'} -{w}^t_k \right\|^2\\
        &=\sum^{N_u}_{{u'}=1}\sum_{k\in S_{u'}}p_k\mathbb{E}\left\|    \left(\bar{v}^{t-1}_{u'} -\eta \sum_{{k'}\in S_{u'}}\frac{p_{k'}/\bar{p}_{u'}}{\mathcal{P}^{\text{edge}}_{k'}} \mathbb{I}_2(\cdot)\nabla F_{k'}(w_{k'}^{t-1})\right)   \right. \\
        &\left.-    \left( {w}^{t-1}_k - \eta  \nabla F_k(w_k^{t-1})   \right)   \right\|^2\\
        &=\eta^2 \sum^{N_u}_{{u'}=1}\sum_{k\in S_{u'}}p_k\mathbb{E}\left\|\sum_{\tau=bE}^{t-1} \left( \nabla F_k(w_k^{\tau}) -\sum_{{k'}\in S_{u'}}\frac{\frac{p_{k'}}{\bar{p}_{u'}}\mathbb{I}_2(\cdot)}{\mathcal{P}^{\text{edge}}_{k'}} \nabla F_{k'}(w_{k'}^{\tau})\right)     \right\|^2\\
        &=\eta^2 \sum^{N_u}_{{u'}=1}\sum_{k\in S_{u'}}p_k\mathbb{E}\left\| \sum_{\tau=bE}^{t-1} \left( \nabla F_k(w_k^{\tau}) - F_k(\bar{v}_{u'}^{\tau})  + F_k(\bar{v}_{u'}^{\tau}) \right.\right.\\
        &\left.\left.- \nabla f_{u'}(\bar{v}_{u'}^\tau) + \nabla f_{u'}(\bar{v}_{u'}^\tau)- \sum_{{k'}\in S_{u'}}\frac{p_{k'}}{\bar{p}_{u'}} \nabla F_{k'}(w_{k'}^{\tau})\right. \right. \\
        &\left. \left.  + \sum_{{k'}\in S_{u'}}\frac{p_{k'}}{\bar{p}_{u'}} \nabla F_{k'}(w_{k'}^{\tau}) -  \sum_{{k'}\in S_{u'}}\frac{p_{k'}/\bar{p}_{u'}}{\mathcal{P}^{\text{edge}}_{k'}} \mathbb{I}_2(\cdot)\nabla F_{k'}(w_{k'}^{\tau})\right)     \right\|^2\\
        &\stackrel{\text{(a)}}{\le} 4\eta^2 \sum^{N_u}_{{u'}=1}\sum_{k\in S_{u'}}p_k\mathbb{E}\left\| \sum_{\tau=bE}^{t-1} \left(   \nabla F_k(w_k^{\tau}) - F_k(\bar{v}_{u'}^{\tau})    \right)   \right\|^2    \\
        & +   4\eta^2 \sum^{N_u}_{{u'}=1}\sum_{k\in S_{u'}}p_k\mathbb{E}\left\| \sum_{\tau=bE}^{t-1} \left(  \nabla F_k(\bar{v}_{u'}^{\tau}) - \nabla f_{u'}(\bar{v}_{u'}^\tau)  \right)   \right\|^2 \\
        &+ \overbrace{4\eta^2 \sum^{N_u}_{{u'}=1}\sum_{k\in S_{u'}}p_k\mathbb{E}\left\| \sum_{\tau=bE}^{t-1} \left(  \nabla f_{u'}(\bar{v}_{u'}^\tau) - \sum_{{k'}\in S_{u'}}\frac{p_{k'}}{\bar{p}_{u'}} \nabla F_{k'}(w_{k'}^{\tau})  \right)   \right\|^2}^{\textcircled{1}}  \\
        &+  \overbrace{4\eta^2 \sum^{N_u}_{{u'}=1}\sum_{k\in S_{u'}}p_k\mathbb{E}\left\| \sum_{\tau=bE}^{t-1} \left(  \sum_{{k'}\in S_{u'}}\frac{p_{k'}}{\bar{p}_{u'}} \nabla F_{k'}(w_{k'}^{\tau})\right.\right.}^{\textcircled{2}}  \\
        &\left. \left. -  \sum_{{k'}\in S_{u'}}\frac{p_{k'}/\bar{p}_{u'}}{\mathcal{P}^{\text{edge}}_{k'}} \mathbb{I}_2(\cdot)\nabla F_{k'}(w_{k'}^{\tau})\right)   \right\|^2,
    \end{aligned}
    \label{eq10}
\end{equation}
where (a) is due to (\ref{a2}). We observe that downward parameter MSE can be bounded by four terms. Two of the terms \textcircled{1} -- \textcircled{2} are upped bounded as follows

\begin{equation}\footnotesize
    \begin{aligned}
         &\textcircled{1}: \\
         & 4\eta^2 \sum^{N_u}_{{u'}=1}\sum_{k\in S_{u'}}p_k\mathbb{E}\left\| \sum_{\tau=bE}^{t-1} \left(  \nabla f_{u'}(\bar{v}_{u'}^\tau) - \sum_{{k'}\in S_{u'}}\frac{p_{k'}}{\bar{p}_{u'}} \nabla F_{k'}(w_{k'}^{\tau})  \right)   \right\|^2 \\
        &\stackrel{\text{(a)}}{\le} 4\eta^2 E \sum^{N_u}_{{u'}=1}\bar{p}_{u'} \mathbb{E}\sum_{\tau=bE}^{t-1} \left\| \left(  \nabla f_{u'}(\bar{v}_{u'}^\tau) - \sum_{{k'}\in S_{u'}}\frac{p_{k'}}{\bar{p}_{u'}} \nabla F_{k'}(w_{k'}^{\tau})  \right)   \right\|^2\\
        &\stackrel{\text{(b)}}{\le} 4\eta^2 \sum^{N_u}_{{u'}=1}\bar{p}_{u'}  E^2 \hat{\epsilon}^2,
    \end{aligned}
\end{equation}
where (a) is due to (\ref{a2}), and (b) is the result of applying the assumption 3.

\begin{equation}\footnotesize
    \begin{aligned}
        &\textcircled{2}: \\
        & 4\eta^2 \sum^{N_u}_{{u'}=1}\sum_{k\in S_{u'}}p_k\mathbb{E}\left\| \sum_{\tau=bE}^{t-1} \left(  \sum_{{k'}\in S_{u'}}\frac{p_{k'}}{\bar{p}_{u'}} \nabla F_{k'}(w_{k'}^{\tau}) \right.\right.\\
        &\left.\left.-  \sum_{{k'}\in S_{u'}}\frac{p_{k'}/\bar{p}_{u'}}{\mathcal{P}^{\text{edge}}_{k'}} \mathbb{I}_2(\cdot)\nabla F_{k'}(w_{k'}^{\tau})\right)   \right\|^2 \\
        &\stackrel{\text{(a)}}{\le}  4\eta^2 \sum^{N_u}_{{u'}=1}\bar{p}_{u'}E\mathbb{E} \sum_{\tau=bE}^{t-1}\sum_{{k'}\in S_{u'}} \frac{p_{k'}}{\bar{p}_{u'}}\left\| \left( \frac{\mathcal{P}^{\text{edge}}_{k'}-\mathbb{I}_2(\cdot)}{\mathcal{P}^{\text{edge}}_{k'}}  \right) \nabla F_{k'}(w_{k'}^{\tau}) \right\|^2 \\
        &\stackrel{\text{(b)}}{\le}  4\eta^2 E^2 A_2^2  \sum^{N_u}_{{u'}=1}\bar{p}_{u'}\sum_{{k'}\in S_{u'}} \frac{p_{k'}}{\bar{p}_{u'}}\left( \frac{1}{\mathcal{P}^{\text{edge}}_{k'}}-1\right) \\
        &= 4\eta^2 E^2 A_2^2 B_3,
    \end{aligned}
\end{equation}
where (a) is due to (\ref{a2}), and (b) is obtained by applying the assumption 4.

Combining the results back into (\ref{eq10}), we have
\begin{equation}\footnotesize
    \begin{aligned}
        & \sum^{N_u}_{{u'}=1}\sum_{k\in S_{u'}}p_k\mathbb{E}\left\|\bar{v}^t_{u'} -{w}^t_k \right\|^2\\
        & \le 8\eta^2 \sum^{N_u}_{{u'}=1}\sum_{k\in S_{u'}}p_k\mathbb{E}\left\| \sum_{\tau=bE}^{t-1} \left(   \nabla F_k(w_k^{\tau}) - F_k(\bar{v}_{u'}^{\tau})    \right)   \right\|^2\\
        &+ 4\eta^2 \sum^{N_u}_{{u'}=1}\bar{p}_{u'}  E^2 \hat{\epsilon}^2  + 4\eta^2 E^2 A_2^2 B_3 \\
         &\stackrel{\text{(a)}}{\le}  8\eta^2 L^2 \sum^{N_u}_{{u'}=1}\sum_{k\in S_{u'}}p_k \sum_{\tau=bE}^{t-1}E \mathbb{E}\left\|  w_k^{\tau} - \bar{v}_{u'}^{\tau}    \right\|^2 + 4\eta^2 \sum^{N_u}_{{u'}=1}\bar{p}_{u'}  E^2 \hat{\epsilon}^2 \\
        &+ 4\eta^2 E^2 A_2^2 B_3 \\
        &\stackrel{\text{(b)}}{\le}8\eta^2 L^2 \sum^{N_u}_{{u'}=1}\sum_{k\in S_{u'}}p_k \sum_{\tau=bE}^{bE+E-1}E \mathbb{E}\left\|  w_k^{\tau} - \bar{v}_{u'}^{\tau}    \right\|^2 \\
        &+ 4\eta^2 \sum^{N_u}_{{u'}=1}\bar{p}_{u'}  E^2 \hat{\epsilon}^2  + 4\eta^2 E^2 A_2^2 B_3, 
    \end{aligned}
    \label{eq11}
\end{equation}
where (a) is due to the use of assumption 1, and (b) holds because of $c\le E-1$. (\ref{eq11}) is also true for $t=bE$ with $\sum_{k\in S_{u'}}p_k\mathbb{E}\left\|\bar{v}^{bE}_{u'} -{w}^{bE}_k \right\|^2=0$.

Since $T-1=dG, \ d=0,1,...,T-1$ is also an integer multiple of $E$. That is , $T-1=f_{u'}E$. Taking the average over time, we have
\begin{equation}\footnotesize
    \begin{aligned}
        &\frac{1}{T}\sum^{T-1}_{t=0} \sum^{N_u}_{{u'}=1}\sum_{k\in S_{u'}}p_k\mathbb{E}\left\|\bar{v}^t_{u'} -{w}^t_k \right\|^2\\
        &\le 4\eta^2 E^2 A_2^2 B_3 + 4\eta^2 \sum^{N_u}_{{u'}=1}\bar{p}_{u'}  E^2 \hat{\epsilon}^2  \\
        & + 8\eta^2 L^2 \sum^{N_u}_{{u'}=1}\sum_{k\in S_{u'}}p_k \sum_{b=0}^{f_{u'}-1} \sum_{\tau=bE}^{bE+E-1}E^2 \mathbb{E}\left\|  w_k^{\tau} - \bar{v}_{u'}^{\tau}    \right\|^2 \\
        &\le 4\eta^2 E^2 A_2^2 B_3 + 4\eta^2 \sum^{N_u}_{{u'}=1}\bar{p}_{u'}  E^2 \hat{\epsilon}^2  \\
        & + \frac{1}{T}\sum_{\tau=0}^{T-1}8\eta^2 L^2E^2 \sum^{N_u}_{{u'}=1}\sum_{k\in S_{u'}}p_k \mathbb{E}\left\|  w_k^{\tau} - \bar{v}_{u'}^{\tau}    \right\|^2.
    \end{aligned}
\end{equation}

Similar to the last step in deriving the upward parameter MSE, we move the last term to the left side divide both sides by the obtained coefficient term. We also suppose $1-8\eta^2 L^2E^2 >0$, which yields:
\begin{equation}\footnotesize
    \begin{aligned}
        &\frac{1}{T}\sum^{T-1}_{t=0} \sum^{N_u}_{{u'}=1}\sum_{k\in S_{u'}}p_k\mathbb{E}\left\|\bar{v}^t_{u'} -{w}^t_k \right\|^2\\
        &\le \frac{4\eta^2 E^2 A_2^2 B_3}{1-8\eta^2 L^2E^2} + \frac{4\eta^2 \sum^{N_u}_{{u'}=1}\bar{p}_{u'}  E^2 \hat{\epsilon}^2}{1-8\eta^2 L^2E^2}.
    \end{aligned}
    \label{eq12}
\end{equation}

Substituting the result of (\ref{eq12}) into (\ref{eq13}), we obtain the bound on the upward parameter MSE, given by
\begin{equation}\footnotesize
    \begin{aligned}
        &\frac{1}{T}\sum^{T-1}_{t=0}\sum^{N_u}_{u=1}\mathbb{E}\left\|\bar{w}^t -\bar{v}^t_u \right\|^2\\
         &\le \frac{4\eta^2 G^2 A_2^2  (B_3+B_2)}{(1-24\eta^2G^2L^2)}  +  \frac{12\eta^2G^2\epsilon^2}{(1-24\eta^2G^2L^2)} \\
         &+ \frac{4\eta^2G^2L^2}{(1-24\eta^2G^2L^2)}\left( \frac{4\eta^2 E^2 A_2^2 B_3}{1-8\eta^2 L^2E^2} + \frac{4\eta^2 \sum^{N_u}_{{u'}=1}\bar{p}_{u'}  E^2 \hat{\epsilon}^2}{1-8\eta^2 L^2E^2}\right).
    \end{aligned}
    \label{eq14}
\end{equation}
We can obtain the final result on the convergence upper bound by taking the average over time and substitute (\ref{eq12}) and (\ref{eq14}) to (\ref{eq8}).

\begin{equation}\footnotesize
    \begin{aligned}
        &\frac{1}{T}\sum^{T-1}_{t=0}\mathbb{E}\left\| \nabla f(\bar{w}^t)\right\|^2\\
        &\le  \frac{2}{\eta T}\left[\mathbb{E} f(\Bar{w}^{0}) - f^* \right]+ 4B_1^2A_1^2+ 4B_1^2A_2^2 \\
        &+\frac{16\eta^2 G^2 A_2^2 L^2 (B_3+B_2)}{(1-24\eta^2G^2L^2)}  +  \frac{48\eta^2G^2\epsilon^2L^2}{(1-24\eta^2G^2L^2)} \\
        & + \left(1+\frac{4\eta^2G^2L^2}{(1-24\eta^2G^2L^2)} \right)\left( \frac{16\eta^2 E^2 A_2^2 B_3L^2}{1-8\eta^2 L^2E^2} + \frac{16\eta^2 \sum^{N_u}_{{u'}=1}\bar{p}_{u'}  E^2 \hat{\epsilon}^2 L^2}{1-8\eta^2 L^2E^2}\right)
    \end{aligned}
    \label{eq15}
\end{equation}

For $0<\eta<\frac{1}{4\sqrt{3}GL}$, we have
\begin{equation}\footnotesize
    \begin{aligned}
        0 < \frac{4G^2L^2\eta^2}{1-24G^2L^2\eta^2} \le \frac{1}{6}
    \end{aligned}
\end{equation}
and 
\begin{equation}\footnotesize
    \begin{aligned}
        0 < \frac{1}{1-8\eta^2L^2E^2} \le \frac{1}{1-8\eta^2L^2G^2} \le \frac{6}{5}.
    \end{aligned}
\end{equation}
It allows us to define a variable $C$ with a value determined as follows
\begin{equation}\footnotesize
    \begin{aligned}
        C &=\left(1+\frac{4\eta^2G^2L^2}{(1-24\eta^2G^2L^2)} \right)\left( \frac{16\eta^2}{1-8\eta^2 L^2E^2}\right) \\
        &= (1+\frac{1}{6})\times16\times\frac{6}{5}=\frac{112}{5}.
    \end{aligned}
\end{equation}

Substituting the obtained values into (\ref{eq15}):
\begin{equation}\footnotesize
    \begin{aligned}
        &\frac{1}{T}\sum^{T-1}_{t=0}\mathbb{E}\left\| \nabla f(\bar{w}^t)\right\|^2\\
        &\le  \frac{2}{\eta T}\left[\mathbb{E} f(\Bar{w}^{0}) - f^* \right]+ 4B_1^2A_1^2+ 4B_1^2A_2^2 \\
        &+\frac{16\eta^2 G^2 A_2^2 L^2 (B_3+B_2)}{(1-24\eta^2G^2L^2)}  +  \frac{48\eta^2G^2\epsilon^2L^2}{(1-24\eta^2G^2L^2)} \\
        & + \left(1+\frac{4\eta^2G^2L^2}{(1-24\eta^2G^2L^2)} \right)\left( \frac{16\eta^2 E^2 A_2^2 B_3L^2}{1-8\eta^2 L^2E^2} + \frac{16\eta^2 \sum^{N_u}_{{u'}=1}\bar{p}_{u'}  E^2 \hat{\epsilon}^2 L^2}{1-8\eta^2 L^2E^2}\right)\\
        &\le  \frac{2}{\eta T}\left[\mathbb{E} f(\Bar{w}^{0}) - f^* \right]+ 4B_1^2A_1^2+ 4B_1^2A_2^2 \\
        & + 32\eta^2 G^2 A_2^2 L^2 (B_3+B_2) + 96\eta^2G^2\epsilon^2L^2 + \frac{112}{5}\eta^2 E^2 A^2_2 L^2 B_3 \\
        & + \frac{112}{5}\eta^2 L^2\sum_{{u'}=1}^{N_u}\bar{p}_{u'} E^2 \hat{\epsilon}^2\\
        &\le  \frac{2}{\eta T}\left[\mathbb{E} f(\Bar{w}^{0}) - f^* \right]+ 4B_1^2A_1^2+ 4B_1^2A_2^2 \\
        & + 2C\eta^2 G^2 A_2^2 L^2 (B_3+B_2) + 5C\eta^2G^2\epsilon^2L^2 + C\eta^2 E^2 A^2_2 L^2 B_3 \\
        & + C\eta^2 L^2\sum_{{u'}=1}^{N_u}\bar{p}_{u'} E^2 \hat{\epsilon}^2.
    \end{aligned}
\end{equation}

\end{document}